\documentclass[english]{article}
\usepackage{mathpazo}
\usepackage[latin9]{inputenc}
\usepackage[letterpaper]{geometry}
\usepackage{babel}
\usepackage{verbatim}
\usepackage{amsmath}
\usepackage{amssymb}
\usepackage{esint}
\usepackage[unicode=true,
 bookmarks=true,bookmarksnumbered=false,bookmarksopen=false,
 breaklinks=false,pdfborder={0 0 1},backref=false,colorlinks=false]
 {hyperref}

\makeatletter

\providecommand{\tabularnewline}{\\}

\usepackage{babel}

\usepackage[svgnames]{xcolor}
\usepackage{tikz}
\usetikzlibrary{decorations.markings}
\usetikzlibrary{shapes.geometric}
\usetikzlibrary{calc}
\usetikzlibrary{arrows}

\pgfdeclarelayer{edgelayer}
\pgfdeclarelayer{nodelayer}
\pgfsetlayers{edgelayer,nodelayer,main}

\tikzstyle{none}=[inner sep=0pt]

\tikzstyle{Vertex}=[circle,fill=black,draw=black]
\tikzstyle{Node}=[rectangle,fill=blue,draw=blue]
\tikzstyle{Point}=[circle,fill=red,draw=red,scale=0.33]

\tikzstyle{Edge}=[-,draw=black]
\tikzstyle{Link}=[-,dashed,draw=blue]
\tikzstyle{Segment}=[-,dotted,very thick,draw=red]
\tikzstyle{LinkArrow}=[-,draw=blue,
  postaction={decorate},decoration={markings,
  mark=at position .66 with {\arrow{latex}}}]
\tikzstyle{SegmentArrow}=[-,draw=red,
  postaction={decorate},decoration={markings,
  mark=at position .66 with {\arrow{latex}}}]
\def\centerarc [#1] (#2) (#3:#4:#5); { \draw[#1] ($(#2)+({#5*cos(#3)},{#5*sin(#3)})$) arc (#3:#4:#5); }
\newcommand{\midarrow}{\tikz \draw[-triangle 90] (0,0) -- +(.1,0);}
\newcommand{\midarrowop}{\tikz \draw[-triangle 90] (0,0) -- +(-.1,0);}

\renewcommand\[{\begin{equation}}
\renewcommand\]{\end{equation}}

\usepackage[margin=1cm]{caption}

\DeclareMathOperator{\e}{e}

\DeclareMathOperator{\ii}{i}

\usepackage[svgnames]{xcolor}
\usepackage{tikz}
\usetikzlibrary{decorations.markings}
\usetikzlibrary{shapes.geometric}

\pgfdeclarelayer{edgelayer}
\pgfdeclarelayer{nodelayer}
\pgfsetlayers{edgelayer,nodelayer,main}

\tikzstyle{none}=[inner sep=0pt]

\tikzstyle{Vertex}=[circle,fill=black,draw=black]
\tikzstyle{Node}=[rectangle,fill=blue,draw=blue]
\tikzstyle{Point}=[circle,fill=red,draw=red,scale=0.33]

\tikzstyle{Edge}=[-,draw=black]
\tikzstyle{Link}=[-,dashed,draw=blue]
\tikzstyle{Segment}=[-,dotted,very thick,draw=red]
\tikzstyle{LinkArrow}=[-,draw=blue,
  postaction={decorate},decoration={markings,
  mark=at position .66 with {\arrow{latex}}}]
\tikzstyle{SegmentArrow}=[-,draw=red,
  postaction={decorate},decoration={markings,
  mark=at position .66 with {\arrow{latex}}}]

\makeatother

\begin{document}

\global\long\def\A{\mathbf{A}}%
\global\long\def\B{\mathbf{B}}%
\global\long\def\C{\mathbf{C}}%
\global\long\def\D{\mathbf{D}}%
\global\long\def\E{\mathbf{E}}%
\global\long\def\F{\mathbf{F}}%
\global\long\def\G{\mathbf{G}}%
\global\long\def\H{\mathbf{H}}%
\global\long\def\I{\mathbf{I}}%
\global\long\def\J{\mathbf{J}}%
\global\long\def\K{\mathbf{K}}%
\global\long\def\LL{\mathbf{L}}%
\global\long\def\M{\mathbf{M}}%
\global\long\def\N{\mathbf{N}}%
\global\long\def\OO{\mathbf{O}}%
\global\long\def\P{\mathbf{P}}%
\global\long\def\Q{\mathbf{Q}}%
\global\long\def\RR{\mathbf{R}}%
\global\long\def\SS{\mathbf{S}}%
\global\long\def\T{\mathbf{T}}%
\global\long\def\U{\mathbf{U}}%
\global\long\def\V{\mathbf{V}}%
\global\long\def\W{\mathbf{W}}%
\global\long\def\X{\mathbf{X}}%
\global\long\def\Y{\mathbf{Y}}%
\global\long\def\Z{\mathbf{Z}}%

\global\long\def\a{\mathbf{a}}%
\global\long\def\b{\mathbf{b}}%
\global\long\def\c{\mathbf{c}}%
\global\long\def\dd{\mathbf{d}}%
\global\long\def\ee{\mathbf{e}}%
\global\long\def\f{\mathbf{f}}%
\global\long\def\g{\mathbf{g}}%
\global\long\def\h{\mathbf{h}}%
\global\long\def\iii{\mathbf{i}}%
\global\long\def\j{\mathbf{j}}%
\global\long\def\k{\mathbf{k}}%
\global\long\def\l{\boldsymbol{l}}%
\global\long\def\el{\boldsymbol{\ell}}%
\global\long\def\m{\mathbf{m}}%
\global\long\def\n{\mathbf{n}}%
\global\long\def\o{\mathbf{o}}%
\global\long\def\p{\mathbf{p}}%
\global\long\def\q{\mathbf{q}}%
\global\long\def\r{\mathbf{r}}%
\global\long\def\s{\mathbf{s}}%
\global\long\def\t{\mathbf{t}}%
\global\long\def\u{\mathbf{u}}%
\global\long\def\v{\mathbf{v}}%
\global\long\def\w{\mathbf{w}}%
\global\long\def\x{\mathbf{x}}%
\global\long\def\y{\mathbf{y}}%
\global\long\def\z{\mathbf{z}}%

\global\long\def\Ga{\boldsymbol{\Gamma}}%
\global\long\def\De{\boldsymbol{\Delta}}%
\global\long\def\Th{\boldsymbol{\Theta}}%
\global\long\def\La{\boldsymbol{\Lambda}}%
\global\long\def\Xii{\boldsymbol{\Xi}}%
\global\long\def\Pii{\boldsymbol{\Pi}}%
\global\long\def\Si{\boldsymbol{\Sigma}}%
\global\long\def\Ph{\boldsymbol{\Phi}}%
\global\long\def\Ps{\boldsymbol{\Psi}}%
\global\long\def\Om{\boldsymbol{\Omega}}%

\global\long\def\al{\boldsymbol{\alpha}}%
\global\long\def\be{\boldsymbol{\beta}}%
\global\long\def\ga{\boldsymbol{\gamma}}%
\global\long\def\de{\boldsymbol{\delta}}%
\global\long\def\ep{\boldsymbol{\epsilon}}%
\global\long\def\vep{\boldsymbol{\varepsilon}}%
\global\long\def\ze{\boldsymbol{\zeta}}%
\global\long\def\et{\boldsymbol{\eta}}%
\global\long\def\th{\boldsymbol{\theta}}%
\global\long\def\io{\boldsymbol{\iota}}%
\global\long\def\ka{\boldsymbol{\kappa}}%
\global\long\def\la{\boldsymbol{\lambda}}%
\global\long\def\muu{\boldsymbol{\mu}}%
\global\long\def\nuu{\boldsymbol{\nu}}%
\global\long\def\xii{\boldsymbol{\xi}}%
\global\long\def\pii{\boldsymbol{\pi}}%
\global\long\def\rhh{\boldsymbol{\rho}}%
\global\long\def\si{\boldsymbol{\sigma}}%
\global\long\def\ta{\boldsymbol{\tau}}%
\global\long\def\ups{\boldsymbol{\upsilon}}%
\global\long\def\ph{\boldsymbol{\phi}}%
\global\long\def\vph{\boldsymbol{\varphi}}%
\global\long\def\ch{\boldsymbol{\chi}}%
\global\long\def\ps{\boldsymbol{\psi}}%
\global\long\def\om{\boldsymbol{\omega}}%

\global\long\def\AAb{\boldsymbol{\mathcal{A}}}%
\global\long\def\BBb{\boldsymbol{\mathcal{B}}}%
\global\long\def\CCb{\boldsymbol{\mathcal{C}}}%
\global\long\def\DDb{\boldsymbol{\mathcal{D}}}%
\global\long\def\EEb{\boldsymbol{\mathcal{E}}}%
\global\long\def\FFb{\boldsymbol{\mathcal{F}}}%
\global\long\def\GGb{\boldsymbol{\mathcal{G}}}%
\global\long\def\HHb{\boldsymbol{\mathcal{H}}}%
\global\long\def\IIb{\boldsymbol{\mathcal{I}}}%
\global\long\def\JJb{\boldsymbol{\mathcal{J}}}%
\global\long\def\KKb{\boldsymbol{\mathcal{K}}}%
\global\long\def\LLb{\boldsymbol{\mathcal{L}}}%
\global\long\def\MMb{\boldsymbol{\mathcal{M}}}%
\global\long\def\NNb{\boldsymbol{\mathcal{N}}}%
\global\long\def\OOb{\boldsymbol{\mathcal{O}}}%
\global\long\def\PPb{\boldsymbol{\mathcal{P}}}%
\global\long\def\QQb{\boldsymbol{\mathcal{Q}}}%
\global\long\def\RRb{\boldsymbol{\mathcal{R}}}%
\global\long\def\SSb{\boldsymbol{\mathcal{S}}}%
\global\long\def\TTb{\boldsymbol{\mathcal{T}}}%
\global\long\def\UUb{\boldsymbol{\mathcal{U}}}%
\global\long\def\VVb{\boldsymbol{\mathcal{V}}}%
\global\long\def\WWb{\boldsymbol{\mathcal{W}}}%
\global\long\def\XXb{\boldsymbol{\mathcal{X}}}%
\global\long\def\YYb{\boldsymbol{\mathcal{Y}}}%
\global\long\def\ZZb{\boldsymbol{\mathcal{Z}}}%

\global\long\def\Ab{\bar{A}}%
\global\long\def\Bb{\bar{B}}%
\global\long\def\Cb{\bar{C}}%
\global\long\def\Db{\bar{D}}%
\global\long\def\Eb{\bar{E}}%
\global\long\def\Fb{\bar{F}}%
\global\long\def\Gb{\bar{G}}%
\global\long\def\Hb{\bar{H}}%
\global\long\def\Ib{\bar{I}}%
\global\long\def\Jb{\bar{J}}%
\global\long\def\Kb{\bar{K}}%
\global\long\def\Lb{\bar{L}}%
\global\long\def\Mb{\bar{M}}%
\global\long\def\Nb{\bar{N}}%
\global\long\def\Ob{\bar{O}}%
\global\long\def\Pb{\bar{P}}%
\global\long\def\Qb{\bar{Q}}%
\global\long\def\Rb{\bar{R}}%
\global\long\def\Sb{\bar{S}}%
\global\long\def\Tb{\bar{T}}%
\global\long\def\Ub{\bar{U}}%
\global\long\def\Vb{\bar{V}}%
\global\long\def\Wb{\bar{W}}%
\global\long\def\Xb{\bar{X}}%
\global\long\def\Yb{\bar{Y}}%
\global\long\def\Zb{\bar{Z}}%

\global\long\def\ab{\bar{a}}%
\global\long\def\bb{\bar{b}}%
\global\long\def\cb{\bar{c}}%
\global\long\def\db{\bar{d}}%
\global\long\def\eb{\bar{e}}%
\global\long\def\fb{\bar{f}}%
\global\long\def\gb{\bar{g}}%
\global\long\def\hb{\bar{h}}%
\global\long\def\ib{\bar{i}}%
\global\long\def\jb{\bar{j}}%
\global\long\def\kb{\bar{k}}%
\global\long\def\lb{\bar{l}}%
\global\long\def\elb{\bar{\ell}}%
\global\long\def\mb{\bar{m}}%
\global\long\def\nb{\bar{n}}%
\global\long\def\ob{\bar{o}}%
\global\long\def\pb{\bar{p}}%
\global\long\def\qb{\bar{q}}%
\global\long\def\rb{\bar{r}}%
\global\long\def\sb{\bar{s}}%
\global\long\def\tb{\bar{t}}%
\global\long\def\ub{\bar{u}}%
\global\long\def\vb{\bar{v}}%
\global\long\def\wb{\bar{w}}%
\global\long\def\xb{\bar{x}}%
\global\long\def\yb{\bar{y}}%
\global\long\def\zb{\bar{z}}%

\global\long\def\Gab{\bar{\Gamma}}%
\global\long\def\Deb{\bar{\Delta}}%
\global\long\def\Thb{\bar{\Theta}}%
\global\long\def\Lab{\bar{\Lambda}}%
\global\long\def\Xib{\bar{\Xi}}%
\global\long\def\Pib{\bar{\Pi}}%
\global\long\def\Sib{\bar{\Sigma}}%
\global\long\def\Phb{\bar{\Phi}}%
\global\long\def\Psb{\bar{\Psi}}%
\global\long\def\Thb{\bar{\Theta}}%

\global\long\def\alb{\bar{\alpha}}%
\global\long\def\beb{\bar{\beta}}%
\global\long\def\gab{\bar{\gamma}}%
\global\long\def\deb{\bar{\delta}}%
\global\long\def\epb{\bar{\epsilon}}%
\global\long\def\vepb{\bar{\varepsilon}}%
\global\long\def\zeb{\bar{\zeta}}%
\global\long\def\etb{\bar{\eta}}%
\global\long\def\thb{\bar{\theta}}%
\global\long\def\iob{\bar{\iota}}%
\global\long\def\kab{\bar{\kappa}}%
\global\long\def\lab{\bar{\lambda}}%
\global\long\def\mub{\bar{\mu}}%
\global\long\def\nub{\bar{\nu}}%
\global\long\def\xib{\bar{\xi}}%
\global\long\def\pib{\bar{\pi}}%
\global\long\def\rhb{\bar{\rho}}%
\global\long\def\sib{\bar{\sigma}}%
\global\long\def\tab{\bar{\tau}}%
\global\long\def\upb{\bar{\upsilon}}%
\global\long\def\phb{\bar{\phi}}%
\global\long\def\vphb{\bar{\varphi}}%
\global\long\def\chb{\bar{\chi}}%
\global\long\def\psb{\bar{\psi}}%
\global\long\def\omb{\bar{\omega}}%

\global\long\def\adt{\dot{a}}%
\global\long\def\add{\ddot{a}}%
\global\long\def\bd{\dot{b}}%
\global\long\def\bdd{\ddot{b}}%
\global\long\def\cd{\dot{c}}%
\global\long\def\cdd{\ddot{c}}%
\global\long\def\ddd{\dot{d}}%
\global\long\def\dddd{\ddot{d}}%
\global\long\def\ed{\dot{e}}%
\global\long\def\edd{\ddot{e}}%
\global\long\def\fd{\dot{f}}%
\global\long\def\fdd{\ddot{f}}%
\global\long\def\gd{\dot{g}}%
\global\long\def\gdd{\ddot{g}}%
\global\long\def\hd{\dot{h}}%
\global\long\def\hdd{\ddot{h}}%
\global\long\def\kd{\dot{k}}%
\global\long\def\kdd{\ddot{k}}%
\global\long\def\ld{\dot{l}}%
\global\long\def\ldd{\ddot{l}}%
\global\long\def\eld{\dot{\ell}}%
\global\long\def\eldd{\ddot{\ell}}%
\global\long\def\md{\dot{m}}%
\global\long\def\mdd{\ddot{m}}%
\global\long\def\nd{\dot{n}}%
\global\long\def\ndd{\ddot{n}}%
\global\long\def\od{\dot{o}}%
\global\long\def\odd{\ddot{o}}%
\global\long\def\pd{\dot{p}}%
\global\long\def\pdd{\ddot{p}}%
\global\long\def\qd{\dot{q}}%
\global\long\def\qdd{\ddot{q}}%
\global\long\def\rd{\dot{r}}%
\global\long\def\rdd{\ddot{r}}%
\global\long\def\sd{\dot{s}}%
\global\long\def\sdd{\ddot{s}}%
\global\long\def\td{\dot{t}}%
\global\long\def\tdd{\ddot{t}}%
\global\long\def\ud{\dot{u}}%
\global\long\def\udd{\ddot{u}}%
\global\long\def\vd{\dot{v}}%
\global\long\def\vdd{\ddot{v}}%
\global\long\def\wdt{\dot{w}}%
\global\long\def\wdd{\ddot{w}}%
\global\long\def\xd{\dot{x}}%
\global\long\def\xdd{\ddot{x}}%
\global\long\def\yd{\dot{y}}%
\global\long\def\ydd{\ddot{y}}%
\global\long\def\zd{\dot{z}}%
\global\long\def\zdd{\ddot{z}}%

\global\long\def\Adt{\dot{A}}%
\global\long\def\Add{\ddot{A}}%
\global\long\def\Bd{\dot{B}}%
\global\long\def\Bdd{\ddot{B}}%
\global\long\def\Cd{\dot{C}}%
\global\long\def\Cdd{\ddot{C}}%
\global\long\def\Dd{\dot{D}}%
\global\long\def\Ddd{\ddot{D}}%
\global\long\def\Ed{\dot{E}}%
\global\long\def\Edd{\ddot{E}}%
\global\long\def\Fd{\dot{F}}%
\global\long\def\Fdd{\ddot{F}}%
\global\long\def\Gd{\dot{G}}%
\global\long\def\Gdd{\ddot{G}}%
\global\long\def\Hd{\dot{H}}%
\global\long\def\Hdd{\ddot{H}}%
\global\long\def\Id{\dot{I}}%
\global\long\def\Idd{\ddot{I}}%
\global\long\def\Jd{\dot{J}}%
\global\long\def\Jdd{\ddot{J}}%
\global\long\def\Kd{\dot{K}}%
\global\long\def\Kdd{\ddot{K}}%
\global\long\def\Ld{\dot{L}}%
\global\long\def\Ldd{\ddot{L}}%
\global\long\def\Md{\dot{M}}%
\global\long\def\Mdd{\ddot{M}}%
\global\long\def\Nd{\dot{N}}%
\global\long\def\Ndd{\ddot{N}}%
\global\long\def\Od{\dot{O}}%
\global\long\def\Odd{\ddot{O}}%
\global\long\def\Pd{\dot{P}}%
\global\long\def\Pdd{\ddot{P}}%
\global\long\def\Qd{\dot{Q}}%
\global\long\def\Qdd{\ddot{Q}}%
\global\long\def\Rd{\dot{R}}%
\global\long\def\Rdd{\ddot{R}}%
\global\long\def\Sd{\dot{S}}%
\global\long\def\Sdd{\ddot{S}}%
\global\long\def\Td{\dot{T}}%
\global\long\def\Tdd{\ddot{T}}%
\global\long\def\Ud{\dot{U}}%
\global\long\def\Udd{\ddot{U}}%
\global\long\def\Vd{\dot{R}}%
\global\long\def\Vdd{\ddot{R}}%
\global\long\def\Wd{\dot{W}}%
\global\long\def\Wdd{\ddot{W}}%
\global\long\def\Xd{\dot{X}}%
\global\long\def\Xdd{\ddot{X}}%
\global\long\def\Yd{\dot{Y}}%
\global\long\def\Ydd{\ddot{Y}}%
\global\long\def\Zd{\dot{Z}}%
\global\long\def\Zdd{\ddot{Z}}%

\global\long\def\Gad{\dot{\Gamma}}%
\global\long\def\Gadd{\ddot{\Gamma}}%
\global\long\def\Ded{\dot{\Delta}}%
\global\long\def\Dedd{\ddot{\Delta}}%
\global\long\def\Thd{\dot{\Theta}}%
\global\long\def\Thdd{\ddot{\Theta}}%
\global\long\def\Lad{\dot{\Lambda}}%
\global\long\def\Ladd{\ddot{\Lambda}}%
\global\long\def\Xid{\dot{\Xi}}%
\global\long\def\Xidd{\ddot{\Xi}}%
\global\long\def\Pid{\dot{\Pi}}%
\global\long\def\Pidd{\ddot{\Pi}}%
\global\long\def\Sid{\dot{\Sigma}}%
\global\long\def\Sidd{\ddot{\Sigma}}%
\global\long\def\Phd{\dot{\Phi}}%
\global\long\def\Phdd{\ddot{\Phi}}%
\global\long\def\Psd{\dot{\Psi}}%
\global\long\def\Psdd{\ddot{\Psi}}%
\global\long\def\Thd{\dot{\Theta}}%
\global\long\def\Thdd{\ddot{\Theta}}%

\global\long\def\ald{\dot{\alpha}}%
\global\long\def\aldd{\ddot{\alpha}}%
\global\long\def\bed{\dot{\beta}}%
\global\long\def\bedd{\ddot{\beta}}%
\global\long\def\gad{\dot{\gamma}}%
\global\long\def\gadd{\ddot{\gamma}}%
\global\long\def\ded{\dot{\delta}}%
\global\long\def\dedd{\ddot{\delta}}%
\global\long\def\epd{\dot{\epsilon}}%
\global\long\def\epdd{\ddot{\epsilon}}%
\global\long\def\vepd{\dot{\varepsilon}}%
\global\long\def\vepdd{\ddot{\varepsilon}}%
\global\long\def\zed{\dot{\zeta}}%
\global\long\def\zedd{\ddot{\zeta}}%
\global\long\def\etd{\dot{\eta}}%
\global\long\def\etdd{\ddot{\eta}}%
\global\long\def\thd{\dot{\theta}}%
\global\long\def\thdd{\ddot{\theta}}%
\global\long\def\iod{\dot{\iota}}%
\global\long\def\iodd{\ddot{\iota}}%
\global\long\def\kad{\dot{\kappa}}%
\global\long\def\kadd{\ddot{\kappa}}%
\global\long\def\lad{\dot{\lambda}}%
\global\long\def\ladd{\ddot{\lambda}}%
\global\long\def\mud{\dot{\mu}}%
\global\long\def\mudd{\ddot{\mu}}%
\global\long\def\nud{\dot{\nu}}%
\global\long\def\nudd{\ddot{\nu}}%
\global\long\def\xid{\dot{\xi}}%
\global\long\def\xidd{\ddot{\xi}}%
\global\long\def\pid{\dot{\pi}}%
\global\long\def\pidd{\ddot{\pi}}%
\global\long\def\rhod{\dot{\rho}}%
\global\long\def\rhodd{\ddot{\rho}}%
\global\long\def\sid{\dot{\sigma}}%
\global\long\def\sidd{\ddot{\sigma}}%
\global\long\def\tad{\dot{\tau}}%
\global\long\def\tadd{\ddot{\tau}}%
\global\long\def\upd{\dot{\upsilon}}%
\global\long\def\updd{\ddot{\upsilon}}%
\global\long\def\phd{\dot{\phi}}%
\global\long\def\phdd{\ddot{\phi}}%
\global\long\def\vpd{\dot{\varphi}}%
\global\long\def\vpdd{\ddot{\varphi}}%
\global\long\def\chd{\dot{\chi}}%
\global\long\def\chdd{\ddot{\chi}}%
\global\long\def\psd{\dot{\psi}}%
\global\long\def\psdd{\ddot{\psi}}%
\global\long\def\omd{\dot{\omega}}%
\global\long\def\omdd{\ddot{\omega}}%

\global\long\def\BBA{\mathbb{A}}%
\global\long\def\BBB{\mathbb{B}}%
\global\long\def\BBC{\mathbb{C}}%
\global\long\def\BBD{\mathbb{D}}%
\global\long\def\BBE{\mathbb{E}}%
\global\long\def\BBF{\mathbb{F}}%
\global\long\def\BBG{\mathbb{G}}%
\global\long\def\BBH{\mathbb{H}}%
\global\long\def\BBI{\mathbb{I}}%
\global\long\def\BBJ{\mathbb{J}}%
\global\long\def\BBK{\mathbb{K}}%
\global\long\def\BBL{\mathbb{L}}%
\global\long\def\BBM{\mathbb{M}}%
\global\long\def\BBN{\mathbb{N}}%
\global\long\def\BBO{\mathbb{O}}%
\global\long\def\BBP{\mathbb{P}}%
\global\long\def\BBQ{\mathbb{Q}}%
\global\long\def\BBR{\mathbb{R}}%
\global\long\def\BBS{\mathbb{S}}%
\global\long\def\BBT{\mathbb{T}}%
\global\long\def\BBU{\mathbb{U}}%
\global\long\def\BBV{\mathbb{V}}%
\global\long\def\BBW{\mathbb{W}}%
\global\long\def\BBX{\mathbb{X}}%
\global\long\def\BBY{\mathbb{Y}}%
\global\long\def\BBZ{\mathbb{Z}}%

\global\long\def\AA{\mathcal{A}}%
\global\long\def\BB{\mathcal{B}}%
\global\long\def\CC{\mathcal{C}}%
\global\long\def\DD{\mathcal{D}}%
\global\long\def\EE{\mathcal{E}}%
\global\long\def\FF{\mathcal{F}}%
\global\long\def\GG{\mathcal{G}}%
\global\long\def\HH{\mathcal{H}}%
\global\long\def\II{\mathcal{I}}%
\global\long\def\JJ{\mathcal{J}}%
\global\long\def\KK{\mathcal{K}}%
\global\long\def\LLL{\mathcal{L}}%
\global\long\def\MM{\mathcal{M}}%
\global\long\def\NN{\mathcal{N}}%
\global\long\def\OOO{\mathcal{O}}%
\global\long\def\PP{\mathcal{P}}%
\global\long\def\QQ{\mathcal{Q}}%
\global\long\def\RRR{\mathcal{R}}%
\global\long\def\SSS{\mathcal{S}}%
\global\long\def\TT{\mathcal{T}}%
\global\long\def\UU{\mathcal{U}}%
\global\long\def\VV{\mathcal{V}}%
\global\long\def\WW{\mathcal{W}}%
\global\long\def\XX{\mathcal{X}}%
\global\long\def\YY{\mathcal{Y}}%
\global\long\def\ZZ{\mathcal{Z}}%

\global\long\def\At{\tilde{A}}%
\global\long\def\Bt{\tilde{B}}%
\global\long\def\Ct{\tilde{C}}%
\global\long\def\Dt{\tilde{D}}%
\global\long\def\Et{\tilde{E}}%
\global\long\def\Ft{\tilde{F}}%
\global\long\def\Gt{\tilde{G}}%
\global\long\def\Ht{\tilde{H}}%
\global\long\def\It{\tilde{I}}%
\global\long\def\Jt{\tilde{J}}%
\global\long\def\Kt{\tilde{K}}%
\global\long\def\Lt{\tilde{L}}%
\global\long\def\Mt{\tilde{M}}%
\global\long\def\Nt{\tilde{N}}%
\global\long\def\Ot{\tilde{O}}%
\global\long\def\Pt{\tilde{P}}%
\global\long\def\Qt{\tilde{Q}}%
\global\long\def\Rt{\tilde{R}}%
\global\long\def\St{\tilde{S}}%
\global\long\def\Tt{\tilde{T}}%
\global\long\def\Ut{\tilde{U}}%
\global\long\def\Vt{\tilde{V}}%
\global\long\def\Wt{\tilde{W}}%
\global\long\def\Xt{\tilde{X}}%
\global\long\def\Yt{\tilde{Y}}%
\global\long\def\Zt{\tilde{Z}}%

\global\long\def\at{\tilde{a}}%
\global\long\def\bt{\tilde{b}}%
\global\long\def\ct{\tilde{c}}%
\global\long\def\dt{\tilde{d}}%
\global\long\def\eet{\tilde{e}}%
\global\long\def\ft{\tilde{f}}%
\global\long\def\gt{\tilde{g}}%
\global\long\def\hht{\tilde{h}}%
\global\long\def\it{\tilde{i}}%
\global\long\def\jt{\tilde{j}}%
\global\long\def\kt{\tilde{k}}%
\global\long\def\lt{\tilde{l}}%
\global\long\def\elt{\tilde{\ell}}%
\global\long\def\mt{\tilde{m}}%
\global\long\def\nt{\tilde{n}}%
\global\long\def\ot{\tilde{o}}%
\global\long\def\pt{\tilde{p}}%
\global\long\def\qt{\tilde{q}}%
\global\long\def\rt{\tilde{r}}%
\global\long\def\st{\tilde{s}}%
\global\long\def\tt{\tilde{t}}%
\global\long\def\ut{\tilde{u}}%
\global\long\def\vt{\tilde{v}}%
\global\long\def\wt{\tilde{w}}%
\global\long\def\xt{\tilde{x}}%
\global\long\def\yt{\tilde{y}}%
\global\long\def\zt{\tilde{z}}%

\global\long\def\mfA{\mathfrak{A}}%
\global\long\def\mfB{\mathfrak{B}}%
\global\long\def\mfC{\mathfrak{C}}%
\global\long\def\mfD{\mathfrak{D}}%
\global\long\def\mfE{\mathfrak{E}}%
\global\long\def\mfF{\mathfrak{F}}%
\global\long\def\mfG{\mathfrak{G}}%
\global\long\def\mfH{\mathfrak{H}}%
\global\long\def\mfI{\mathfrak{I}}%
\global\long\def\mfJ{\mathfrak{J}}%
\global\long\def\mfK{\mathfrak{K}}%
\global\long\def\mfL{\mathfrak{L}}%
\global\long\def\mfM{\mathfrak{M}}%
\global\long\def\mfN{\mathfrak{N}}%
\global\long\def\mfO{\mathfrak{O}}%
\global\long\def\mfP{\mathfrak{P}}%
\global\long\def\mfQ{\mathfrak{Q}}%
\global\long\def\mfR{\mathfrak{R}}%
\global\long\def\mfS{\mathfrak{S}}%
\global\long\def\mfT{\mathfrak{T}}%
\global\long\def\mfU{\mathfrak{U}}%
\global\long\def\mfV{\mathfrak{V}}%
\global\long\def\mfW{\mathfrak{W}}%
\global\long\def\mfX{\mathfrak{X}}%
\global\long\def\mfY{\mathfrak{Y}}%
\global\long\def\mfZ{\mathfrak{Z}}%
\global\long\def\mfa{\mathfrak{a}}%
\global\long\def\mfb{\mathfrak{b}}%
\global\long\def\mfc{\mathfrak{c}}%
\global\long\def\mfd{\mathfrak{d}}%
\global\long\def\mfe{\mathfrak{e}}%
\global\long\def\mff{\mathfrak{f}}%
\global\long\def\mfg{\mathfrak{g}}%
\global\long\def\mfh{\mathfrak{h}}%
\global\long\def\mfi{\mathfrak{i}}%
\global\long\def\mfj{\mathfrak{j}}%
\global\long\def\mfk{\mathfrak{k}}%
\global\long\def\mfl{\mathfrak{l}}%
\global\long\def\mfm{\mathfrak{m}}%
\global\long\def\mfn{\mathfrak{n}}%
\global\long\def\mfo{\mathfrak{o}}%
\global\long\def\mfp{\mathfrak{p}}%
\global\long\def\mfq{\mathfrak{q}}%
\global\long\def\mfr{\mathfrak{r}}%
\global\long\def\mfs{\mathfrak{s}}%
\global\long\def\mft{\mathfrak{t}}%
\global\long\def\mfu{\mathfrak{u}}%
\global\long\def\mfv{\mathfrak{v}}%
\global\long\def\mfw{\mathfrak{w}}%
\global\long\def\mfx{\mathfrak{x}}%
\global\long\def\mfy{\mathfrak{y}}%
\global\long\def\mfz{\mathfrak{z}}%

\global\long\def\d{\mathrm{d}}%
\global\long\def\DDD{\mathrm{D}}%
\global\long\def\EEE{\mathrm{E}}%
\global\long\def\i{\ii}%
\global\long\def\MMM{\mathrm{M}}%
\global\long\def\OOOO{\mathrm{O}}%
\global\long\def\RRRR{\mathrm{R}}%
\global\long\def\TTT{\mathrm{T}}%
\global\long\def\UUU{\mathrm{U}}%

\global\long\def\GL{\mathrm{GL}}%
\global\long\def\ISU{\mathrm{ISU}}%
\global\long\def\ISUT{\mathrm{ISU}\left(2\right)}%
\global\long\def\SL{\mathrm{SL}}%
\global\long\def\SO{\mathrm{SO}}%
\global\long\def\SOH{\mathrm{SO}\left(3\right)}%
\global\long\def\SOT{\mathrm{SO}\left(2\right)}%
\global\long\def\Sp{\mathrm{Sp}}%
\global\long\def\SU{\mathrm{SU}}%
\global\long\def\SUT{\mathrm{SU}\left(2\right)}%
\global\long\def\UO{\mathrm{U}\left(1\right)}%
\global\long\def\gl{\mathfrak{gl}}%
\global\long\def\sl{\mathfrak{sl}}%
\global\long\def\sso{\mathfrak{so}}%
\global\long\def\soh{\mathfrak{so}\left(3\right)}%
\global\long\def\su{\mathfrak{su}}%
\global\long\def\sut{\mathfrak{su}\left(2\right)}%
\global\long\def\isut{\mathfrak{isu}\left(2\right)}%

\global\long\def\so{\Rightarrow}%
\global\long\def\os{\Leftarrow}%
\global\long\def\to{\rightarrow}%
\global\long\def\ot{\leftarrow}%
\global\long\def\soo{\Longrightarrow}%
\global\long\def\oos{\Longleftarrow}%
\global\long\def\too{\longrightarrow}%
\global\long\def\oot{\longleftarrow}%
\global\long\def\sos{\Leftrightarrow}%
\global\long\def\tot{\leftrightarrow}%
\global\long\def\soos{\Longleftrightarrow}%
\global\long\def\toot{\longleftrightarrow}%
\global\long\def\mt{\mapsto}%
\global\long\def\mtt{\longmapsto}%
\global\long\def\dn{\downarrow}%
\global\long\def\up{\uparrow}%
\global\long\def\updn{\updownarrow}%
\global\long\def\sea{\searrow}%
\global\long\def\nea{\nearrow}%
\global\long\def\nwa{\nwarrow}%
\global\long\def\swa{\swarrow}%
\global\long\def\hk{\hookrightarrow}%
\global\long\def\kh{\hookleftarrow}%
\global\long\def\soosp{\quad\Longrightarrow\quad}%
\global\long\def\oossp{\quad\Longleftarrow\quad}%
\global\long\def\soossp{\quad\Longleftrightarrow\quad}%

\global\long\def\multibrl#1{\left(#1\right.}%
\global\long\def\multibrr#1{\left.#1\right)}%
\global\long\def\multisql#1{\left[#1\right.}%
\global\long\def\multisqr#1{\left.#1\right]}%
\global\long\def\multicul#1{\left\{  #1\right.}%
\global\long\def\multicur#1{\left.#1\right\}  }%

\global\long\def\bl{\bigl|}%
\global\long\def\bll{\Bigl|}%
\global\long\def\blll{\biggl|}%
\global\long\def\bllll{\Biggl|}%

\global\long\def\ma#1#2{\left\langle #1\thinspace\middle|\thinspace#2\right\rangle }%
\global\long\def\mma#1#2#3{\left\langle #1\thinspace\middle|\thinspace#2\thinspace\middle|\thinspace#3\right\rangle }%
\global\long\def\mc#1#2{\left\{  #1\thinspace\middle|\thinspace#2\right\}  }%
\global\long\def\mmc#1#2#3{\left\{  #1\thinspace\middle|\thinspace#2\thinspace\middle|\thinspace#3\right\}  }%
\global\long\def\mr#1#2{\left(#1\thinspace\middle|\thinspace#2\right) }%
\global\long\def\mmr#1#2#3{\left(#1\thinspace\middle|\thinspace#2\thinspace\middle|\thinspace#3\right)}%

\global\long\def\pr{\parallel}%
\global\long\def\xx{\times}%
\global\long\def\dg{\lyxmathsym{\textdegree}}%
\global\long\def\sp{,\qquad}%
\global\long\def\sq{\square}%
\global\long\def\pt{\propto}%
\global\long\def\lrc{\lrcorner\thinspace}%
\global\long\def\pexp{\overrightarrow{\exp}}%
\global\long\def\dui#1#2#3{#1_{#2}{}^{#3}}%
\global\long\def\udi#1#2#3{#1^{#2}{}_{#3}}%
\global\long\def\pab{\bar{\partial}}%
\global\long\def\zr{\mathbf{0}}%
\global\long\def\on{\mathbf{1}}%
\global\long\def\na{\boldsymbol{\nabla}}%
\global\long\def\hf{\frac{1}{2}}%
\global\long\def\trd{\frac{1}{3}}%
\global\long\def\fr{\frac{1}{4}}%
\global\long\def\ap{\approx}%
\global\long\def\eqm{\overset{?}{=}}%
\global\long\def\fa{\forall}%
\global\long\def\ex{\exists}%
\global\long\def\XXt{\tilde{\mathbf{X}}}%

\title{Dual 2+1D Loop Quantum Gravity on the Edge}
\author{\textbf{Barak Shoshany}\thanks{bshoshany@perimeterinstitute.ca}\\
\emph{ Perimeter Institute for Theoretical Physics,}\\
\emph{31 Caroline Street North, Waterloo, Ontario, Canada, N2L 2Y5}}
\maketitle
\begin{abstract}
In a recent paper, we introduced a new discretization scheme for gravity
in 2+1 dimensions. Starting from the continuum theory, this new scheme
allowed us to rigorously obtain the discrete phase space of loop gravity,
coupled to particle-like ``edge mode'' degrees of freedom. In this
work, we expand on that result by considering the most general choice
of integration during the discretization process. We obtain a family
of polarizations of the discrete phase space. In particular, one member
of this family corresponds to the usual loop gravity phase space,
while another corresponds to a new polarization, dual to the usual
one in several ways. We study its properties, including the relevant
constraints and the symmetries they generate. Furthermore, we motivate
a relation between the dual polarization and teleparallel gravity.
\end{abstract}
\tableofcontents{}

\section{Introduction}

The theory of general relativity famously describes gravity as a result
of the curvature of spacetime itself. Furthermore, the geometry of
spacetime is assumed to be torsionless by employing the \emph{Levi-Civita
connection}, which is torsionless by definition. While this is the
most popular formulation, there exists an alternative but mathematically
equivalent formulation called \emph{teleparallel gravity}\cite{2005physics...3046U,Maluf:2013gaa,Ferraro:2016wht},
differing from general relativity only by a boundary term. In this
formulation, one instead uses the \emph{Weitzenböck connection}, which
is flat by definition. The gravitational degrees of freedom are then
encoded in the torsion of the spacetime geometry.

\emph{Loop quantum gravity}\cite{Rovelli:2004tv}\emph{ }is a popular
approach towards the formulation of a consistent and physically relevant
theory of quantum gravity. In the canonical version of the theory\cite{thiemann_2007},
one starts by rewriting general relativity in the Hamiltonian formulation
and quantizing using the familiar Dirac procedure\cite{Henneaux:1992ig}.
One finds a fully constrained system, that is, the Hamiltonian is
simply a sum of constraints.

In 2+1 spacetime dimensions, where gravity is topological\cite{Carlip:1998uc},
there are two such constraints:
\begin{itemize}
\item The Gauss (or torsion) constraint, which imposes zero torsion everywhere,
\item The curvature (or flatness) constraint, which imposes zero curvature
everywhere.
\end{itemize}
In the classical theory, it does not matter which constraint is imposed
first. However, in the quantum theory, it does matter, since the Hilbert
space is defined in terms of representations of the symmetries generated
by the constraints.. The first constraint that we impose is used to
define the kinematics of the theory, while the second constraint will
encode the dynamics. Thus, it seems natural to identify general relativity
with the quantization in which the Gauss constraint is imposed first,
and teleparallel gravity with that in which the curvature constraint
is imposed first.

Indeed, in loop quantum gravity, which is a quantization of general
relativity, the Gauss constraint is imposed first. This is done by
selecting, as the basis for the kinematical Hilbert space, the spin
network basis\cite{Ashtekar:1991kc} of rotation-invariant states.
Then, the curvature constraint is imposed at the dynamical level in
order to obtain the Hilbert space of physical states.

In \cite{clement}, an alternative choice was suggested where the
order of constraints is reversed. The curvature constraint is imposed
first by employing the \emph{group network }basis of translation-invariant
states, and the Gauss constraint is the one which encodes the dynamics.
This \emph{dual loop quantum gravity} quantization is the quantum
counterpart of teleparallel gravity, and could be used to study the
dual vacua proposed in \cite{Dittrich:2014wpa,Dittrich:2014wda}.

In this paper, we will only deal with the classical theory. We will
explore a family of discretizations which includes, in particular,
three cases of interest:
\begin{itemize}
\item The loop gravity phase space, which is the classical version of the
spin network basis\cite{Freidel:2010aq}. This case was studied in
detail in our paper \cite{FirstPaper} and is related to 2+1D general
relativity. We will provide a more rigorous derivation of some results,
in particular the discrete curvature constraint, and additional subtle
details which were missing in our initial treatment. The phase space
obtained in this case contains the phase space of spin networks, plus
curvature and torsion excitations corresponding to edge modes which
do not cancel.
\item Dual loop gravity, which is the classical version of the group network
basis. This case was first studied in \cite{Dupuis:2017otn} in the
simple case where there are no curvature or torsion excitations. It
is intuitively related to teleparallel gravity. Here, we will study
this case carefully, incorporating the edge modes as was done in \cite{FirstPaper}
for the loop gravity case. We will rigorously derive the discrete
constraints and the symmetry transformations they generate. The resulting
phase space will contain the phase space of group networks, plus the
same curvature and torsion obtained in the previous case.
\item A mixed phase space, containing both loop gravity and its dual, which
is intuitively related to Chern-Simons theory\cite{Horowitz:1989ng},
as we will motivate below. In this case our formalism should be related
to existing results\cite{Alekseev:1993rj,Alekseev:1994pa,Alekseev:1994au,Meusburger:2003hc,Meusburger:2005mg,Meusburger:2003ta,Meusburger:2008dc}.
\end{itemize}
Crucial to our formalism is the separation of discretization into
two steps. This procedure was first utilized, in the 3+1-dimensional
case, in \cite{Freidel:2011ue,Freidel:2013bfa}, but without considering
any curvature and torsion. The steps are as follows:
\begin{enumerate}
\item Subdivision, or decomposition into subsystems. More precisely, we
define a \emph{cellular decomposition}\footnote{The cells in this decomposition can take any shape.}\emph{
}on our 2-dimensional spatial manifold. This structure has a dual
structure, which as we will see, will be the spin network graph.
\item Truncation, or coarse-graining of the subsystems. In this step, we
assume that there is arbitrary curvature and torsion inside each loop
of the spin network. We then ``compress'' the information about
the geometry into a single point, or vertex, inside the loop. Since
the only way to probe the geometry is by looking at the holonomies
and fluxes on the loops of the spin network, the observables before
and after this truncation are the same.
\end{enumerate}
The \emph{edge modes}, mentioned earlier, are the final piece of our
formalism. When discretizing gauge theories, and gravity in particular,
a major problem is preserving gauge invariance despite the discreteness
of the resulting theory. The presence of boundaries can be shown to
introduce new degrees of freedom, called edge modes\cite{Donnelly:2016auv,Geiller:2017xad,Geiller:2017whh}\footnote{See also \cite{Rovelli:2013fga} for a more intuitive discussion and
\cite{Freidel:2015gpa,Freidel:2016bxd,Freidel:2018pvm} for the case
of 3+1-dimensional gravity.}, which may be used to \emph{dress }observables and make them gauge-invariant.
These edge modes are associated to new boundary symmetries, which
transform them and control the gluing map between subsystems.

As we will see below, the edge modes at the boundaries of the cells
in our cellular decomposition will mostly cancel with the edge modes
on the boundaries of the adjacent cells. However, there will also
be edge modes at the vertices of the cells, which will not have anything
to cancel with. These degrees of freedom will survive the discretization
process, and introduce a particle-like phase space\cite{Kirillov,Rempel:2015foa}
for the curvature and torsion, which we then interpret as mass and
spin respectively.

One might expect that the geometry will be encoded in the constraints
alone, by imposing that a loop of holonomies sees the curvature inside
it and a loop of fluxes sees the torsion inside it. As we will see,
while the constraints do indeed encode the geometry, the presence
of the edge modes enforces the inclusion of the curvature and torsion
themselves as additional phase space variables.

\subsection{Basic Definitions and Notation}

Consider a group $G\ltimes\mfg^{*}\cong T^{*}G$, which is\footnote{The notation $T^{*}G$ signifies the \emph{cotangent bundle }of $G$.}
a generalization of the Euclidean or Poincaré group. One possible
option is
\[
\ISUT\cong\SUT\ltimes\BBR^{3},
\]
but we will keep it general. The algebra for this group is given by
\[
\left[\P_{i},\P_{j}\right]=0\sp\left[\J_{i},\J_{j}\right]=\dui f{ij}k\J_{k}\sp\left[\J_{i},\P_{j}\right]=\dui f{ij}k\P_{k},
\]
where $\dui f{ij}k$ are the structure constants\footnote{They satisfy anti-symmetry $\dui f{ij}k=-\dui f{ji}k$ and the Jacobi
identity $\dui f{[ij}l\dui f{k]l}m=0$. For $\sut$ we have $\dui f{ij}k=\dui{\epsilon}{ij}k$
where $\dui{\epsilon}{ij}k$ is the Levi-Civita symbol.}. The algebra indices $i,j,k$ go from 1 to $\dim\mfg$, which is
e.g. 3 for $\sut$. The generators $\J_{i}$ are the \emph{rotation
}generators, and they correspond to a non-Abelian group $G$, while
the generators $\P_{i}$ are the \emph{translation }generators, and
they correspond to an Abelian normal subgroup $\mfg^{*}$.

Notation-wise, all Lie algebra elements and Lie-algebra-valued forms
will be written in \textbf{bold font} to distinguish them from Lie
group element or Lie-group-valued forms. Furthermore, we will use
calligraphic font for $G\ltimes\mfg^{*}$ or $\mfg\oplus\mfg^{*}$-valued
forms (which will rarely be of interest) and Roman font for $G$,
$\mfg$ or $\mfg^{*}$-valued forms.

Given any two Lie-algebra-valued forms $\A,\B$ of degrees $\deg\A$
and $\deg\B$ respectively, we define the \emph{graded commutator}:
\[
\left[\A,\B\right]\equiv\A\wedge\B-\left(-1\right)^{\deg\A\deg\B}\B\wedge\A.
\]
We also define a \emph{dot (inner) product}, also known as the \emph{Killing
form}, on the generators as follows:
\begin{equation}
\J_{i}\cdot\P_{j}=\delta_{ij}\sp\J_{i}\cdot\J_{j}=\P_{i}\cdot\P_{j}=0.\label{eq:dot-product}
\end{equation}
Given two Lie-algebra-valued forms, the dot product is defined to
include a wedge product. Thus, if $\A\equiv A^{i}\J_{i}$ is a pure
rotation and $\B\equiv B^{i}\P_{i}$ is a pure translation, which
will usually be the case\footnote{In the general case, which will only be relevant for our discussion
of Chern-Simons theory in the next subsection, for $\mfg\oplus\mfg^{*}$-valued
forms $\AAb\equiv\AA_{J}^{i}\J_{i}+\AA_{P}^{i}\P_{i}$ and $\BBb\equiv\BB_{J}^{i}\J_{i}+\BB_{P}^{i}\P_{i}$
we have
\[
\AAb\cdot\BBb=\delta_{ij}\left(\AA_{J}^{i}\wedge\BB_{P}^{j}+\AA_{P}^{i}\wedge\BB_{J}^{j}\right).
\]
}, we have
\[
\A\cdot\B\equiv A^{i}\wedge B_{i}.
\]
Finally, in addition to the exterior derivative $\d$ and the interior
product $\iota$ on spacetime, we introduce a \emph{variational exterior
derivative} $\delta$ and a \emph{variational interior product} $I$
on field space. These operators act analogously to $\d$ and $\iota$,
and in particular they are nilpotent, e.g. $\delta^{2}=0$, and satisfy
the graded Leibniz rule. Degrees of differential forms are counted
with respect to spacetime and field space separately; for example,
if $f$ is a 0-form then $\d\delta f$ is a 1-form on spacetime, due
to $\d$, and independently also a 1-form on field space, due to $\delta$.
The dot product defined above also includes an implicit wedge product
with respect to field-space forms, such that e.g. $\delta\f\cdot\delta\g=-\delta\g\cdot\delta\f$
if $\f$ and $\g$ are 0-forms.

\subsection{The Chern-Simons Action and 2+1D Gravity}

Let $M$ be a 2+1-dimensional spacetime manifold and let $\Sigma$
be a 2-dimensional spatial manifold such that $M=\Sigma\xx\BBR$ where
$\BBR$ represents time. Let us also define the \emph{Chern-Simons
connection 1-form} $\AAb$, valued in $\mfg\oplus\mfg^{*}$:
\begin{equation}
\AAb\equiv\A+\E\equiv A^{i}\J_{i}+E^{i}\P_{i},\label{eq:CS-split}
\end{equation}
where $\A\equiv A^{i}\J_{i}$ is the $\mfg$-valued connection 1-form
and $\E\equiv E^{i}\P_{i}$ is the $\mfg^{*}$-valued frame field
1-form. The $\mfg\oplus\mfg^{*}$-valued \emph{curvature 2-form} $\FFb$
is then defined as:
\[
\FFb\equiv\d\AAb+\hf\left[\AAb,\AAb\right],
\]
and it may be split into
\begin{equation}
\FFb\equiv\F+\T\equiv F^{i}\J_{i}+T^{i}\P_{i},\label{eq:CS-con-split}
\end{equation}
where $\F\equiv F^{i}\J_{i}$ is the $\mfg$-valued curvature 2-form
and $\T\equiv T^{i}\P_{i}$ is the $\mfg^{*}$-valued torsion 2-form,
and they are defined in terms of $\A$ and $\E$ as
\[
\F\equiv\d\A+\hf\left[\A,\A\right]\sp\T\equiv\d_{\A}\E\equiv\d\E+\left[\A,\E\right],
\]
where $\d_{\A}\equiv\d+\left[\A,\cdot\right]$ is the covariant exterior
derivative.

In our notation, the Chern-Simons action is given by
\[
S\left[\AAb\right]=\hf\int_{M}\AAb\cdot\left(\d\AAb+\trd\left[\AAb,\AAb\right]\right),
\]
and its variation is
\[
\delta S\left[\AAb\right]=\int_{M}\left(\FFb\cdot\delta\AAb-\hf\d\left(\AAb\cdot\delta\AAb\right)\right).
\]
From this we can read the equation of motion
\begin{equation}
\FFb=0,\label{eq:eom-CS}
\end{equation}
 and, from the boundary term, the symplectic potential
\begin{equation}
\Theta\left[\AAb\right]\equiv-\hf\int_{\Sigma}\AAb\cdot\delta\AAb,\label{eq:Theta-CS}
\end{equation}
which gives us the symplectic form
\[
\Omega\left[\AAb\right]\equiv\delta\Theta\left[\AAb\right]=-\hf\int_{\Sigma}\delta\AAb\cdot\delta\AAb.
\]
Furthermore, we can write the action\footnote{Here we use the following identities, derived from the properties
of the dot product \eqref{eq:dot-product}and the graded commutator:
\[
\A\cdot\d\A=\E\cdot\d\E=\left[\E,\E\right]=\A\cdot\left[\A,\A\right]=\E\cdot\left[\A,\E\right]=0.
\]
} in terms of $\A$ and $\E$:
\begin{equation}
S\left[\A,\E\right]=\int_{M}\left(\E\cdot\F-\hf\d\left(\A\cdot\E\right)\right).\label{eq:action-dec}
\end{equation}
This is the action for 2+1D gravity, with an additional boundary term
(which is usually disregarded by assuming $M$ has no boundary). Using
the identity $\delta\F=\d_{\A}\delta\A$, we find the variation of
the action is
\[
\delta S\left[\A,\E\right]=\int_{M}\left(\F\cdot\delta\E+\T\cdot\delta\A-\hf\d\left(\E\cdot\delta\A+\A\cdot\delta\E\right)\right),
\]
and thus we see that the equations of motion are
\begin{equation}
\F=0\sp\T=0,\label{eq:eom-dec}
\end{equation}
and the symplectic potential is
\begin{equation}
\Theta\left[\A,\E\right]\equiv-\hf\int_{\Sigma}\left(\E\cdot\delta\A+\A\cdot\delta\E\right).\label{eq:Theta-dec}
\end{equation}
Of course, \eqref{eq:eom-dec} and \eqref{eq:Theta-dec} may be easily
derived from \eqref{eq:eom-CS} and \eqref{eq:Theta-CS}.

\subsection{Phase Space Polarizations and Teleparallel Gravity}

The symplectic potential \eqref{eq:Theta-dec} results in the symplectic
form
\[
\Omega\equiv\delta\Theta=-\int_{\Sigma}\delta\E\cdot\delta\A.
\]
In fact, one may obtain the same symplectic form using a \emph{family
of potentials }of the form
\begin{equation}
\Theta_{\lambda}=-\int_{\Sigma}\left(\left(1-\lambda\right)\E\cdot\delta\A+\lambda\A\cdot\delta\E\right),\label{eq:Theta-general}
\end{equation}
where the parameter $\lambda\in\left[0,1\right]$ determines the \emph{polarization
}of the phase space. This potential may be obtained from a family
of actions of the form
\begin{equation}
S_{\lambda}=\int_{M}\left(\E\cdot\F-\lambda\d\left(\A\cdot\E\right)\right),\label{eq:S_lambda}
\end{equation}
where the difference lies only in the boundary term and thus does
not affect the physics. Hence the choice of polarization does not
matter in the continuum, but it will be very important in the discrete
theory, as we will see below.

The equations of motion for any action of the form \eqref{eq:S_lambda}
(or constraints, in the Hamiltonian formulation) are, as we have seen:
\begin{itemize}
\item The torsion (or Gauss) constraint $\T=0$,
\item The curvature constraint $\F=0$.
\end{itemize}
Now, recall that general relativity is formulated using the \emph{Levi-Civita
connection}, which is torsionless by definition. Thus, the torsion
constraint $\T=0$ can really be seen as \emph{defining }the connection
$\A$ to be torsionless, and thus selecting the theory to be general
relativity. In this case, $\F=0$ is the true equation of motion,
describing the dynamics of the theory.

In the \emph{teleparallel formulation} of gravity we instead use the
\emph{Weitzenböck connection}, which is defined to be flat but not
necessarily torsionless. In this formulation, we interpret the curvature
constraint $\F=0$ as defining the connection $\A$ to be flat, while
$\T=0$ is the true equation of motion.

There are three cases of interest when considering the choice of the
parameter $\lambda$. The case $\lambda=0$ is the one most suitable
for 2+1D general relativity:
\[
S_{\lambda=0}=\int_{M}\E\cdot\F\sp\Theta_{\lambda=0}=-\int_{\Sigma}\E\cdot\delta\A,
\]
since it indeed produces the familiar action for 2+1D gravity. The
case $\lambda=1/2$ is one most suitable for 2+1D Chern-Simons theory:
\[
S_{\lambda=\hf}=\int_{M}\left(\E\cdot\F-\hf\d\left(\A\cdot\E\right)\right)\sp\Theta_{\lambda=\hf}=-\hf\int_{\Sigma}\left(\E\cdot\delta\A+\A\cdot\delta\E\right),
\]
since it corresponds to the Chern-Simons action \eqref{eq:action-dec}.
Finally, the case $\lambda=1$ is one most suitable for 2+1D teleparallel
gravity:
\[
S_{\lambda=1}=\int_{M}\left(\E\cdot\F-\d\left(\A\cdot\E\right)\right)\sp\Theta_{\lambda=1}=-\int_{\Sigma}\A\cdot\delta\E,
\]
as explained in \cite{Teleparallel}.

Further details about the different polarizations may be found in
\cite{Dupuis:2017otn}. However, the discretization procedure in that
paper did not take into account possible curvature and torsion degrees
of freedom. In the rest of this paper, we will include these degrees
of freedom in the discussion by generalizing our results in \cite{FirstPaper}
to include all possible polarizations of the phase space.

\section{The Discrete Geometry}

\subsection{\label{subsec:The-Cellular-Decomposition}The Cellular Decomposition
and Its Dual}

We embed a cellular decomposition $\Delta$ and a dual cellular decomposition
$\Delta^{*}$ in our 2-dimensional spatial manifold $\Sigma$. These
structures consist of the following elements, where each element of
$\Delta$ is uniquely dual to an element of $\Delta^{*}$:
\begin{center}
\begin{tabular}{|c|c|c|}
\hline 
$\Delta$ &  & $\Delta^{*}$\tabularnewline
\hline 
\hline 
0-cells (\emph{vertices}) $v$ & dual to & 2-cells (\emph{faces}) $f_{v}$\tabularnewline
\hline 
1-cells (\emph{edges}) $e$ & dual to & 1-cells (\emph{links}) $e^{*}$\tabularnewline
\hline 
2-cells (\emph{cells}) $c$ & dual to & 0-cells (\emph{nodes}) $c^{*}$\tabularnewline
\hline 
\end{tabular}
\par\end{center}

\begin{figure}[!h]
\begin{centering}
\begin{tikzpicture}[scale=0.8]
	\begin{pgfonlayer}{nodelayer}
		\node [style=none] (0) at (0.25, -1.9) {$c^{*}$};
		\node [style=none] (1) at (0, -0.45) {$v$};
		\node [style=none] (2) at (1.5, 0.6) {$c^{\prime*}$};
		\node [style=none] (3) at (5.25, -3.4) {$v'$};
		\node [style=Vertex] (4) at (0, -0) {};
		\node [style=Vertex] (5) at (0, 4) {};
		\node [style=Vertex] (6) at (-5, -3) {};
		\node [style=Vertex] (7) at (5, -3) {};
		\node [style=Node] (8) at (0, -1.5) {};
		\node [style=Node] (9) at (1.25, 1) {};
		\node [style=Node] (10) at (-1.25, 1) {};
		\node [style=none] (11) at (-3.5, 3.25) {};
		\node [style=none] (12) at (3.5, 3.25) {};
		\node [style=none] (13) at (0, -4.5) {};
	\end{pgfonlayer}
	\begin{pgfonlayer}{edgelayer}
		\draw [style=Edge] (5) to (6);
		\draw [style=Edge] (6) to (7);
		\draw [style=Edge] (7) to (5);
		\draw [style=Edge] (5) to (4);
		\draw [style=Edge] (4) to (7);
		\draw [style=Edge] (4) to (6);
		\draw [style=Link] (8) to (9);
		\draw [style=Link] (9) to (10);
		\draw [style=Link] (10) to (8);
		\draw [style=Link] (10) to (11.center);
		\draw [style=Link] (9) to (12.center);
		\draw [style=Link] (8) to (13.center);
	\end{pgfonlayer}
\end{tikzpicture}
\par\end{centering}
\caption{\label{fig:Triangle}A simple piece of the cellular decomposition
$\Delta$, in black, and its dual spin network $\Gamma^{*}$, in blue.
The vertices $v$ of the 1-skeleton $\Gamma\subset\Delta$ are shown
as black circles, while the nodes $c^{*}$ of $\Gamma^{*}$ are shown
as blue squares. The edges $e\in\Gamma$ are shown as black solid
lines, while the links $e^{*}\in\Gamma^{*}$ are shown as blue dashed
lines. In particular, two nodes $c^{*}$ and $c^{\prime*}$, connected
by a link $e^{*}=\left(cc'\right)^{*}$, are labeled, as well as two
vertices $v$ and $v'$, connected by an edge $e=\left(vv'\right)=\left(cc'\right)=c\cap c'$,
which is dual to the link $e^{*}$. There is one face in the illustration,
$f_{v}$, which is the triangle enclosed by the three blue links at
the center.}
\end{figure}
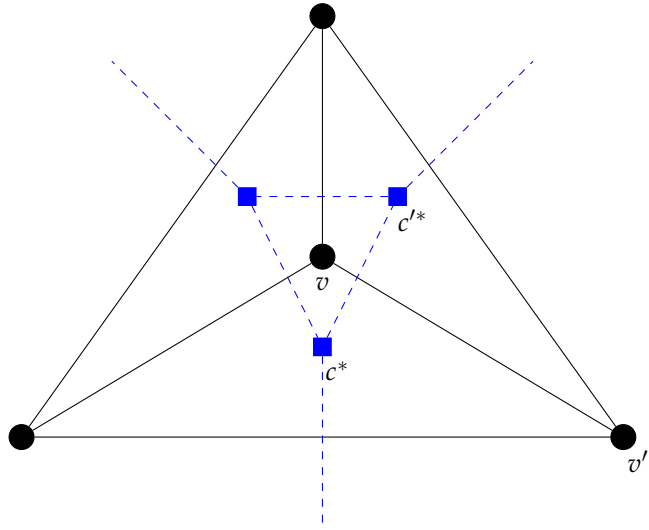

The \emph{1-skeleton graph} $\Gamma\subset\Delta$ is the set of all
vertices and edges of $\Delta$. Its dual is the \emph{spin network
graph} $\Gamma^{*}\subset\Delta^{*}$, the set of all nodes and links
of $\Delta^{*}$. Both graphs are oriented, and we write $e=\left(vv'\right)$
to indicate that the edge $e$ starts at the vertex $v$ and ends
at $v'$, and $e^{*}=\left(cc'\right)^{*}$ to indicate that the link
$e^{*}$ starts at the node $c^{*}$ and ends at $c^{\prime*}$. Furthermore,
since edges are where two cells intersect, we write $e=\left(cc'\right)\equiv\partial c\cap\partial c'$
to denote that the edge $e$ is the intersection of the boundaries
$\partial c$ and $\partial c'$ of the cells $c$ and $c'$ respectively.
If the link $e^{*}$ is dual to the edge $e$, then we have that $e=\left(cc'\right)$
and $e^{*}=\left(cc'\right)^{*}$; therefore the notation is consistent.
This construction is illustrated in Fig. \ref{fig:Triangle} (taken
from \cite{FirstPaper}).

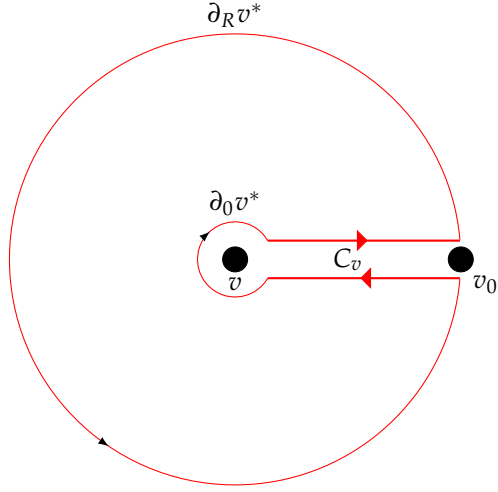
\begin{figure}[!h]
\begin{centering}
\begin{tikzpicture}
	\node [style=Vertex] at (0, 0) {};
	\node [style=none] at (0, -0.33) {$v$};
	\node [style=none] at (0, 0.75) {$\partial_{0}v^{*}$};
	\node [style=none] at (0, 3.25) {$\partial_{R}v^{*}$};
	\node [style=none] at (1.5, 0) {$C_{v}$};
	\centerarc [style=SegmentArrow] (0,0) (330:30:0.5);
	\centerarc [style=SegmentArrow] (0,0) (5:355:3);
	\draw [red,thick] ($({0.5*cos(30)},{0.5*sin(30)})$) -- node {\midarrow} ($({3*cos(5)},{0.5*sin(30)})$);
	\draw [red,thick]  ($({0.5*cos(330)},{0.5*sin(330)})$) -- node {\midarrowop} ($({3*cos(355)},{0.5*sin(330)})$);
	\node [style=Vertex] () at (3, 0) {};
	\node [style=none] () at (3.33, -0.33) {$v_{0}$};
\end{tikzpicture}
\par\end{centering}
\caption{\label{fig:Disk}The punctured disk $v^{*}$. The figure shows the
vertex $v$, cut $C_{v}$, inner boundary $\partial_{0}v^{*}$, outer
boundary $\partial_{R}v^{*}$, and reference point $v_{0}$.}
\end{figure}

For the purpose of doing calculations, it will prove useful to introduce
\emph{disks} $D_{v}$ around each vertex $v$. The disks have a radius
$R$, small enough that the entire disk $D_{v}$ is inside the face
$f_{v}$ for every $v$. We also define \emph{punctured disks} $v^{*}$,
which are obtained from the full disks $D_{v}$ by removing the vertex
$v$, which is at the center, and a \emph{cut }$C_{v}$, connecting
$v$ to an arbitrary point $v_{0}$ on the boundary $\partial D_{v}$.
Thus
\[
v^{*}\equiv D_{v}\backslash\left(\left\{ v\right\} \cup C_{v}\right).
\]
The punctured disks are equipped with a cylindrical coordinate system
$\left(r_{v},\phi_{v}\right)$ such that $r_{v}\in\left(0,R\right)$
and $\phi_{v}\in\left(\alpha_{v}-\hf,\alpha_{v}+\hf\right)$; note
that $\phi_{v}$ is scaled by $2\pi$, so it has a period of 1, for
notational brevity. The boundary of the punctured disk is such that
\[
\partial v^{*}=\partial_{0}v^{*}\cup C_{v}\cup\partial_{R}v^{*},
\]
where $\partial_{0}v^{*}$ is the \emph{inner boundary }at $r_{v}=0$,
$C_{v}$ is the cut at $\phi_{v}=\alpha_{v}-\hf$, and $\partial_{R}v^{*}$
is the \emph{outer boundary }at $r_{v}=R$, and the point where the
cut meets the outer boundary is $v_{0}\equiv\left(R,\alpha_{v}-\hf\right)$.
Note that $\partial_{R}v^{*}=\partial D_{v}$. The punctured disk
is illustrated in Fig. \ref{fig:Disk} (taken from \cite{FirstPaper}).

The outer boundary $\partial_{R}v^{*}$ of each disk is composed of
arcs $\left(vc_{i}\right)$ such that
\[
\partial_{R}v^{*}=\bigcup_{i=1}^{N_{v}}\left(vc_{i}\right),
\]
where $N_{v}$ is the number of cells around $v$ and the cells are
enumerated $c_{1},\ldots,c_{N_{v}}$. Similarly, the boundary $\partial c$
of the cell $c$ is composed of edges $\left(cc_{i}\right)$ and arcs
$\left(cv_{i}\right)$ such that
\[
\partial c=\bigcup_{i=1}^{N_{c}}\left(\left(cc_{i}\right)\cup\left(cv_{i}\right)\right),
\]
where $N_{c}$ is the number of cells adjacent to $c$ or, equivalently,
the number of vertices around $c$. We will use these decompositions
during the discretization process.

\subsection{Truncating the Geometry to the Vertices}

\subsubsection{Motivation}

Before the equations of motion (i.e. the curvature and torsion constraints
$\F=\T=0$) are applied, the geometry on $\Sigma$ can have arbitrary
curvature and torsion. We would like to capture the ``essence''
of the curvature and torsion and encode them on codimension 2 defects.

For this purpose, we can imagine looking at every possible loop on
the spin network graph $\Gamma^{*}$ and taking a holonomy in $G\ltimes\mfg^{*}$
around it. This holonomy will have a part valued in $\mfg$, which
will encode the curvature, and a part valued in $\mfg^{*}$, which
will encode the torsion.

A loop of the spin network is the boundary $\partial f_{v}$ of a
face $f_{v}$. Since the face is dual to a vertex $v$, the natural
place to encode the geometry would be at the vertex. Thus, we will
place the defects at the vertices, and give them the appropriate values
in $\mfg\oplus\mfg^{*}$ obtained by the holonomies.

The disks $D_{v}$ defined above are in a 1-to-1 correspondence with
the faces $f_{v}$. In fact, we can imagine deforming the disks such
that they cover the faces, and their boundaries $\partial D_{v}$
are exactly the loops $\partial f_{v}$. Thus, we may perform calculations
on the disks instead on the faces.

This intuitive and qualitative motivation will be made precise in
the following subsections.

\subsubsection{The Chern-Simons Connection on the Disks}

We define the Chern-Simons\footnote{Recall that we use calligraphic font to denote forms valued in the
double $G\ltimes\mfg^{*}$, and bold calligraphic font for forms valued
in its Lie algebra $\mfg\oplus\mfg^{*}$.} connection on the punctured disk $v^{*}$ as follows:
\begin{equation}
\AAb\bl_{v^{*}}\equiv\mathring{\HH}_{v}^{-1}\d\mathring{\HH}_{v}\equiv\HH_{v}^{-1}\d\HH_{v}+\HH_{v}^{-1}\MMb_{v}\HH_{v}\thinspace\d\phi_{v},\label{eq:A-CS}
\end{equation}
where:
\begin{itemize}
\item $\mathring{\HH}_{v}$ is a non-periodic $G\ltimes\mfg^{*}$-valued
0-form defined as $\mathring{\HH}_{v}\equiv\e^{\MMb_{v}\phi_{v}}\HH_{v}$,
\item $\HH_{v}$ is a periodic\footnote{By ``periodic'' we mean that, under $\phi\mt\phi+1$, the non-periodic
variable $\mathring{\HH}_{v}$ gets an additional factor of $\e^{\MMb_{v}}$
due to the term $\e^{\MMb_{v}\phi_{v}}$, while the periodic variable
$\HH_{v}$ is invariant. (Recall that we are scaling $\phi$ by $2\pi$,
so the period is $1$ and not $2\pi$.)} $\mfg\oplus\mfg^{*}$-valued 0-form,
\item $\MMb_{v}$ is a constant element of the Cartan subalgebra $\mfh\oplus\mfh^{*}$
of $\mfg\oplus\mfg^{*}$.
\end{itemize}
Note that this connection is related by a gauge transformation of
the form $\AAb_{0}\mt\HH_{v}^{-1}\d\HH_{v}+\HH_{v}^{-1}\AAb_{0}\HH_{v}$
to a connection $\AAb_{0}$ defined as follows:
\[
\AAb_{0}\equiv\MMb_{v}\thinspace\d\phi_{v}.
\]
The connection $\AAb_{0}$ satisfies $\left[\AAb_{0},\AAb_{0}\right]=0$,
so its curvature is $\FFb_{0}\equiv\d\AAb_{0}$. This curvature vanishes
everywhere on the punctured disk (which excludes the point $v$),
since $\d^{2}\phi_{v}=0$. However, at the origin of our coordinate
system, i.e. the vertex $v$, $\phi_{v}$ is not well-defined, so
we cannot guarantee that $\FFb_{0}$ vanishes at $v$ itself.

In fact, we can show that it must not vanish there. If we integrate
the curvature on the full disk $D_{v}$ using Stokes' theorem, we
get:
\[
\int_{D_{v}}\FFb_{0}=\oint_{\partial D_{v}}\AAb_{0}=\MMb_{v}\oint_{\partial D_{v}}\d\phi_{v}=\MMb_{v},
\]
where $\oint_{\partial D_{v}}\d\phi_{v}=1$ since we are using coordinates
scaled by $2\pi$, and we used the fact that $\MMb_{v}$ is constant.
We conclude that, since $\FFb_{0}$ vanishes everywhere on $v^{*}$,
and yet it integrates to a finite value at $D_{v}$, the curvature
$\FFb_{0}$ must take the form of a Dirac delta function centered
at $v$:
\[
\FFb_{0}=\MMb_{v}\thinspace\delta\left(v\right),
\]
where $\delta\left(v\right)$ is a distributional 2-form such that
for any 0-form $f$,
\[
\int_{\Sigma}f\thinspace\delta\left(v\right)\equiv f\left(v\right).
\]
The final step is to gauge-transform back from $\AAb_{0}$ to the
initial connection $\AAb$ defined in \eqref{eq:A-CS}. The curvature
transforms in the usual way, $\FFb_{0}\mt\HH_{v}^{-1}\FFb_{0}\HH_{v}\equiv\FFb$,
so we get
\[
\FFb\bl_{D_{v}}=\HH_{v}^{-1}\MMb_{v}\HH_{v}\thinspace\delta\left(v\right)\equiv\PPb_{v}\thinspace\delta\left(v\right),
\]
where we defined
\[
\PPb_{v}\equiv\HH_{v}^{-1}\MMb_{v}\HH_{v}.
\]
Note again that, while $\FFb\bl_{D_{v}}$ (on the full disk) does
not vanish, $\FFb\bl_{v^{*}}$ (on the punctured disk) does vanish.

\subsubsection{The Connection and Frame Field on the Disks}

Now that we have defined the Chern-Simons connection 1-form $\AAb$
and found its curvature $\FFb$ on the disks, we split $\AAb$ into
a $\mfg$-valued connection 1-form $\A$ a $\mfg^{*}$-valued frame
field 1-form $\E$ as defined in \eqref{eq:CS-split}. Similarly,
we split $\FFb$ into a $\mfg$-valued curvature 2-form $\F$ and
a $\mfg^{*}$-valued torsion 2-form $\T$ as defined in \eqref{eq:CS-con-split}.

From \eqref{eq:CS-split} we get:
\begin{equation}
\A\bl_{v^{*}}=\mathring{h}_{v}^{-1}\d\mathring{h}_{v}\sp\E\bl_{v^{*}}=\mathring{h}_{v}^{-1}\d\mathring{\x}_{v}\mathring{h}_{v},\label{eq:A-E-disks}
\end{equation}
where:
\begin{itemize}
\item $\mathring{h}_{v}$ is a non-periodic $G$-valued 0-form and $\mathring{\x}_{v}$
is a non-periodic $\mfg^{*}$-valued 0-form such that\footnote{This notation differs from the one we used in \cite{FirstPaper}.
For the periodic variables, we used $h$ and $\y$ in \cite{FirstPaper}.
Here, we still use $h$, but instead of $\y$ we use $\x$ due to
this variable's relation to the flux $\X$, as shown below. For the
non-periodic variables, we used $u$ and $\w$ in \cite{FirstPaper}.
Here we use $\mathring{h}$ and $\mathring{\x}$ in order to avoid
introducing additional letters, which might be confusing. The circle
above the letter conveys that it involves the angular variable $\phi$
and is thus non-periodic.}
\begin{equation}
\mathring{h}_{v}\equiv\e^{\M_{v}\phi_{v}}h_{v}\sp\mathring{\x}_{v}\equiv\e^{\M_{v}\phi_{v}}\left(\x_{v}+\SS_{v}\phi_{v}\right)\e^{-\M_{v}\phi_{v}},\label{eq:u-v-def}
\end{equation}
\item $h_{v}$ is a periodic $G$-valued 0-form,
\item $\x_{v}$ is a periodic $\mfg^{*}$-valued 0-form,
\item $\M_{v}$ is a constant element of the Cartan subalgebra $\mfh$ of
$\mfg$, such that $\M_{v}\equiv M_{v}\J_{1}$ where $\J_{1}$ is
the Cartan generator,
\item $\SS_{v}$ is a constant element of the Cartan subalgebra $\mfh^{*}$
of $\mfg^{*}$, such that $\SS_{v}\equiv S_{v}\P_{1}$ where $\P_{1}$
is the Cartan generator,
\item By construction $\left[\M_{v},\SS_{v}\right]=0$.
\end{itemize}
The full expressions for $\A$ and $\E$ on $v^{*}$ in terms of $h_{v}$
and $\x_{v}$ are as follows:
\begin{equation}
\A\bl_{v^{*}}=h_{v}^{-1}\d h_{v}+h_{v}^{-1}\M_{v}h_{v}\thinspace\d\phi_{v}\sp\E\bl_{v^{*}}=h_{v}^{-1}\d\x_{v}h_{v}+h_{v}^{-1}\left(\SS_{v}+\left[\M_{v},\x_{v}\right]\right)h_{v}\thinspace\d\phi_{v}.\label{eq:A-E-v}
\end{equation}
Furthermore, from \eqref{eq:CS-con-split} we get:
\[
\F\bl_{D_{v}}=\p_{v}\thinspace\delta\left(v\right),\qquad\T\bl_{D_{v}}=\j_{v}\,\delta\left(v\right),
\]
where $\p_{v},\j_{v}$ represent the \emph{momentum }and \emph{angular
momentum }respectively:
\[
\p_{v}\equiv h_{v}^{-1}\M_{v}h_{v},\qquad\j_{v}\equiv h_{v}^{-1}\left(\SS_{v}+\left[\M_{v},\x_{v}\right]\right)h_{v}.
\]
In terms of $\p_{v}$ and $\j_{v}$, we may write $\A$ and $\E$
on the disk as follows:
\[
\A\bl_{v^{*}}=h_{v}^{-1}\d h_{v}+\p_{v}\thinspace\d\phi_{v}\sp\E\bl_{v^{*}}=h_{v}^{-1}\d\x_{v}h_{v}+\j_{v}\thinspace\d\phi_{v}.
\]
It is clear that the first term in each definition is flat and torsionless,
while the second term (involving $\p_{v}$ and $\j_{v}$ respectively)
is the one which contributes to the curvature and torsion at $v$.
Since the punctured disk $v^{*}$ does not include $v$ itself, the
curvature and torsion vanish everywhere on it: 
\[
\F\bl_{v^{*}}=0\sp\T\bl_{v^{*}}=0.
\]
As before, while $\F$ and $\T$ do not vanish on the full disk $D_{v}$,
they do vanish on $v^{*}$. We call this type of geometry a \emph{piecewise
flat and torsionless geometry}\footnote{The question of whether the geometry we have defined here has a notion
of a ``continuum limit'', e.g. by shrinking the loops to points
such that the discrete defects at the vertices become continuous curvature
and torsion, is left for future work.}. Given a particular spin network $\Gamma^{*}$, and assuming that
information about the curvature and torsion may only be obtained by
taking holonomies along the loops of this spin network, the piecewise
flat and torsionless geometry carries, at least intuitively, the exact
same information as the arbitrary geometry we had before.

\subsubsection{The Connection and Frame Field on the Cells}

Now that we have defined $\A$ and $\E$ on the punctured disks $v^{*}$,
defining them on the cells $c$ is a piece of cake. The geometry inside
the cells is flat and torsionless everywhere, not distributional.
Thus, the expressions for $\A$ and $\E$ on $c$ are analogous to
the first term in each of the expressions in \eqref{eq:A-E-v}, which
is the flat and torsionless term:
\begin{equation}
\A\bl_{c}=h_{c}^{-1}\d h_{c}\sp\E\bl_{c}=h_{c}^{-1}\d\x_{c}h_{c},\label{eq:A-E-cells}
\end{equation}
where $h_{c}$ is a $G$-valued 0-form and $\x_{c}$ is a $\mfg^{*}$-valued
0-form. Of course, by construction, the curvature and torsion associated
to this connection and frame field vanish everywhere on the cell:
\[
\F\bl_{c}=0\sp\T\bl_{c}=0.
\]

\subsection{Dressed Holonomies and Edge Modes}

Consider the definition $\A\bl_{c}=h_{c}^{-1}\d h_{c}$ for $\A$
in terms of $h_{c}$. Note that $\A$ is invariant under the left
action transformation $h_{c}\mt g_{c}h_{c}$ for some constant $g_{c}\in G$.
Thus, inverting the definition $\A\bl_{c}=h_{c}^{-1}\d h_{c}$ to
find $h_{c}$ in terms of $\A$, we get
\[
h_{c}\left(x\right)=h_{c}\left(c^{*}\right)\pexp\int_{c^{*}}^{x}\A,
\]
where $\pexp$ is a \emph{path-ordered exponential}, and $h_{c}\left(c^{*}\right)$
is a new degree of freedom which does not exist in $\A$. The notation
suggests that it is the holonomy ``from $c^{*}$ to itself'', but
it is in general not the identity! The notation $h_{c}\left(c^{*}\right)$
is just a placeholder for the \emph{edge mode }which ``dresses''
the holonomy.

For the ``undressed'' holonomy -- which is simply the path-ordered
exponential from the node $c^{*}$ to some point $x$ -- we thus
have
\begin{equation}
\pexp\int_{c^{*}}^{x}\A=h_{c}^{-1}\left(c^{*}\right)h_{c}\left(x\right).\label{eq:undressed-c}
\end{equation}
Similarly, the definition $\A\bl_{v^{*}}=h_{v}^{-1}\d h_{v}+h_{v}^{-1}\M_{v}h_{v}\thinspace\d\phi_{v}$
is invariant under $h_{v}\mt g_{v}h_{v}$, but only if $g_{v}$ is
in $H$, the Cartan subgroup of $G$, since it must commute with $\M_{v}$.
Inverting the relation $\A\bl_{v^{*}}=\mathring{h}_{v}^{-1}\d\mathring{h}_{v}$,
we get
\[
\mathring{h}_{v}\left(x\right)=h_{v}\left(v\right)\pexp\int_{v}^{x}\A,
\]
where again the edge mode $h_{v}\left(v\right)$ is a new degree of
freedom. The undressed holonomy is then
\begin{equation}
\pexp\int_{v}^{x}\A=h_{v}^{-1}\left(v\right)\mathring{h}_{v}\left(x\right)=h_{v}^{-1}\left(v\right)\e^{\M_{v}\phi_{v}\left(x\right)}h_{v}\left(x\right).\label{eq:undressed-v}
\end{equation}
From \eqref{eq:undressed-c} and \eqref{eq:undressed-v}, we may construct
general path-ordered exponentials from some point $x$ to another
point $y$ by breaking the path from $x$ to $y$ such that it passes
through an intermediate point. If that point is the node $c^{*}$,
then we get
\[
\pexp\int_{x}^{y}\A=\left(\pexp\int_{x}^{c^{*}}\A\right)\left(\pexp\int_{c^{*}}^{y}\A\right)=\left(h_{c}^{-1}\left(x\right)h_{c}\left(c^{*}\right)\right)\left(h_{c}^{-1}\left(c^{*}\right)h_{c}\left(y\right)\right)=h_{c}^{-1}\left(x\right)h_{c}\left(y\right),
\]
and if it's the vertex $v$, we similarly get
\begin{equation}
\pexp\int_{x}^{y}\A=\left(\pexp\int_{x}^{v}\A\right)\left(\pexp\int_{v}^{y}\A\right)=h_{v}^{-1}\left(x\right)\e^{\M_{v}\left(\phi_{v}\left(y\right)-\phi_{v}\left(x\right)\right)}h_{v}\left(y\right).\label{eq:exp-xy-phi}
\end{equation}
Furthermore, we may use the continuity relations \eqref{eq:continuity-ccp}
and \eqref{eq:continuity-cv} (to be discussed later) to obtain a
relation between the path-ordered integrals and the holonomies $h_{cc'}$
and $h_{cv}$. If $y\in\left(cc'\right)$ then we can write
\begin{equation}
\pexp\int_{x}^{y}\A=h_{c}^{-1}\left(x\right)h_{cc'}h_{c'}\left(y\right),\label{eq:exp-xy-h}
\end{equation}
and if $y\in\left(cv\right)$ then we can write
\[
\pexp\int_{x}^{y}\A=h_{c}^{-1}\left(x\right)h_{cv}\mathring{h}_{v}\left(y\right)=h_{c}^{-1}\left(x\right)h_{cv}\e^{\M_{v}\phi_{v}\left(y\right)}h_{v}\left(y\right).
\]
Note that, in particular,
\begin{equation}
\pexp\int_{c^{*}}^{c^{\prime*}}\A=h_{c}^{-1}\left(c^{*}\right)h_{cc'}h_{c'}\left(c^{\prime*}\right).\label{eq:exp-ccp-h}
\end{equation}
A similar discussion applies to the translational holonomies $\x_{c}$
and $\x_{v}$, and one finds two new degrees of freedom, $\x_{c}\left(c^{*}\right)$
and $\x_{v}\left(v\right)$.

\section{Discretizing the Symplectic Potential}

\subsection{The Choice of Polarization}

Recall that there is a family of symplectic potential given by \eqref{eq:Theta-general}:
\begin{equation}
\Theta_{\lambda}=-\int_{\Sigma}\left(\left(1-\lambda\right)\E\cdot\delta\A+\lambda\A\cdot\delta\E\right).\label{eq:Theta-general-2}
\end{equation}
We would like to replace $\A$ and $\E$ by their discretized expressions
given by \eqref{eq:A-E-cells} and \eqref{eq:A-E-disks}. Before we
do this for each cell and disk individually, let us consider a toy
model where we simply take $\A=h^{-1}\d h$ and $\E=h^{-1}\d\x h$
for some $G$-valued 0-form $h$ and $\mfg^{*}$-valued 0-form $\x$
over the entire manifold $\Sigma$. We begin by calculating the variations
of these expressions, obtaining
\[
\delta\A=\delta\left(h^{-1}\d h\right)=h^{-1}\left(\d\De h\right)h,
\]
\[
\delta\E=\delta\left(h^{-1}\d\x h\right)=h^{-1}\left(\d\delta\x+\left[\d\x,\De h\right]\right)h,
\]
where we have defined the notation $\De h\equiv\delta hh^{-1}$ for
the Maurer-Cartan form on field space. Thus, we have
\[
\Theta_{\lambda}=-\int_{\Sigma}\left(\left(1-\lambda\right)\d\x\cdot\d\De h+\lambda\d hh^{-1}\cdot\left(\d\delta\x+\left[\d\x,\De h\right]\right)\right),
\]
where we used the cyclicity of the dot product to cancel some group
elements. Now, the first term is very simple; in fact, it is clearly
an exact 2-form, and thus may be easily integrated. However, the second
term is complicated, and it is unclear if it can be integrated. Nevertheless,
we know that every choice of $\lambda$ leads to the same symplectic
form:
\[
\Omega=\delta\Theta_{\lambda}=-\int_{\Sigma}\delta\E\cdot\delta\A=-\int_{\Sigma}\left(\d\delta\x+\left[\d\x,\De h\right]\right)\cdot\d\De h.
\]
Furthermore, we have seen from \eqref{eq:S_lambda} that the difference
between different polarizations amounts to the addition of a boundary
term and is equivalent to an integration by parts. Thus, we employ
the following trick. First we take $\lambda=0$ in $\Theta_{\lambda}$,
so that it becomes the 2+1D gravity polarization:
\[
\Theta=-\int_{\Sigma}\E\cdot\delta\A.
\]
Then, in the discretization process, we obtain
\[
\Theta=-\int_{\Sigma}\d\x\cdot\d\De h.
\]
The integrand in an exact 2-form, and thus may be integrated in two
equivalent ways:
\[
\d\x\cdot\d\De h=\d\left(\x\cdot\d\De h\right)=-\d\left(\d\x\cdot\De h\right).
\]
Note that the 1-forms $\x\cdot\d\De h$ and $\d\x\cdot\De h$ differ
only by a boundary term of the form $\d\left(\x\cdot\De h\right)$,
and they may be obtained from each other with integration by parts,
just as for the different polarizations. In fact, we may write:
\begin{equation}
\E\cdot\delta\A=\d\x\cdot\d\De h=\lambda\d(\x\cdot\d\De h)-\left(1-\lambda\right)\d(\d\x\cdot\De h).\label{eq:dxdDh}
\end{equation}
We claim that, even though technically both options are equivalent
discretizations of the $\lambda=0$ polarization in \eqref{eq:Theta-general-2},
there is in fact reason to believe that the choice of $\lambda$ in
\eqref{eq:Theta-general-2} corresponds to the same choice of $\lambda$
in \eqref{eq:dxdDh}! We will motivate this by showing that the choice
$\lambda=0$ corresponds to the usual loop gravity polarization, which
is associated with usual general relativity, while the choice $\lambda=1$
corresponds to a dual polarization which, as we will see, is associated
with teleparallel gravity.

\subsection{Decomposing the Spatial Manifold}

As we have seen, the spatial manifold $\Sigma$ is decomposed into
cells $c$ and disks $v^{*}$. The whole manifold $\Sigma$ may be
recovered by taking the union of the cells with the \emph{closures
}of the disks (recall that the vertices $v$ are not in $v^{*}$,
they are on their boundaries):
\[
\Sigma=\left(\bigcup_{c}c\right)\cup\left(\bigcup_{v}v^{*}\cup\partial v^{*}\right).
\]
Here, we are assuming that the cells and punctured disks are disjoint;
the disks ``eat into'' the cells. We can thus split $\Theta$ into
contributions from each cell $c$ and punctured disk $v^{*}$:
\[
\Theta=\sum_{c}\Theta_{c}+\sum_{v}\Theta_{v^{*}},
\]
where
\[
\Theta_{c}=-\int_{c}\E\cdot\delta\A\sp\Theta_{v^{*}}=-\int_{v^{*}}\E\cdot\delta\A.
\]
Given the discretizations \eqref{eq:A-E-cells} and \eqref{eq:A-E-disks},
we replace $h,\x$ in \eqref{eq:dxdDh} with $h_{c},\x_{c}$ or $\mathring{h}_{v},\mathring{\x}_{v}$
respectively, and then integrate using Stokes' theorem to obtain:
\[
\Theta_{c}=\int_{\partial c}\left(\left(1-\lambda\right)\d\x_{c}\cdot\De h_{c}-\lambda\x_{c}\cdot\d\De h_{c}\vphantom{\bll}\right),
\]
\[
\Theta_{v^{*}}=\int_{\partial v^{*}}\left(\left(1-\lambda\right)\d\mathring{\x}_{v}\cdot\De\mathring{h}_{v}-\lambda\mathring{\x}_{v}\cdot\d\De\mathring{h}_{v}\vphantom{\bll}\right).
\]
In the next few subsections, we will manipulate these expressions
so that they can be integrated once again to obtain truly discrete
symplectic potentials.

\subsection{The Vertex and Cut Contributions}

The boundary $\partial v^{*}$ splits into three contributions: one
from the inner boundary $\partial_{0}v^{*}$ (which is the vertex
$v$), one from the cut $C_{v}$, and one from the outer boundary
$\partial_{R}v^{*}$. Thus we have
\[
\Theta_{v^{*}}=-\Theta_{\partial_{0}v^{*}}-\Theta_{C_{v}}+\Theta_{\partial_{R}v^{*}},
\]
where the minus sign comes from the fact that orientation of the outer
boundary is opposite to that of the inner boundary. Here we will discuss
the first two terms, while the contribution from the outer boundary
$\partial_{R}v^{*}$ will be calculated in Sec. \ref{subsec:Edge-Arc}.

Writing the terms in the integrand explicitly in terms of $\x_{v},h_{v}$
using \eqref{eq:u-v-def}, and making use of the identities
\[
\d\mathring{\x}_{v}=\e^{\M_{v}\phi_{v}}\left(\d\x_{v}+\left(\SS_{v}+\left[\M_{v},\x_{v}\right]\right)\d\phi_{v}\right)\e^{-\M_{v}\phi_{v}},
\]
\[
\De\mathring{h}_{v}=\e^{\M_{v}\phi_{v}}\left(\delta\M_{v}\phi_{v}+\De h_{v}\right)\e^{-\M_{v}\phi_{v}},
\]
\[
\d\De\mathring{h}_{v}=\e^{\M_{v}\phi_{v}}\left(\d\De h_{v}+\left(\delta\M_{v}+\left[\M_{v},\De h_{v}\right]\right)\d\phi_{v}\right)\e^{-\M_{v}\phi_{v}},
\]
we get
\[
\d\mathring{\x}_{v}\cdot\De\mathring{h}_{v}=\left(\d\x_{v}+\left(\SS_{v}+\left[\M_{v},\x_{v}\right]\right)\d\phi_{v}\right)\cdot\left(\delta\M_{v}\phi_{v}+\De h_{v}\right),
\]
\[
\mathring{\x}_{v}\cdot\d\De\mathring{h}_{v}=\left(\x_{v}+\SS_{v}\phi_{v}\right)\cdot\left(\d\De h_{v}+\left(\delta\M_{v}+\left[\M_{v},\De h_{v}\right]\right)\d\phi_{v}\right).
\]
The integral on the inner boundary $\partial_{0}v^{*}$ is easily
calculated, since $\x_{v}$ and $h_{v}$ obtain the constant values
$\x_{v}\left(v\right)$ and $h_{v}\left(v\right)$ on the inner boundary.
Hence $\d\x_{v}\left(v\right)=\d\De h_{v}\left(v\right)=0$, and these
expressions simplify to\footnote{Here we used the identity $\left[\A,\B\right]\cdot\C=\A\cdot\left[\B,\C\right]$
to get $\left[\M_{v},\x_{v}\right]\cdot\delta\M_{v}=\x_{v}\cdot\left[\delta\M_{v},\M_{v}\right]=0$
and $\SS_{v}\cdot\left[\M_{v},\De h_{v}\right]=\De h_{v}\cdot\left[\SS_{v},\M_{v}\right]=0$.}
\[
\d\mathring{\x}_{v}\cdot\De\mathring{h}_{v}\bl_{\partial_{0}v^{*}}=\left(\phi_{v}\SS_{v}\cdot\delta\M_{v}+\left(\SS_{v}+\left[\M_{v},\x_{v}\left(v\right)\right]\right)\cdot\De h_{v}\left(v\right)\right)\d\phi_{v},
\]
\[
\mathring{\x}_{v}\cdot\d\De\mathring{h}_{v}\bl_{\partial_{0}v^{*}}=\left(\phi_{v}\SS_{v}\cdot\delta\M_{v}+\x_{v}\left(v\right)\cdot\left(\delta\M_{v}+\left[\M_{v},\De h_{v}\left(v\right)\right]\right)\right)\d\phi_{v}.
\]
To evaluate the contribution from the inner boundary, we integrate
from $\phi_{v}=\alpha_{v}-1/2$ to $\phi_{v}=\alpha_{v}+1/2$. Then
since
\[
\int_{\alpha_{v}-1/2}^{\alpha_{v}+1/2}\d\phi_{v}=1\sp\int_{\alpha_{v}-1/2}^{\alpha_{v}+1/2}\phi_{v}\thinspace\d\phi_{v}=\alpha_{v},
\]
we get:
\[
\Theta_{\partial_{0}v^{*}}=\left(1-2\lambda\right)\alpha_{v}\SS_{v}\cdot\delta\M_{v}+\left(1-\lambda\right)\left(\SS_{v}+\left[\M_{v},\x_{v}\left(v\right)\right]\right)\cdot\De h_{v}\left(v\right)-\lambda\x_{v}\left(v\right)\cdot\left(\delta\M_{v}+\left[\M_{v},\De h_{v}\left(v\right)\right]\right),
\]
which may be simplified to
\[
\Theta_{\partial_{0}v^{*}}=\left(1-2\lambda\right)\alpha_{v}\SS_{v}\cdot\delta\M_{v}+\left(1-\lambda\right)\SS_{v}\cdot\De h_{v}\left(v\right)-\lambda\x_{v}\left(v\right)\cdot\delta\M_{v}+\left[\M_{v},\x_{v}\left(v\right)\right]\cdot\De h_{v}\left(v\right).
\]
Next, we have the cut $C_{v}$. Since $\d\phi_{v}=0$ on the cut,
we have a significant simplification:
\[
\d\mathring{\x}_{v}\cdot\De\mathring{h}_{v}\bl_{C_{v}}=\d\x_{v}\cdot\left(\delta\M_{v}\phi_{v}+\De h_{v}\right),
\]
\[
\mathring{\x}_{v}\cdot\d\De\mathring{h}_{v}\bl_{C_{v}}=\left(\x_{v}+\SS_{v}\phi_{v}\right)\cdot\d\De h_{v}.
\]
In fact, the cut has two sides: one at $\phi_{v}=\alpha_{v}-1/2$
and another at $\phi_{v}=\alpha_{v}+1/2$, with opposite orientation.
Let us label them $C_{v}^{-}$ and $C_{v}^{+}$ respectively. Any
term that does not depend explicitly on $\phi_{v}$ will vanish when
we take the difference between both sides of the cut, since they only
differ by the value of $\phi_{v}$. Thus only the terms $\d\x_{v}\cdot\delta\M_{v}\phi_{v}$
and $\SS_{v}\cdot\d\De h_{v}\phi_{v}$ survive. The relevant contribution
from each side of the cut is therefore:
\begin{align*}
\Theta_{C_{v}^{\pm}} & =\int_{r=0}^{R}\left(\left(1-\lambda\right)\d\x_{v}\cdot\delta\M_{v}\phi_{v}-\lambda\SS_{v}\cdot\d\De h_{v}\phi_{v}\right)\blll_{\phi_{v}=\alpha_{v}\pm1/2}\\
 & =\left(\alpha_{v}\pm\hf\right)\left(\left(1-\lambda\right)\delta\M_{v}\cdot\int_{r=0}^{R}\d\x_{v}-\lambda\SS_{v}\cdot\int_{r=0}^{R}\d\De h_{v}\right)\\
 & =\left(\alpha_{v}\pm\hf\right)\left(\left(1-\lambda\right)\delta\M_{v}\cdot\left(\x_{v}\left(v_{0}\right)-\x_{v}\left(v\right)\right)-\lambda\SS_{v}\cdot\left(\De h_{v}\left(v_{0}\right)-\De h_{v}\left(v\right)\right)\right),
\end{align*}
where the point at $r=0$ is the vertex $v$, and the point at $r=R$
and $\phi_{v}=\alpha_{v}\pm1/2$ is labeled $v_{0}$. Taking the difference
between both sides of the cut, we thus get the total contribution:
\begin{align*}
\Theta_{C_{v}} & =\Theta_{C_{v}^{+}}-\Theta_{C_{v}^{-}}\\
 & =\left(\left(\alpha_{v}+\hf\right)-\left(\alpha_{v}-\hf\right)\right)\left(\left(1-\lambda\right)\left(\x_{v}\left(v_{0}\right)-\x_{v}\left(v\right)\right)\cdot\delta\M_{v}-\lambda\SS_{v}\cdot\left(\De h_{v}\left(v_{0}\right)-\De h_{v}\left(v\right)\right)\right)\\
 & =\left(1-\lambda\right)\left(\x_{v}\left(v_{0}\right)-\x_{v}\left(v\right)\right)\cdot\delta\M_{v}-\lambda\SS_{v}\cdot\left(\De h_{v}\left(v_{0}\right)-\De h_{v}\left(v\right)\right).
\end{align*}
Adding up the contributions from the inner boundary and the cut, we
obtain the vertex symplectic potential $\Theta_{v}\equiv-\left(\Theta_{\partial_{0}v^{*}}+\Theta_{C_{v}}\right)$:
\begin{align}
\Theta_{v} & =-\left(1-2\lambda\right)\alpha_{v}\SS_{v}\cdot\delta\M_{v}-\SS_{v}\cdot\left(\De h_{v}\left(v\right)-\lambda\De h_{v}\left(v_{0}\right)\right)+\label{eq:Theta_v}\\
 & \qquad+\left(\x_{v}\left(v\right)-\left(1-\lambda\right)\x_{v}\left(v_{0}\right)\right)\cdot\delta\M_{v}-\left[\M_{v},\x_{v}\left(v\right)\right]\cdot\De h_{v}\left(v\right).
\end{align}

\subsection{The ``Particle'' Potential}

Let $\x_{v}^{\parallel}\left(v_{0}\right)$ be the component of $\x_{v}\left(v_{0}\right)$
parallel to $\SS_{v}$:
\begin{equation}
\x_{v}\left(v_{0}\right)\equiv\x_{v}^{\parallel}\left(v_{0}\right)+\x_{v}^{\perp}\left(v_{0}\right)\sp\x_{v}^{\parallel}\left(v_{0}\right)\equiv\left(\x_{v}\left(v_{0}\right)\cdot\J_{1}\right)\P_{1},
\end{equation}
where $\J_{1}$ and $\P_{1}$ are the Cartan generator of rotations
and translations respectively, and we remind the reader that the dot
product is defined in \eqref{eq:dot-product} as $\J_{i}\cdot\P_{j}=\delta_{ij}$
and $\J_{i}\cdot\J_{j}=\P_{i}\cdot\P_{j}=0$. Similarly, let $\De^{\parallel}h_{v}\left(v_{0}\right)$
be the component of $\De h_{v}\left(v_{0}\right)$ parallel to $\M_{v}$:
\[
\De h_{v}\left(v_{0}\right)\equiv\De^{\parallel}h_{v}\left(v_{0}\right)+\De^{\bot}h_{v}\left(v_{0}\right)\sp\De^{\parallel}h_{v}\left(v_{0}\right)\equiv\left(\De h_{v}\left(v_{0}\right)\cdot\P_{1}\right)\J_{1}.
\]
Let us now define a $\mfg$-valued 0-form $\De H_{v}$, which is a
1-form on field space (i.e. a variation\footnote{\label{fn:H_v-foot}Despite the suggestive notation, in principle
$\De H_{v}$ need not be of the form $\delta H_{v}H_{v}^{-1}$ for
some $G$-valued 0-form $H_{v}$. It can instead be of the form $\delta\h_{v}$
for some $\mfg$-valued 0-form $\h_{v}$. Its precise form is left
implicit, and we merely assume that there is a solution for either
$H_{v}$ or $\h_{v}$ in terms of $h_{v}\left(v\right)$ and $h_{v}\left(v_{0}\right)$.}):
\begin{equation}
\De H_{v}\equiv\De h_{v}\left(v\right)-\lambda\De^{\parallel}h_{v}\left(v_{0}\right),\label{eq:DeltaH_v}
\end{equation}
and a $\mfg^{*}$-valued 0-form $\X_{v}$ called the \emph{vertex
flux}:
\[
\X_{v}\equiv\x_{v}\left(v\right)-\left(1-\lambda\right)\x_{v}^{\parallel}\left(v_{0}\right)-\left(1-2\lambda\right)\alpha_{v}\SS_{v}.
\]
Then since $\SS_{v}\cdot\De h_{v}\left(v_{0}\right)=\SS_{v}\cdot\De^{\parallel}h_{v}\left(v_{0}\right)$
we have
\[
\SS_{v}\cdot\left(\De h_{v}\left(v\right)-\lambda\De h_{v}\left(v_{0}\right)\right)=\SS_{v}\cdot\De H_{v},
\]
and since $\x_{v}\left(v_{0}\right)\cdot\delta\M_{v}=\x_{v}^{\parallel}\left(v_{0}\right)\cdot\delta\M_{v}$
we have
\[
\left(\x_{v}\left(v\right)-\left(1-\lambda\right)\x_{v}\left(v_{0}\right)-\left(1-2\lambda\right)\alpha_{v}\SS_{v}\right)\cdot\delta\M_{v}=\X_{v}\cdot\delta\M_{v}.
\]
Furthermore, since $\left[\M_{v},\x_{v}^{\parallel}\left(v_{0}\right)\right]=\left[\M_{v},\SS_{v}\right]=0$
and $\left[\M_{v},\X_{v}\right]\cdot\De^{\parallel}h_{v}\left(v_{0}\right)=0$
we have
\[
\left[\M_{v},\x_{v}\left(v\right)\right]\cdot\De h_{v}\left(v\right)=\left[\M_{v},\X_{v}\right]\cdot\De H_{v}.
\]
Therefore \eqref{eq:Theta_v} becomes
\begin{equation}
\Theta_{v}=\X_{v}\cdot\delta\M_{v}-\left(\SS_{v}+\left[\M_{v},\X_{v}\right]\right)\cdot\De H_{v}.\label{eq:Theta_v-simplified}
\end{equation}
This potential resembles that of a point particle with mass $\M_{v}$
and spin $\SS_{v}$. Note that the free parameter $\lambda$ has been
absorbed into $\X_{v}$ and $\De H_{v}$, so this potential is obtained
independently of the value of $\lambda$ and thus the choice of polarization!

\subsection{\label{subsec:Edge-Arc}The Edge and Arc Contributions}

To summarize our progress so far, we now have
\[
\Theta=\sum_{c}\Theta_{c}+\sum_{v}\Theta_{\partial_{R}v^{*}}+\sum_{v}\Theta_{v},
\]
where
\[
\Theta_{c}=\int_{\partial c}\left(\left(1-\lambda\right)\d\x_{c}\cdot\De h_{c}-\lambda\x_{c}\cdot\d\De h_{c}\right),
\]
\[
\Theta_{\partial_{R}v^{*}}=\int_{\partial_{R}v^{*}}\left(\left(1-\lambda\right)\d\mathring{\x}_{v}\cdot\De\mathring{h}_{v}-\lambda\mathring{\x}_{v}\cdot\d\De\mathring{h}_{v}\right),
\]
and $\Theta_{v}$ is given by \eqref{eq:Theta_v-simplified}. In order
to simplify $\Theta_{\partial_{R}v^{*}}$, we recall from Sec. \ref{subsec:The-Cellular-Decomposition}
that the boundary $\partial c$ of the cell $c$ is composed of edges
$\left(cc_{i}\right)$ and arcs $\left(cv_{i}\right)$ such that
\[
\partial c=\bigcup_{i=1}^{N_{c}}\left(\left(cc_{i}\right)\cup\left(cv_{i}\right)\right),
\]
while the outer boundary $\partial_{R}v^{*}$ of the disk $v^{*}$
is composed of arcs $\left(vc_{i}\right)$ such that
\[
\partial_{R}v^{*}=\bigcup_{i=1}^{N_{v}}\left(vc_{i}\right),
\]
where $N_{v}$ is the number of cells around $v$. Importantly, in
terms of orientation, $\left(cc'\right)=\left(c'c\right)^{-1}$ and
$\left(cv\right)=\left(vc\right)^{-1}$. We thus see that each edge
$\left(cc'\right)$ is integrated over exactly twice, once from the
integral over $\partial c$ and once from the integral over $\partial c'$
with opposite orientation, and similarly each arc $\left(cv\right)$
is integrated over twice, once from $\partial c$ and once from $\partial_{R}v^{*}$
with opposite orientation. Hence we may rearrange the sums and integrals
as follows:
\[
\Theta=\sum_{\left(cc'\right)}\Theta_{cc'}+\sum_{\left(vc\right)}\Theta_{vc}+\sum_{v}\Theta_{v},
\]
where
\[
\Theta_{cc'}\equiv\int_{\left(cc'\right)}\left(\left(1-\lambda\right)(\d\x_{c}\cdot\De h_{c}-\d\x_{c'}\cdot\De h_{c'})-\lambda(\x_{c}\cdot\d\De h_{c}-\x_{c'}\cdot\d\De h_{c'})\vphantom{\mathring{h}_{v}}\right),
\]
\[
\Theta_{vc}\equiv\int_{\left(vc\right)}\left(\left(1-\lambda\right)(\d\mathring{\x}_{v}\cdot\De\mathring{h}_{v}-\d\x_{c}\cdot\De h_{c})-\lambda(\mathring{\x}_{v}\cdot\d\De\mathring{h}_{v}-\x_{c}\cdot\d\De h_{c})\right).
\]
Next, we note that the connection $\A$ and frame field $\E$ are
defined using different variables on each cell and disk, but overall
they must be continuous on the entire spatial manifold $\Sigma$.
This implies that the variables from each cell and disk, when evaluated
on the edges and arcs, must be related via \emph{continuity relations},
which are, for the edges $\left(cc'\right)$,
\begin{equation}
h_{c'}=h_{c'c}h_{c}\sp\x_{c'}=h_{c'c}(\x_{c}-\x_{c}^{c'})h_{cc'}\sp\textrm{on }\left(cc'\right),\label{eq:continuity-ccp}
\end{equation}
and for the arcs $\left(vc\right)$
\begin{equation}
h_{c}=h_{cv}\mathring{h}_{v}\sp\x_{c}=h_{cv}(\mathring{\x}_{v}-\x_{v}^{c})h_{vc}\sp\textrm{on }\left(vc\right),\label{eq:continuity-cv}
\end{equation}
where $h_{cc'}$, $h_{cv}$, $\x_{c}^{c'}$ and $\x_{c}^{v}$ are
all constant and satisfy
\begin{equation}
h_{cc'}=h_{c'c}^{-1}\sp h_{vc}=h_{cv}^{-1}\sp\x_{c}^{c'}=-h_{cc'}\x_{c'}^{c}h_{c'c}\sp\x_{c}^{v}=-h_{cv}\x_{v}^{c}h_{vc}.\label{eq:h-y-inverse}
\end{equation}
By plugging these relations into $\Theta_{cc'}$ and $\Theta_{vc}$
and simplifying, using the identities
\[
\De h_{c'}=h_{c'c}\left(\De h_{c}-\De h_{c}^{c'}\right)h_{cc'}\sp\De h_{c}=h_{cv}\left(\De\mathring{h}_{v}-\De h_{v}^{c}\right)h_{vc},
\]
where $\De h_{c}^{c'}\equiv\delta h_{cc'}h_{c'c}$ and $\De h_{v}^{c}\equiv\delta h_{vc}h_{cv}$,
we find:
\begin{equation}
\Theta_{cc'}=\left(1-\lambda\right)\De h_{c}^{c'}\cdot\int_{\left(cc'\right)}\d\x_{c}-\lambda\x_{c}^{c'}\cdot\int_{\left(cc'\right)}\d\De h_{c},\label{eq:Theta_ccp}
\end{equation}
\begin{equation}
\Theta_{vc}=\left(1-\lambda\right)\De h_{v}^{c}\cdot\int_{\left(vc\right)}\d\mathring{\x}_{v}-\lambda\x_{v}^{c}\cdot\int_{\left(vc\right)}\d\De\mathring{h}_{v}.\label{eq:Theta_vc}
\end{equation}

\subsection{Holonomies and Fluxes}

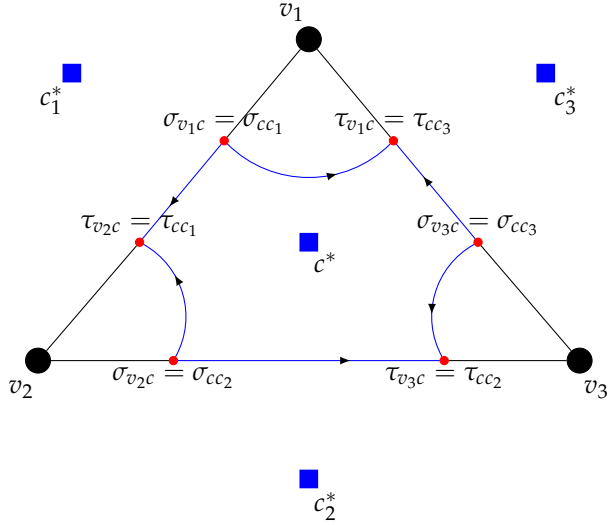
\begin{figure}[!h]
\begin{centering}
\begin{tikzpicture}[scale=0.9]
	\begin{pgfonlayer}{nodelayer}
		\node [style=Node] (0) at (0, -0) {};
		\node [style=Vertex] (1) at (0, 3) {};
		\node [style=Vertex] (2) at (-4, -1.75) {};
		\node [style=Vertex] (3) at (4, -1.75) {};
		\node [style=none] (4) at (0.25, -0.3) {$c^{*}$};
		\node [style=none] (5) at (-0.25, 3.4) {$v_1$};
		\node [style=Point] (6) at (-1.25, 1.5) {};
		\node [style=Point] (7) at (1.25, 1.5) {};
		\node [style=Point] (8) at (2.5, -0) {};
		\node [style=Point] (9) at (2, -1.75) {};
		\node [style=Point] (10) at (-2.5, -0) {};
		\node [style=Point] (11) at (-2, -1.75) {};
		\node [style=none] (12) at (-4.25, -2.17) {$v_2$};
		\node [style=none] (13) at (4.25, -2.17) {$v_3$};
		\node [style=Node] (14) at (3.5, 2.5) {};
		\node [style=Node] (15) at (-3.5, 2.5) {};
		\node [style=Node] (16) at (0, -3.5) {};
		\node [style=none] (17) at (3.8, 2.1) {$c_3^{*}$};
		\node [style=none] (18) at (-3.8, 2.1) {$c_1^{*}$};
		\node [style=none] (19) at (0.25, -3.9) {$c_2^{*}$};
		\node [style=none] (20) at (1.25, 1.75) {$\tau_{v_1c}=\tau_{cc_3}$};
		\node [style=none] (21) at (-1.25, 1.75) {$\sigma_{v_1c}=\sigma_{cc_1}$};
		\node [style=none] (22) at (-2.5, 0.25) {$\tau_{v_2c}=\tau_{cc_1}$};
		\node [style=none] (23) at (-2, -2) {$\sigma_{v_2c}=\sigma_{cc_2}$};
		\node [style=none] (24) at (2, -2) {$\tau_{v_3c}=\tau_{cc_2}$};
		\node [style=none] (25) at (2.5, 0.25) {$\sigma_{v_3c}=\sigma_{cc_3}$};
	\end{pgfonlayer}
	\begin{pgfonlayer}{edgelayer}
		\draw [style=LinkArrow, bend right=45, looseness=1.00] (6) to (7);
		\draw [style=LinkArrow, bend right=45, looseness=1.00] (11) to (10);
		\draw [style=LinkArrow, bend right=45, looseness=1.00] (8) to (9);
		\draw [style=LinkArrow] (8) to (7);
		\draw [style=LinkArrow] (6) to (10);
		\draw [style=LinkArrow] (11) to (9);
		\draw [style=Edge] (1) to (6);
		\draw [style=Edge] (10) to (2);
		\draw [style=Edge] (2) to (11);
		\draw [style=Edge] (9) to (3);
		\draw [style=Edge] (3) to (8);
		\draw [style=Edge] (7) to (1);
	\end{pgfonlayer}
\end{tikzpicture}
\par\end{centering}
\caption{\label{fig:IntersectionPoints}The intersection points (red circles)
of truncated edges and arcs along the oriented boundary $\partial c$
(blue arrows).}
\end{figure}

Let us label the source and target points of the edge $\left(cc'\right)$
as $\sigma_{cc'}$ and $\tau_{cc'}$ respectively, and the source
and target points of the arc $\left(vc\right)$ as $\sigma_{vc}$
and $\tau_{vc}$ respectively, where $\sigma$ stands for ``source''
and $\tau$ for ``target'':
\[
\left(cc'\right)\equiv\left(\sigma_{cc'}\tau_{cc'}\right)\sp\left(vc\right)\equiv\left(\sigma_{vc}\tau_{vc}\right).
\]
This labeling is illustrated in Fig. \ref{fig:IntersectionPoints}
(taken from \cite{FirstPaper}). We now define holonomies and fluxes
on the edges and their dual links, and on the arcs and their dual
line segments.

\subsubsection{Holonomies on the Links and Segments}

The rotational\footnote{Recall that we are dealing with a generalized Euclidean or Poincaré
group $G\ltimes\mfg^{*}$ where $G$ represents rotations and $\mfg^{*}$
represents translations (or generalizations thereof). $h_{cc'}$ is
valued in $G$ and is thus a rotational holonomy, while $\x_{c}^{c'}$
is valued in $\mfg^{*}$ and is thus a translational holonomy.} holonomy $h_{cc'}$ comes from the continuity relations \eqref{eq:continuity-ccp}.
Its role is relating the variables $h_{c},\x_{c}$ on the cell $c$
to the variables $h_{c'},\x_{c'}$ on the cell $c'$. Now, in the
relation $h_{c}\left(x\right)=h_{cc'}h_{c'}\left(x\right)$, the holonomy
on the left-hand side is from the node $c^{*}$ to a point $x$ on
the edge $\left(cc'\right)$. Therefore, the holonomy on the right-hand
side should also take us from $c^{*}$ to $x$. Since $h_{c'}\left(x\right)$
is the holonomy from $c^{\prime*}$ to $x$, we see that $h_{cc'}$
must take us from $c^{*}$ to $c^{\prime*}$. In other words, the
holonomy $h_{cc'}$ is exactly the holonomy from $c^{*}$ to $c^{\prime*}$,
along\footnote{Since the geometry is flat, the actual path taken does not matter,
only that it starts at $c^{*}$ and ends at $c^{\prime*}$. We may
therefore assume without loss of generality that the path taken by
$h_{cc'}$ is, in fact, along the link $\left(cc'\right)^{*}$.} the link $\left(cc'\right)^{*}$.

Thus we define\footnote{The change from lower-case $h$ to upper-case $H$ is only symbolic
here, but it will become more meaningful when we define other holonomies
and fluxes below.} \emph{holonomies along the links $\left(cc'\right)^{*}$}:
\begin{equation}
H_{cc'}\equiv h_{cc'}\sp\De H_{c}^{c'}\equiv\delta H_{cc'}H_{c'c}.\label{eq:hol-link}
\end{equation}
Similarly, the holonomy $h_{vc}$ comes from the continuity relations
\eqref{eq:continuity-cv}, and it takes us from the vertex $v$ to
the node $c^{*}$. We define $\left(vc\right)^{*}$ to be the line
segment connecting $v$ to $c^{*}$; it is dual to the arc $\left(vc\right)$
and its inverse is $\left(cv\right)^{*}$. We then define \emph{holonomies
along the segments $\left(vc\right)^{*}$}:
\begin{equation}
H_{vc}\equiv h_{vc}\sp\De H_{v}^{c}\equiv\delta H_{vc}H_{cv}.\label{eq:hol-seg}
\end{equation}
The inverse holonomies follow immediately from the relations $h_{cc'}^{-1}=h_{c'c}$
and $h_{vc}^{-1}=h_{cv}$:
\[
H_{cc'}^{-1}=H_{c'c}\sp H_{vc}^{-1}=H_{cv}.
\]

\subsubsection{Fluxes on the Edges and Arcs}

From the integral in the first term of \eqref{eq:Theta_ccp}, we are
inspired to define \emph{fluxes along the edges }$\left(cc'\right)$:
\begin{equation}
\XXt_{c}^{c'}\equiv\int_{\left(cc'\right)}\d\x_{c}=\x_{c}\left(\tau_{cc'}\right)-\x_{c}\left(\sigma_{cc'}\right).\label{eq:flux-edge}
\end{equation}
The tilde specifies that the flux $\XXt_{c}^{c'}$ is on the edge
$\left(cc'\right)$ dual to the link $\left(cc'\right)^{*}$; the
flux $\X_{c}^{c'}$, to be defined below, is on the link, and similarly
we will define $\Ht_{cc'}$ to be the holonomy on the edge, while
$H_{cc'}$ is the holonomy on the link.

The flux $\XXt_{c}^{c'}$ is a composition of two translational holonomies.
The holonomy $-\x_{c}\left(\sigma_{cc'}\right)$ takes us from the
point $\sigma_{cc'}$ to the node $c^{*}$, and then the holonomy
$\x_{c}\left(\tau_{cc'}\right)$ takes us from $c^{*}$ to $\tau_{cc'}$.
Hence, the composition of these holonomies is a translational holonomy
from $\sigma_{cc'}$ to $\tau_{cc'}$, that is, along\footnote{Again, since the geometry is flat, the path passing through the node
$c^{*}$ is equivalent to the path going along the edge $\left(cc'\right)$.} the edge $\left(cc'\right)$, as claimed.

To find the inverse flux we use $\left(cc'\right)=\left(c'c\right)^{-1}$,
$\sigma_{cc'}=\tau_{c'c}$ and \eqref{eq:continuity-ccp}:
\[
\XXt_{c'}^{c}\equiv\int_{\left(c'c\right)}\d\x_{c'}=\x_{c'}\left(\tau_{c'c}\right)-\x_{c'}\left(\sigma_{c'c}\right)=h_{c'c}\left(\x_{c}\left(\sigma_{cc'}\right)-\x_{c}\left(\tau_{cc'}\right)\right)h_{cc'}=-H_{c'c}\XXt_{c}^{c'}H_{cc'}.
\]
Similarly, from the first integral in \eqref{eq:Theta_vc} we are
inspired to define \emph{fluxes along the arcs} $\left(vc\right)$:
\begin{equation}
\XXt_{v}^{c}\equiv\int_{\left(vc\right)}\d\mathring{\x}_{v}=\mathring{\x}_{v}\left(\tau_{vc}\right)-\mathring{\x}_{v}\left(\sigma_{vc}\right).\label{eq:flux-arc-vc}
\end{equation}
Note that this time, the two translational holonomies are composed
at $v$. As for the inverse, we define $\XXt_{c}^{v}$ as follows
and use \eqref{eq:continuity-cv} to find a relation with $\XXt_{v}^{c}$,
taking into account the fact that $\left(cv\right)=\left(vc\right)^{-1}$
and $\sigma_{cv}=\tau_{vc}$:
\begin{equation}
\XXt_{c}^{v}\equiv\int_{\left(cv\right)}\d\x_{c}=\x_{c}\left(\tau_{cv}\right)-\x_{c}\left(\sigma_{cv}\right)=h_{cv}\left(\mathring{\x}_{v}\left(\sigma_{vc}\right)-\mathring{\x}_{v}\left(\tau_{vc}\right)\right)h_{vc}=-H_{cv}\XXt_{v}^{c}H_{vc}.\label{eq:flux-arc-cv}
\end{equation}
In conclusion, we have the relations
\[
\XXt_{c'}^{c}=-H_{c'c}\XXt_{c}^{c'}H_{cc'}\sp\XXt_{c}^{v}=-H_{cv}\XXt_{v}^{c}H_{vc}.
\]

\subsubsection{Holonomies on the Edges and Arcs}

The holonomies and fluxes defined thus far will be used in the $\lambda=0$
polarization. In the $\lambda=1$ (dual) polarization, let us define
\emph{holonomies along the edges} $\left(cc'\right)$ and \emph{holonomies
along the arcs} $\left(vc\right)$:
\begin{equation}
\Ht_{cc'}\equiv h_{c}^{-1}\left(\sigma_{cc'}\right)h_{c}\left(\tau_{cc'}\right)\sp\De\Ht_{c}^{c'}\equiv\delta\Ht_{cc'}\Ht_{c'c},\label{eq:Ht-ccp}
\end{equation}
\begin{equation}
\Ht_{vc}\equiv\mathring{h}_{v}^{-1}\left(\sigma_{vc}\right)\mathring{h}_{v}\left(\tau_{vc}\right)\sp\De\Ht_{v}^{c}\equiv\delta\Ht_{vc}\Ht_{cv}.\label{eq:Ht-vc}
\end{equation}
As with $\XXt_{c}^{c'}$, the holonomy $\Ht_{cc'}$ starts from $\sigma_{cc'}$,
goes to $c^{*}$ via $h_{c}^{-1}\left(\sigma_{cc'}\right)$, and then
goes to $\tau_{cc'}$ via $h_{c}\left(\tau_{cc'}\right)$. Therefore
it is indeed a holonomy along the edge $\left(cc'\right)$. Similarly,
the holonomy $\Ht_{vc}$ starts from $\sigma_{vc}$, goes to $v$
via $\mathring{h}_{v}^{-1}\left(\sigma_{vc}\right)$, and then goes
to $\tau_{vc}$ via $\mathring{h}_{v}\left(\tau_{vc}\right)$. Therefore
it is indeed a holonomy along the arc $\left(vc\right)$.

The difference compared to $\XXt_{c}^{c'}$ is that in $\Ht_{cc'}$
we have rotational instead of translational holonomies, and the composition
of holonomies is (non-Abelian) multiplication instead of addition.
As before, the tilde specifies that the holonomy is on the edges or
arcs and not the dual links or segments.

The variations of these holonomies are:
\begin{equation}
\De\Ht_{c}^{c'}=h_{c}^{-1}\left(\sigma_{cc'}\right)\left(\De h_{c}\left(\tau_{cc'}\right)-\De h_{c}\left(\sigma_{cc'}\right)\right)h_{c}\left(\sigma_{cc'}\right)=h_{c}^{-1}\left(\sigma_{cc'}\right)\left(\int_{\left(cc'\right)}\d\De h_{c}\right)h_{c}\left(\sigma_{cc'}\right),\label{eq:DeltaHccp}
\end{equation}
\begin{equation}
\De\Ht_{v}^{c}=\mathring{h}_{v}^{-1}\left(\sigma_{vc}\right)\left(\De\mathring{h}_{v}\left(\tau_{vc}\right)-\De\mathring{h}_{v}\left(\sigma_{vc}\right)\right)\mathring{h}_{v}\left(\sigma_{vc}\right)=\mathring{h}_{v}^{-1}\left(\sigma_{vc}\right)\left(\int_{\left(vc\right)}\d\De\mathring{h}_{v}\right)\mathring{h}_{v}\left(\sigma_{vc}\right).\label{eq:DeltaHvc}
\end{equation}
Thus, we see that they relate to the integrals in the second terms
of \eqref{eq:Theta_ccp} and \eqref{eq:Theta_vc}.

Since $\left(cc'\right)=\left(c'c\right)^{-1}$, it is obvious that
$\Ht_{cc'}^{-1}=\Ht_{c'c}$. Furthermore, by combining \eqref{eq:Ht-vc}
with \eqref{eq:continuity-cv} we may obtain an expression for $\Ht_{vc}$
in terms of $h_{c}$:
\[
\Ht_{vc}=h_{c}^{-1}\left(\sigma_{vc}\right)h_{c}\left(\tau_{vc}\right).
\]
If we now define
\begin{equation}
\Ht_{cv}\equiv h_{c}^{-1}\left(\sigma_{cv}\right)h_{c}\left(\tau_{cv}\right),\label{eq:Ht-vc-h_c}
\end{equation}
then using the relations $\sigma_{cv}=\tau_{vc}$ and $\tau_{cv}=\sigma_{vc}$,
which come from the fact that $\left(vc\right)=\left(cv\right)^{-1}$,
it is easy to see that $\Ht_{vc}^{-1}=\Ht_{cv}$. In conclusion, the
inverses of these holonomies satisfy the relationships
\[
\Ht_{cc'}^{-1}=\Ht_{c'c}\sp\Ht_{vc}^{-1}=\Ht_{cv}.
\]

\subsubsection{Fluxes on the Links and Segments}

Just as we defined the holonomies on the links and segments from the
variables $h_{cc'}$ and $h_{vc}$, which were used in the continuity
relations \eqref{eq:continuity-ccp} and \eqref{eq:continuity-cv},
we can similarly define the fluxes on the links and segments from
the variables $\x_{c}^{c'}$ and $\x_{v}^{c}$. These will, again,
be used in the dual polarization.

Let us define \emph{fluxes along the links $\left(cc'\right)^{*}$
and segments $\left(vc\right)^{*}$}:
\begin{equation}
\X_{c}^{c'}\equiv h_{c}^{-1}\left(\sigma_{cc'}\right)\x_{c}^{c'}h_{c}\left(\sigma_{cc'}\right)\sp\X_{v}^{c}\equiv\mathring{h}_{v}^{-1}\left(\sigma_{vc}\right)\x_{v}^{c}\mathring{h}_{v}\left(\sigma_{vc}\right).\label{eq:fluxes-links}
\end{equation}
The factors of $h_{c}\left(\sigma_{cc'}\right)$ and $\mathring{h}_{v}\left(\sigma_{vc}\right)$
are needed because they appear alongside the integrals in the variations
\eqref{eq:DeltaHccp} and \eqref{eq:DeltaHvc}. Thus, if we want the
second terms in \eqref{eq:Theta_ccp} and \eqref{eq:Theta_vc} to
look like we want them to, we must include these extra factors in
the definition of the fluxes. The fluxes are still translational holonomies
between two cells (in the case of $\x_{c}^{c'}$) or a cell and a
disk (in the case of $\x_{v}^{c}$), but they contain an extra rotation
at the starting point.

The inverse link flux $\X_{c'}^{c}$ follows from \eqref{eq:continuity-ccp},
\eqref{eq:h-y-inverse} and $\sigma_{cc'}=\tau_{c'c}$, while the
inverse segment flux $\X_{c}^{v}\equiv h_{c}^{-1}\left(\sigma_{cv}\right)\x_{c}^{v}h_{c}\left(\sigma_{cv}\right)$
follows from \eqref{eq:continuity-cv}, \eqref{eq:h-y-inverse} and
$\sigma_{cv}=\tau_{vc}$:
\[
\X_{c'}^{c}=-\Ht_{cc'}^{-1}\X_{c}^{c'}\Ht_{cc'}\sp\X_{c}^{v}=-\Ht_{vc}^{-1}\X_{v}^{c}\Ht_{vc}.
\]

\subsubsection{The Symplectic Potential in Terms of the Holonomies and Fluxes}

With the holonomies and fluxes defined above, we find that we can
write the symplectic potential on the edges and arcs, \eqref{eq:Theta_ccp}
and \eqref{eq:Theta_vc}, as:
\[
\Theta_{cc'}=\left(1-\lambda\right)\XXt_{c}^{c'}\cdot\De H_{c}^{c'}-\lambda\X_{c}^{c'}\cdot\De\Ht_{c}^{c'},
\]
\[
\Theta_{vc}=\left(1-\lambda\right)\XXt_{v}^{c}\cdot\De H_{v}^{c}-\lambda\X_{v}^{c}\cdot\De\Ht_{v}^{c}.
\]
The full symplectic potential becomes:
\begin{align*}
\Theta & =\sum_{\left(cc'\right)}\left(\left(1-\lambda\right)\XXt_{c}^{c'}\cdot\De H_{c}^{c'}-\lambda\X_{c}^{c'}\cdot\De\Ht_{c}^{c'}\right)+\\
 & \qquad+\sum_{\left(vc\right)}\left(\left(1-\lambda\right)\XXt_{v}^{c}\cdot\De H_{v}^{c}-\lambda\X_{v}^{c}\cdot\De\Ht_{v}^{c}\right)+\\
 & \qquad+\sum_{v}\left(\X_{v}\cdot\delta\M_{v}-\left(\SS_{v}+\left[\M_{v},\X_{v}\right]\right)\cdot\De H_{v}\right).
\end{align*}
Notice how the holonomies and fluxes are always dual to each other:
one with tilde (on the edges/arcs) and one without tilde (on the links/segments).
For the $\lambda=0$ polarization, the holonomies are on the links
$\left(cc'\right)^{*}$ and segments $\left(vc\right)^{*}$ and the
fluxes are on their dual edges $\left(cc'\right)$ and arcs $\left(vc\right)$.
This is the polarization considered in \cite{FirstPaper}, and corresponds
to the usual loop gravity picture. For the $\lambda=1$ (dual) polarization,
we have the opposite case: the fluxes are on the links $\left(cc'\right)^{*}$
and segments $\left(vc\right)^{*}$ and the holonomies are on their
dual edges $\left(cc'\right)$ and arcs $\left(vc\right)$. For any
other choice of $\lambda$, we have a combination of both polarizations.

The phase space corresponding to $\X\cdot\De H$ for some flux $\X$
and holonomy $H$ is called the \emph{holonomy-flux phase space},
and it is the classical phase space of the spin networks which appear
in loop quantum gravity.

\section{The Gauss and Curvature Constraints}

We have seen that, in the continuum, the constraints are $\F=\T=0$.
Let us see how they translate to constraints on the discrete phase
space. There will be two types of constraints: the \emph{curvature
constraints }which corresponds to $\F=0$, and the \emph{Gauss constraints
}which correspond to $\T=0$. The constraints will be localized in
three different types of places: on the cells, on the disks, and on
the faces. After deriving all of the constraints and showing that
they are identically satisfied in our construction, we will summarize
and interpret them. The reader who is not interested in the details
of the calculation may wish to skip to Sec. \ref{subsec:Summary-and-Interpretation}.

\subsection{Derivation of the Constraints on the Cells}

\subsubsection{The Gauss Constraint on the Cells}

The cell Gauss constraint $\G_{c}$ will impose the torsionlessness
condition $\T\equiv\d_{\A}\E=0$ inside the cells:
\[
0=\G_{c}\equiv\int_{c}h_{c}\left(\d_{\A}\E\right)h_{c}^{-1}=\int_{c}\d\left(h_{c}\E h_{c}^{-1}\right)=\int_{\partial c}h_{c}\E h_{c}^{-1}=\int_{\partial c}\d\x_{c}.
\]
As we have seen, $\partial c$ is composed of edges $\left(cc_{i}\right)$
and arcs $\left(cv_{i}\right)$ such that
\[
\partial c=\bigcup_{i=1}^{N_{c}}\left(\left(cc_{i}\right)\cup\left(cv_{i}\right)\right).
\]
Therefore we can split the integral as follows:
\[
\G_{c}=\sum_{c'\ni c}\int_{\left(cc'\right)}\d\x_{c}+\sum_{v\ni c}\int_{\left(cv\right)}\d\x_{c},
\]
where $c'\ni c$ means ``all cells $c'$ adjacent to $c$'' and
$v\ni c$ means ``all vertices $v$ adjacent to $c$''.

Using the fluxes defined in \eqref{eq:flux-edge} and \eqref{eq:flux-arc-cv},
we get \footnote{Note that in \cite{FirstPaper} we used a different convention for
$\X_{c}^{v}$. This resulted in a relative minus sign between the
two terms, which does not appear in this paper.}
\begin{equation}
\G_{c}=\sum_{c'\ni c}\XXt_{c}^{c'}+\sum_{v\ni c}\XXt_{c}^{v}=0.\label{eq:Gauss-cell}
\end{equation}
This constraint is satisfied identically in our construction. Indeed,
from \eqref{eq:flux-edge} and \eqref{eq:flux-arc-cv} we have
\[
\XXt_{c}^{c'}=\x_{c}\left(\tau_{cc'}\right)-\x_{c}\left(\sigma_{cc'}\right)\sp\XXt_{c}^{v}=\x_{c}\left(\tau_{cv}\right)-\x_{c}\left(\sigma_{cv}\right).
\]
Since $\tau_{cc_{i}}=\sigma_{cv_{i}}$ and $\tau_{cv_{i}}=\sigma_{cc_{i+1}}$
(the end of an edge is the beginning of an arc and the end of an arc
is the beginning of an edge), and $\tau_{cv_{N_{c}}}=\sigma_{cc_{1}}$
(the end of the last arc is the beginning of the first edge), it is
easy to see that the sum $\sum_{c'\ni c}\XXt_{c}^{c'}+\sum_{v\ni c}\XXt_{c}^{v}$
evaluates to zero.

\subsubsection{The Curvature Constraint on the Cells}

The cell curvature constraint $F_{c}$ will impose that $\F\equiv\d\A+\hf\left[\A,\A\right]=0$
inside the cells. An equivalent condition is that the holonomy around
the cell evaluates to the identity:
\[
1=F_{c}\equiv\pexp\int_{\partial c}\A.
\]
Since $\partial c=\bigcup_{i=1}^{N_{c}}\left(\left(cc_{i}\right)\cup\left(cv_{i}\right)\right)$,
we may decompose this as a product of path-ordered exponentials over
edges and arcs:
\[
F_{c}=\prod_{i=1}^{N_{c}}\left(\pexp\int_{\left(cc_{i}\right)}\A\right)\left(\pexp\int_{\left(cv_{i}\right)}\A\right).
\]
Furthermore, since the geometry is flat, we may deform the paths so
that instead of going along the edges and arcs, it passes through
the node $c^{*}$. From \eqref{eq:undressed-c} we have that
\[
\pexp\int_{c^{*}}^{x}\A=h_{c}^{-1}\left(c^{*}\right)h_{c}\left(x\right),
\]
so
\[
\pexp\int_{\left(cc_{i}\right)}\A=\pexp\int_{\sigma_{cc_{i}}}^{\tau_{cc_{i}}}\A=\left(\pexp\int_{\sigma_{cc_{i}}}^{c^{*}}\A\right)\left(\pexp\int_{c^{*}}^{\tau_{cc_{i}}}\A\right)=h_{c}^{-1}\left(\sigma_{cc_{i}}\right)h_{c}\left(\tau_{cc_{i}}\right)=\Ht_{cc_{i}},
\]
where we used the definition \eqref{eq:Ht-ccp} of the holonomy on
the edge. Note that the contribution from $h_{c}\left(c^{*}\right)$
cancels. Similarly, we find
\begin{equation}
\pexp\int_{\left(cv_{i}\right)}\A=h_{c}^{-1}\left(\sigma_{cv_{i}}\right)h_{c}\left(\tau_{cv_{i}}\right)=\Ht_{cv_{i}},\label{eq:exp-cv}
\end{equation}
where we used \eqref{eq:Ht-vc-h_c}. Hence we obtain
\begin{equation}
F_{c}=\prod_{i=1}^{N_{c}}\Ht_{cc_{i}}\Ht_{cv_{i}}=1.\label{eq:curv-con-cells}
\end{equation}
This is the curvature constraint on the cells. It is easy to show
that it is satisfied identically in our construction. Indeed, using
again the relations $\tau_{cc_{i}}=\sigma_{cv_{i}}$, $\tau_{cv_{i}}=\sigma_{cc_{i+1}}$
and $\tau_{cv_{N_{c}}}=\sigma_{cc_{1}}$, we immediately see that
\[
\prod_{i=1}^{N_{c}}\Ht_{cc_{i}}\Ht_{cv_{i}}=\prod_{i=1}^{N_{c}}\left(h_{c}^{-1}\left(\sigma_{cc_{i}}\right)h_{c}\left(\tau_{cc_{i}}\right)\right)\left(h_{c}^{-1}\left(\sigma_{cv_{i}}\right)h_{c}\left(\tau_{cv_{i}}\right)\right)=1,
\]
as desired.

\subsection{Derivation of the Constraints on the Disks}

Since we have places the curvature and torsion excitations inside
the disks, the constraints on the disks must involve these excitations
-- namely, $\M_{v}$ and $\SS_{v}$. We will now see that this is
indeed the case.

\subsubsection{The Gauss Constraint on the Disks}

The disk Gauss constraint $\G_{v}$ will impose the torsionlessness
condition $\T\equiv\d_{\A}\E=0$ inside the \textit{punctured}\footnote{As we have seen, we only have $\T=0$ inside the \emph{punctured }disk
$v^{*}$; at the vertex $v$ itself there is torsion, but $v$ is
not part of $v^{*}$. Instead, it is on its (inner) boundary. As can
be seen from Fig. \ref{fig:Disk}, the path we take here, as given
by \eqref{eq:breakdown}, does not enclose the vertex, and therefore
the interior of the path is indeed torsionless.} disks:
\[
0=\G_{v}\equiv\int_{v^{*}}\mathring{h}_{v}\left(\d_{\A}\E\right)\mathring{h}_{v}^{-1}=\int_{v^{*}}\d\left(\mathring{h}_{v}\E\mathring{h}_{v}^{-1}\right)=\int_{\partial v^{*}}\mathring{h}_{v}\E\mathring{h}_{v}^{-1}=\int_{\partial v^{*}}\d\mathring{\x}_{v}.
\]
The boundary $\partial v^{*}$ is composed of the inner boundary $\partial_{0}v^{*}$,
the outer boundary $\partial_{R}v^{*}$, and the cut $C_{v}$:
\begin{equation}
\partial v^{*}=\partial_{0}v^{*}\cup\partial_{R}v^{*}\cup C_{v}.\label{eq:breakdown}
\end{equation}
Hence
\[
\G_{v}=\int_{\partial_{R}v^{*}}\d\mathring{\x}_{v}-\int_{\partial_{0}v^{*}}\d\mathring{\x}_{v}-\int_{C_{v}}\d\mathring{\x}_{v},
\]
where the minus signs represent the relative orientations of each
piece. On the inner boundary $\partial_{0}v^{*}$, we use the fact
that $\x_{v}$ takes the constant value $\x_{v}\left(v\right)$ to
obtain
\begin{align*}
\int_{\partial_{0}v^{*}}\d\mathring{\x}_{v} & =\e^{\M_{v}\phi_{v}}\left(\x_{v}\left(v\right)+\SS_{v}\phi_{v}\right)\e^{-\M_{v}\phi_{v}}\bll_{\phi_{v}=\alpha_{v}-\hf}^{\alpha_{v}+\hf}\\
 & =\SS_{v}+\e^{\M_{v}\left(\alpha_{v}-\hf\right)}\left(\e^{\M_{v}}\x_{v}\left(v\right)\e^{-\M_{v}}-\x_{v}\left(v\right)\right)\e^{-\M_{v}\left(\alpha_{v}-\hf\right)}.
\end{align*}
The outer boundary $\partial_{R}v^{*}$ splits into arcs, and we use
the definition \eqref{eq:flux-arc-vc} of the flux:
\[
\int_{\partial_{R}v^{*}}\d\mathring{\x}_{v}=\sum_{c\in v}\int_{\left(vc\right)}\d\mathring{\x}_{v}=\sum_{c\in v}\XXt_{v}^{c}.
\]
On the cut $C_{v}$, we have contributions from both sides, one at
$\phi_{v}=\alpha_{v}-\hf$ and another at $\phi_{v}=\alpha_{v}+\hf$
with opposite orientation. Since $\d\phi_{v}=0$ on the cut, we have:
\[
\d\mathring{\x}_{v}\bl_{C_{v}}=\e^{\M_{v}\phi_{v}}\d\x_{v}\e^{-\M_{v}\phi_{v}},
\]
and thus 
\begin{align*}
\int_{C_{v}}\d\mathring{\x}_{v} & =\int_{r=0}^{R}\left(\e^{\M_{v}\phi_{v}}\d\x_{v}\e^{-\M_{v}\phi_{v}}\bll_{\phi_{v}=\alpha_{v}-\hf}^{\phi_{v}=\alpha_{v}+\hf}\right)\\
 & =\e^{\M_{v}\left(\alpha_{v}-\hf\right)}\left(\e^{\M_{v}}\left(\x_{v}\left(v_{0}\right)-\x_{v}\left(v\right)\right)\e^{-\M_{v}}-\left(\x_{v}\left(v_{0}\right)-\x_{v}\left(v\right)\right)\right)\e^{-\M_{v}\left(\alpha_{v}-\hf\right)},
\end{align*}
since $\x_{v}$ has the value $\x_{v}\left(v_{0}\right)$ at $r=R$
and $\x_{v}\left(v\right)$ at $r=0$ on the cut.

Adding up the integrals, we find that the Gauss constraint on the
disk is 
\begin{equation}
\G_{v}=\sum_{c\in v}\XXt_{v}^{c}-\SS_{v}-\e^{\M_{v}\left(\alpha_{v}-\hf\right)}\left(\e^{\M_{v}}\x_{v}\left(v_{0}\right)\e^{-\M_{v}}-\x_{v}\left(v_{0}\right)\right)\e^{-\M_{v}\left(\alpha_{v}-\hf\right)}=0.\label{eq:G_v}
\end{equation}
In fact, since this constraint is used as a generator of symmetries
(as we will see below), it automatically comes dotted with a Cartan
element $\be_{v}$, which commutes with $\e^{\M_{v}}$. Therefore,
the last term may be ignored, and the constraint simplifies to
\[
\be_{v}\cdot\G_{v}=\be_{v}\cdot\left(\sum_{c\in v}\XXt_{v}^{c}-\SS_{v}\right)=0.
\]
Thus it may also be written
\begin{equation}
\sum_{c\in v}\XXt_{v}^{c}=\SS_{v}.\label{eq:Gauss-disk}
\end{equation}
To see that this constraint is satisfied identically in our construction,
let us combine \eqref{eq:flux-arc-vc} with \eqref{eq:u-v-def} to
obtain
\[
\XXt_{v}^{c}=\SS_{v}\left(\tau_{vc}-\sigma_{vc}\right)+\e^{\M_{v}\tau_{vc}}\x_{v}\left(\tau_{vc}\right)\e^{-\M_{v}\tau_{vc}}-\e^{\M_{v}\sigma_{vc}}\x_{v}\left(\sigma_{vc}\right)\e^{-\M_{v}\sigma_{vc}},
\]
where we used a slight abuse of notation by using $\sigma_{vc}$ and
$\tau_{vc}$ to denote the corresponding angles, $\sigma_{vc}\equiv\phi_{v}\left(\sigma_{vc}\right)$
and $\tau_{vc}\equiv\phi_{v}\left(\tau_{vc}\right)$. Let us now sum
over the fluxes for each arc. Since $\tau_{vc_{i}}=\sigma_{vc_{i+1}}$
(each arc ends where the next one starts) and $\tau_{vc_{N_{v}}}=\sigma_{vc_{1}}+1$
(the last arc ends a full circle after the first arc began\footnote{Recall that we are using scaled angles such that a full circle corresponds
to 1 instead of $2\pi$!}), we get
\[
\sum_{i=1}^{N_{v}}\XXt_{v}^{c_{i}}=\SS_{v}+\e^{\M_{v}\sigma_{vc_{1}}}\left(\e^{\M_{v}}\x_{v}\left(\sigma_{vc_{1}}\right)\e^{-\M_{v}}-\x_{v}\left(\sigma_{vc_{1}}\right)\right)\e^{-\M_{v}\sigma_{vc_{1}}}.
\]
Choosing without loss of generality the point $v_{0}$ to be at the
beginning of the first edge, $v_{0}=\sigma_{vc_{1}}$, and recalling
that this point corresponds to the angle $\phi_{v}=\alpha_{v}-\hf$,
we indeed obtain precisely the constraint \eqref{eq:G_v}.

\subsubsection{The Curvature Constraint on the Disks}

The disk curvature constraint $F_{v}$ will impose that $\F\equiv\d\A+\hf\left[\A,\A\right]=0$
inside the punctured disks\footnote{Again, we only have $\F=0$ inside the \emph{punctured }disk $v^{*}$;
at the vertex $v$ itself, there is curvature. However, the path of
integration does not enclose the vertex, and therefore the interior
of the path is indeed flat.}. An equivalent condition is that the holonomy around the punctured
disk evaluates to the identity:
\[
1=F_{v}\equiv\pexp\int_{\partial v^{*}}\A=\pexp\left(\int_{C_{v}^{-}}\A\right)\pexp\left(\int_{\partial_{R}v^{*}}\A\right)\pexp\left(\int_{C_{v}^{+}}\A\right)\pexp\left(\int_{\partial_{0}v^{*}}\A\right).
\]
Let us describe the path of integration step by step, referring to
Fig. \ref{fig:Disk}:
\begin{itemize}
\item We start at $v$, at the polar coordinates $r_{v}=0$ and $\phi_{v}=\alpha_{v}-1/2$.
\item We take the path $C_{v}^{-}$ along the cut at $\phi_{v}=\alpha_{v}-1/2$
from $r_{v}=0$ to $r_{v}=R$.
\item We go around the outer boundary $\partial_{R}v^{*}$ of the disk at
$r_{v}=R$ from $\phi_{v}=\alpha_{v}-1/2$ to $\phi_{v}=\alpha_{v}+1/2$.
\item We take the path $C_{v}^{+}$ along the cut at $\phi_{v}=\alpha_{v}+1/2$
from $r_{v}=R$ to $r_{v}=0$.
\item Finally, we go around the inner boundary $\partial_{0}v^{*}$ of the
disk at $r_{v}=0$ from $\phi_{v}=\alpha_{v}+1/2$ to $\phi_{v}=\alpha_{v}-1/2$,
back to our starting point.
\end{itemize}
Let us evaluate each term individually. On $C_{v}^{-}$ and $C_{v}^{+}$
we have\footnote{Note that the angle $\phi_{v}\left(x\right)$ in the term $\e^{\M_{v}\phi_{v}\left(x\right)}$
in \eqref{eq:undressed-v} refers to the \emph{difference in angles}
between the starting point and the final point, therefore it vanishes
in this case since the path along the cut is purely radial.} from \eqref{eq:undressed-v}:
\[
\pexp\left(\int_{C_{v}^{-}}\A\right)=\pexp\int_{v}^{v_{0}}\A=h_{v}^{-1}\left(v\right)h_{v}\left(v_{0}\right),
\]
\[
\pexp\left(\int_{C_{v}^{+}}\A\right)=\pexp\int_{v_{0}}^{v}\A=h_{v}^{-1}\left(v_{0}\right)h_{v}\left(v\right).
\]

On the inner boundary we have, again using \eqref{eq:undressed-v},
\[
\pexp\int_{\partial_{0}v^{*}}\A=\pexp\int_{v\left(\phi_{v}=\alpha_{v}+1/2\right)}^{v\left(\phi_{v}=\alpha_{v}-1/2\right)}\A=h_{v}^{-1}\left(v\right)\e^{-\M_{v}}h_{v}\left(v\right),
\]
since $h_{v}$ is periodic. The minus sign comes from the fact that
we are going from a larger angle to a smaller angle. Finally, on the
outer boundary we have, splitting into arcs and then using \eqref{eq:exp-cv}
and $\left(vc\right)=\left(cv\right)^{-1}$,
\[
\pexp\int_{\partial_{R}v^{*}}\A=\prod_{c\in v}\left(\pexp\int_{\left(vc\right)}\A\right)=\prod_{c\in v}\Ht_{vc}.
\]
In conclusion, the curvature constraint on the disks is
\[
F_{v}=h_{v}^{-1}\left(v\right)h_{v}\left(v_{0}\right)\left(\prod_{c\in v}\Ht_{vc}\right)h_{v}^{-1}\left(v_{0}\right)\e^{-\M_{v}}h_{v}\left(v\right)=1.
\]
In fact, we can multiply both sides by $h_{v}^{-1}\left(v_{0}\right)h_{v}\left(v\right)$
from the left and $h_{v}^{-1}\left(v\right)h_{v}\left(v_{0}\right)$
and obtain, after redefining $F_{v}$,
\[
F_{v}\equiv\left(\prod_{c\in v}\Ht_{vc}\right)h_{v}^{-1}\left(v_{0}\right)\e^{-\M_{v}}h_{v}\left(v_{0}\right)=1.
\]
This may be written more suggestively as
\[
\prod_{c\in v}\Ht_{vc}=h_{v}^{-1}\left(v_{0}\right)\e^{\M_{v}}h_{v}\left(v_{0}\right).
\]
Let us now show that this constraint is satisfied identically in our
construction. From \eqref{eq:Ht-vc} we have
\[
\Ht_{vc}\equiv\mathring{h}_{v}^{-1}\left(\sigma_{vc}\right)\mathring{h}_{v}\left(\tau_{vc}\right),
\]
and using the definition $\mathring{h}_{v}\equiv\e^{\M_{v}\phi_{v}}h_{v}$
from \eqref{eq:u-v-def} we get
\[
\Ht_{vc}=h_{v}^{-1}\left(\sigma_{vc}\right)\e^{\M_{v}\left(\phi_{v}\left(\tau_{vc}\right)-\phi_{v}\left(\sigma_{vc}\right)\right)}h_{v}\left(\tau_{vc}\right).
\]
Now, consider the product
\[
\prod_{c\in v}\Ht_{vc}=\prod_{i=1}^{N_{v}}h_{v}^{-1}\left(\sigma_{vc_{i}}\right)\e^{\M_{v}\left(\phi_{v}\left(\tau_{vc_{i}}\right)-\phi_{v}\left(\sigma_{vc_{i}}\right)\right)}h_{v}\left(\tau_{vc_{i}}\right).
\]
This is a telescoping product; the term $h_{v}\left(\tau_{vc_{i}}\right)$
always cancels the term $h_{v}^{-1}\left(\sigma_{vc_{i+1}}\right)$
in the next factor in the product. After the cancellations take place,
we are left only with $h_{v}^{-1}\left(\sigma_{vc_{1}}\right)$, the
product of exponents 
\[
\prod_{i=1}^{N_{v}}\e^{\M_{v}\left(\phi_{v}\left(\tau_{vc_{i}}\right)-\phi_{v}\left(\sigma_{vc_{i}}\right)\right)}=\e^{\M_{v}},
\]
where we used the fact that the angles sum to 1, and $h_{v}\left(\tau_{vc_{N_{v}}}\right)=h_{v}\left(\sigma_{vc_{1}}\right)$.
In conclusion:
\[
\prod_{c\in v}\Ht_{vc}=h_{v}^{-1}\left(\sigma_{vc_{1}}\right)\e^{\M_{v}}h_{v}\left(\sigma_{vc_{1}}\right).
\]
If we then choose, without loss of generality, the point $v_{0}$
(which defines the cut $C_{v}$) to be at $\sigma_{vc_{1}}$ (where
$c_{1}$ is an arbitrarily chosen cell), we get
\[
\prod_{c\in v}\Ht_{vc}=h_{v}^{-1}\left(v_{0}\right)\e^{\M_{v}}h_{v}\left(v_{0}\right),
\]
and we see that the constraint is indeed identically satisfied.

\subsection{Derivation of the Constraints on the Faces}

We have seen that the Gauss constraints, as we have defined them,
involve the fluxes on the edges and arcs. Since these fluxes are not
part of the phase space for $\lambda=1$, these constraints cannot
be imposed in that case. Similarly, the curvature constraints involve
the holonomies on the edges and arcs and therefore will not work for
the case $\lambda=0$. This is a result of formulating both constraints
on the cells and disks, which then requires us to use the holonomies
and fluxes on the edges and arcs which are on their boundaries.

Alternatively, instead of demanding that the torsion and curvature
vanish on the cells and disks, we may demand that they vanish on the
faces $f_{v}$ created by the spin network links. Since the (closures
of the) faces cover the entire spatial manifold $\Sigma$, this is
entirely equivalent.

This alternative form is obtained by deforming (or expanding) the
disks such that they coincide with the faces. The inner boundary $\partial_{0}v^{*}\to\partial_{0}f_{v}$
is still the vertex $v$. The outer boundary $\partial_{R}v^{*}\to\partial_{R}f_{v}$
now consists of links $\left(c_{i}c_{i+1}\right)^{*}$, where $i\in\left\{ 1,\ldots,N_{v}\right\} $
and $c_{N_{v}+1}\equiv c_{1}$. The point $v_{0}$ on the outer boundary
can now be identified, without loss of generality, with the node $c_{1}^{*}$.
Thus, the cut $C_{v}\to C_{f_{v}}$ now extends from $v$ to $c_{1}^{*}$.

Since the spatial manifold $\Sigma$ is now composed solely of the
union of the closures of the faces, and not cells and disks, we only
need one type of Gauss constraint and one type of curvature constraint.
Let us derive them now.

\subsubsection{The Gauss Constraint on the Faces}

The face Gauss constraint $\G_{f_{v}}$ will impose the torsionlessness
condition $\T\equiv\d_{\A}\E=0$ inside the faces:
\[
0=\G_{f_{v}}\equiv\int_{f_{v}}\mathring{h}_{v}\left(\d_{\A}\E\right)\mathring{h}_{v}^{-1}=\int_{f_{v}}\d\left(\mathring{h}_{v}\E\mathring{h}_{v}^{-1}\right)=\int_{\partial f_{v}}\mathring{h}_{v}\E\mathring{h}_{v}^{-1}=\int_{\partial f_{v}}\d\mathring{\x}_{v}.
\]
The boundary $\partial f_{v}$ is composed of the inner boundary $\partial_{0}f_{v}$,
the outer boundary $\partial_{R}f_{v}$, and the cut $C_{f_{v}}$:
\begin{equation}
\G_{f_{v}}=\int_{\partial_{R}f_{v}}\d\mathring{\x}_{v}-\int_{\partial_{0}f_{v}}\d\mathring{\x}_{v}-\int_{C_{f_{v}}}\d\mathring{\x}_{v},\label{eq:G_f_v-decomp}
\end{equation}
where the minus signs represent the relative orientations of each
piece. On the inner boundary $\partial_{0}f_{v}$, we use the fact
that $\x_{v}$ takes the constant value $\x_{v}\left(v\right)$ to
obtain as for $\partial_{0}v^{*}$ above:
\[
\int_{\partial_{0}f_{v}}\d\mathring{\x}_{v}=\SS_{v}+\e^{\M_{v}\left(\alpha_{v}-\hf\right)}\left(\e^{\M_{v}}\x_{v}\left(v\right)\e^{-\M_{v}}-\x_{v}\left(v\right)\right)\e^{-\M_{v}\left(\alpha_{v}-\hf\right)}.
\]
On the cut $C_{v}$, we have as before
\[
\int_{C_{v}}\d\mathring{\x}_{v}=\e^{\M_{v}\left(\alpha_{v}-\hf\right)}\left(\e^{\M_{v}}\left(\x_{v}\left(v_{0}\right)-\x_{v}\left(v\right)\right)\e^{-\M_{v}}-\left(\x_{v}\left(v_{0}\right)-\x_{v}\left(v\right)\right)\right)\e^{-\M_{v}\left(\alpha_{v}-\hf\right)}.
\]
The outer boundary $\partial_{R}f_{v}$ splits into links:
\begin{equation}
\int_{\partial_{R}f_{v}}\d\mathring{\x}_{v}=\sum_{i=1}^{N_{v}}\int_{c_{i}^{*}}^{c_{i+1}^{*}}\d\mathring{\x}_{v}=\sum_{i=1}^{N_{v}}\left(\mathring{\x}_{v}\left(c_{i+1}^{*}\right)-\mathring{\x}_{v}\left(c_{i}^{*}\right)\right).\label{eq:outer-1}
\end{equation}
Now, \eqref{eq:continuity-cv} can be inverted\footnote{Note that \eqref{eq:continuity-cv} is only valid on the arc $\left(vc\right)$,
which is the boundary between $c$ and $v^{*}$. However, since we
have expanded the disks, the arcs now coincide with the links, with
every arc $\left(vc\right)$ intersecting the two links connected
to the node $c^{*}$. Thus the equation is still valid at $c^{*}$
itself.} to get
\[
\mathring{\x}_{v}=h_{vc}\x_{c}h_{cv}+\x_{v}^{c}.
\]
Plugging into \eqref{eq:outer-1}, we get
\[
\int_{\partial_{R}f_{v}}\d\mathring{\x}_{v}=\sum_{i=1}^{N_{v}}\left(h_{vc_{i+1}}\x_{c_{i+1}}\left(c_{i+1}^{*}\right)h_{c_{i+1}v}-h_{vc_{i}}\x_{c_{i}}\left(c_{i}^{*}\right)h_{c_{i}v}+\x_{v}^{c_{i+1}}-\x_{v}^{c_{i}}\right).
\]
In fact, we can get rid of the first two terms, since the sum is telescoping:
each term of the from $h_{vc_{i}}\x_{c_{i}}\left(c_{i}^{*}\right)h_{c_{i}v}$
for $i=j$ is canceled\footnote{Of course, $\x_{v}^{c_{i+1}}$ and $\x_{v}^{c_{i}}$ also cancel each
other, but we choose to leave them.} by a term of the form $h_{vc_{i+1}}\x_{c_{i+1}}\left(c_{i+1}^{*}\right)h_{c_{i+1}v}$
for $i=j-1$. Thus we get
\begin{equation}
\int_{\partial_{R}f_{v}}\d\mathring{\x}_{v}=\sum_{i=1}^{N_{v}}\left(\x_{v}^{c_{i+1}}-\x_{v}^{c_{i}}\right).\label{eq:outer-2}
\end{equation}

Next, we note that from \eqref{eq:continuity-ccp} we have
\[
h_{cc'}=h_{c}h_{c'}^{-1}\sp\x_{c}^{c'}=\x_{c}-h_{cc'}\x_{c'}h_{c'c},
\]
and if we plug in \eqref{eq:continuity-cv} for $h_{c}$, $h_{c'}$,
$\x_{c}$ and $\x_{c'}$ we get 
\begin{equation}
h_{cc'}=h_{cv}h_{vc'},\label{eq:hccp-decomp}
\end{equation}
\begin{equation}
\x_{c}^{c'}=h_{cv}(\mathring{\x}_{v}-\x_{v}^{c})h_{vc}-h_{cc'}h_{c'v}(\mathring{\x}_{v}-\x_{v}^{c'})h_{vc'}h_{c'c}.\label{eq:xccp-decomp}
\end{equation}
From \eqref{eq:hccp-decomp} we see that $h_{cc'}h_{c'v}=h_{cv}$.
Plugging this into \eqref{eq:xccp-decomp}, we get the simplified
expression
\begin{equation}
\x_{c}^{c'}=h_{cv}\left(\x_{v}^{c'}-\x_{v}^{c}\right)h_{vc}.\label{eq:xccp-decomp-simp}
\end{equation}
Therefore, we may rewrite \eqref{eq:outer-2} as:
\begin{equation}
\int_{\partial_{R}f_{v}}\d\mathring{\x}_{v}=\sum_{i=1}^{N_{v}}h_{vc_{i}}\x_{c_{i}}^{c_{i+1}}h_{c_{i}v}.\label{eq:outer-3}
\end{equation}
Finally, we recall from \eqref{eq:fluxes-links} the definition of
the fluxes on the links:
\[
\X_{c}^{c'}\equiv h_{c}^{-1}\left(\sigma_{cc'}\right)\x_{c}^{c'}h_{c}\left(\sigma_{cc'}\right)=h_{c}^{-1}\left(v_{0}\right)\x_{c}^{c'}h_{c}\left(v_{0}\right).
\]
In the second equality we use the fact that, since we have deformed
the disks, the source point $\sigma_{cc'}$ of the edge $\left(cc'\right)$
lies on the spin network itself, and we can further deform the edge
such that $\sigma_{cc'}=v_{0}$. Plugging into \eqref{eq:outer-3},
we obtain
\[
\int_{\partial_{R}f_{v}}\d\mathring{\x}_{v}=\sum_{i=1}^{N_{v}}h_{vc_{i}}h_{c_{i}}\left(v_{0}\right)\X_{c_{i}}^{c_{i+1}}h_{c_{i}}^{-1}\left(v_{0}\right)h_{c_{i}v}.
\]
Finally, from \eqref{eq:continuity-cv} we have $h_{vc}h_{c}=\mathring{h}_{v}$,
and we get
\begin{equation}
\int_{\partial_{R}f_{v}}\d\mathring{\x}_{v}=\mathring{h}_{v}\left(v_{0}\right)\left(\sum_{i=1}^{N_{v}}\X_{c_{i}}^{c_{i+1}}\right)\mathring{h}_{v}^{-1}\left(v_{0}\right).\label{eq:outer-4}
\end{equation}
Adding up the integrals in \eqref{eq:G_f_v-decomp}, we obtain the
Gauss constraint on the faces:
\begin{equation}
\G_{f_{v}}=\mathring{h}_{v}\left(v_{0}\right)\left(\sum_{i=1}^{N_{v}}\X_{c_{i}}^{c_{i+1}}\right)\mathring{h}_{v}^{-1}\left(v_{0}\right)-\SS_{v}-\e^{\M_{v}\left(\alpha_{v}-\hf\right)}\left(\e^{\M_{v}}\x_{v}\left(v_{0}\right)\e^{-\M_{v}}-\x_{v}\left(v_{0}\right)\right)\e^{-\M_{v}\left(\alpha_{v}-\hf\right)}=0.\label{eq:Gauss-faces}
\end{equation}
Just like the Gauss constraint on the disks, this can be simplified
by noting that the constraint comes dotted with an element $\be_{f_{v}}$
of the Cartan subalgebra, which commutes with $\M_{v}$:
\[
\be_{f_{v}}\cdot\G_{f_{v}}=\be_{f_{v}}\cdot\left(h_{v}\left(v_{0}\right)\left(\sum_{i=1}^{N_{v}}\X_{c_{i}}^{c_{i+1}}\right)h_{v}^{-1}\left(v_{0}\right)-\SS_{v}\right)=0,
\]
where we used the fact that $\mathring{h}_{v}=\e^{\M_{v}\phi_{v}}h_{v}$
and the $\e^{\M_{v}\phi_{v}}$ part commutes with $\be_{f_{v}}$.
Thus, Gauss constraint on the faces may be rewritten in a simplified
way:
\[
\G_{f_{v}}\equiv\sum_{i=1}^{N_{v}}\X_{c_{i}}^{c_{i+1}}-h_{v}^{-1}\left(v_{0}\right)\SS_{v}h_{v}\left(v_{0}\right)=0.
\]
Let us now show that this constraint is satisfied identically. We
have from the definition of $\mathring{\x}_{v}$:
\begin{align*}
\int_{\partial_{R}f_{v}}\d\mathring{\x}_{v} & =\sum_{i=1}^{N_{v}}\int_{c_{i}^{*}}^{c_{i+1}^{*}}\d\mathring{\x}_{v}=\sum_{i=1}^{N_{v}}\left(\mathring{\x}_{v}\left(c_{i+1}^{*}\right)-\mathring{\x}_{v}\left(c_{i}^{*}\right)\right)\\
 & =\sum_{i=1}^{N_{v}}\left(\e^{\M_{v}\phi_{v}\left(c_{i+1}^{*}\right)}\x_{v}\left(c_{i+1}^{*}\right)\e^{-\M_{v}\phi_{v}\left(c_{i+1}^{*}\right)}-\e^{\M_{v}\phi_{v}\left(c_{i}^{*}\right)}\x_{v}\left(c_{i}^{*}\right)\e^{-\M_{v}\phi_{v}\left(c_{i}^{*}\right)}+\SS_{v}\left(\phi_{v}\left(c_{i+1}^{*}\right)-\phi_{v}\left(c_{i}^{*}\right)\right)\right).
\end{align*}
The sum is telescoping, and every term cancels the previous one. However,
in the term with $i=N_{v}$, we have
\[
\phi_{v}\left(c_{N_{v}+1}^{*}\right)=\phi_{v}\left(c_{1}^{*}\right)+1,
\]
since $\phi_{v}$, unlike $\x_{v}$, is not periodic. Therefore, the
first and last terms do not cancel each other. If we furthermore choose
$v_{0}\equiv c_{1}^{*}$, we get
\[
\int_{\partial_{R}f_{v}}\d\mathring{\x}_{v}=\SS_{v}+\e^{\M_{v}\phi_{v}\left(v_{0}\right)}\left(\e^{\M_{v}}\x_{v}\left(v_{0}\right)\e^{-\M_{v}}-\x_{v}\left(v_{0}\right)\right)\e^{-\M_{v}\phi_{v}\left(v_{0}\right)}.
\]
Then, using \eqref{eq:outer-4} we immediately obtain \eqref{eq:Gauss-faces},
as desired.

\subsubsection{The Curvature Constraint on the Faces}

The face curvature constraint $F_{f_{v}}$ will impose that $\F\equiv\d\A+\hf\left[\A,\A\right]=0$
inside the faces. As before, an equivalent condition is that the holonomy
around the face evaluates to the identity:
\[
1=F_{f_{v}}\equiv\pexp\int_{\partial f_{v}}\A=\pexp\left(\int_{C_{v}^{-}}\A\right)\pexp\left(\int_{\partial_{R}f_{v}}\A\right)\pexp\left(\int_{C_{v}^{+}}\A\right)\pexp\left(\int_{\partial_{0}f_{v}}\A\right).
\]
On $C_{v}^{-}$ and $C_{v}^{+}$ we have as before
\[
\pexp\left(\int_{C_{v}^{-}}\A\right)=\pexp\int_{v}^{v_{0}}\A=h_{v}^{-1}\left(v\right)h_{v}\left(v_{0}\right),
\]
\[
\pexp\left(\int_{C_{v}^{+}}\A\right)=\pexp\int_{v_{0}}^{v}\A=h_{v}^{-1}\left(v_{0}\right)h_{v}\left(v\right).
\]

On the inner boundary we have
\[
\pexp\int_{\partial_{0}f_{v}}\A=\pexp\int_{v\left(\phi_{v}=\alpha_{v}+1/2\right)}^{v\left(\phi_{v}=\alpha_{v}-1/2\right)}\A=h_{v}^{-1}\left(v\right)\e^{-\M_{v}}h_{v}\left(v\right).
\]
Finally, we decompose the outer boundary (which is now a loop on the
spin network) into links:
\[
\pexp\int_{\partial_{R}f_{v}}\A=\prod_{i=1}^{N_{v}}\left(\pexp\int_{c_{i}^{*}}^{c_{i+1}^{*}}\A\right).
\]
From \eqref{eq:exp-ccp-h} we know that
\[
\pexp\int_{c^{*}}^{c^{\prime*}}\A=h_{c}^{-1}\left(c^{*}\right)h_{cc'}h_{c'}\left(c^{\prime*}\right),
\]
and therefore
\[
\pexp\int_{\partial_{R}f_{v}}\A=\prod_{i=1}^{N_{v}}h_{c_{i}}^{-1}\left(c_{i}^{*}\right)h_{c_{i}c_{i+1}}h_{c_{i+1}}\left(c_{i+1}^{*}\right)=h_{c_{1}}^{-1}\left(v_{0}\right)\left(\prod_{i=1}^{N_{v}}h_{c_{i}c_{i+1}}\right)h_{c_{1}}\left(v_{0}\right),
\]
where we used the choice $v_{0}\equiv c_{1}^{*}$ and the fact that
the product is telescoping, that is, each term $h_{c_{i+1}}\left(c_{i+1}^{*}\right)$
cancels the term $h_{c_{i+1}}^{-1}\left(c_{i+1}^{*}\right)$ which
follows it, except the first and last terms, which have nothing to
cancel with.

Joining the integrals, we get
\[
h_{v}^{-1}\left(v\right)h_{v}\left(v_{0}\right)h_{c_{1}}^{-1}\left(v_{0}\right)\left(\prod_{i=1}^{N_{v}}h_{c_{i}c_{i+1}}\right)h_{c_{1}}\left(v_{0}\right)h_{v}^{-1}\left(v_{0}\right)\e^{-\M_{v}}h_{v}\left(v\right)=1.
\]
From \eqref{eq:continuity-cv} we find that
\[
h_{c_{1}}\left(v_{0}\right)h_{v}^{-1}\left(v_{0}\right)=h_{c_{1}v},
\]
and thus
\[
h_{v}^{-1}\left(v\right)h_{vc_{1}}\left(\prod_{i=1}^{N_{v}}h_{c_{i}c_{i+1}}\right)h_{c_{1}v}\e^{-\M_{v}}h_{v}\left(v\right)=1.
\]
For the last step, since we have the identity on the right-hand side,
we may cycle the group elements and rewrite the constraint as follows:
\[
F_{f_{v}}\equiv\left(\prod_{i=1}^{N_{v}}h_{c_{i}c_{i+1}}\right)h_{c_{1}v}\e^{-\M_{v}}h_{vc_{1}}=1.
\]
Switching to the notation of \eqref{eq:hol-link} and \eqref{eq:hol-seg},
we rewrite this as
\[
F_{f_{v}}\equiv\left(\prod_{i=1}^{N_{v}}H_{c_{i}c_{i+1}}\right)H_{c_{1}v}\e^{-\M_{v}}H_{vc_{1}}=1.
\]
An even nicer form of this constraint is
\begin{equation}
\prod_{i=1}^{N_{v}}H_{c_{i}c_{i+1}}=H_{c_{1}v}\e^{\M_{v}}H_{vc_{1}}.\label{eq:curv-con-faces}
\end{equation}
In other words, the loop of holonomies on the left-hand side would
be the identity if there is no curvature, that is, $\M_{v}=0$.

To show that this constraint is satisfied identically, we use \eqref{eq:exp-xy-phi}
with $x=c^{*}$ and $y=c^{\prime*}$:
\[
\pexp\int_{c^{*}}^{c^{\prime*}}\A=h_{v}^{-1}\left(c^{*}\right)\e^{\M_{v}\left(\phi_{v}\left(c^{\prime*}\right)-\phi_{v}\left(c^{*}\right)\right)}h_{v}\left(c^{\prime*}\right).
\]
Comparing with \eqref{eq:exp-ccp-h}, we see that
\[
h_{c}^{-1}\left(c^{*}\right)h_{cc'}h_{c'}\left(c^{\prime*}\right)=h_{v}^{-1}\left(c^{*}\right)\e^{\M_{v}\left(\phi_{v}\left(c^{\prime*}\right)-\phi_{v}\left(c^{*}\right)\right)}h_{v}\left(c^{\prime*}\right),
\]
and therefore
\[
h_{cc'}=h_{cv}\e^{\M_{v}\left(\phi_{v}\left(c^{\prime*}\right)-\phi_{v}\left(c^{*}\right)\right)}h_{vc'}.
\]
We now use this to rewrite the left-hand side of \eqref{eq:curv-con-faces}
as follows:
\[
\prod_{i=1}^{N_{v}}h_{c_{i}c_{i+1}}=\prod_{i=1}^{N_{v}}h_{c_{i}v}\e^{\M_{v}\left(\phi_{v}\left(c_{i+1}^{*}\right)-\phi_{v}\left(c_{i}^{*}\right)\right)}h_{vc_{i+1}}.
\]
Again, we have a telescoping product, and after canceling terms we
are left with
\[
\prod_{i=1}^{N_{v}}h_{c_{i}c_{i+1}}=h_{c_{1}v}\e^{\M_{v}}h_{vc_{1}},
\]
which is exactly \eqref{eq:curv-con-faces} after using \eqref{eq:hol-link}
and \eqref{eq:hol-seg}.

\subsection{\label{subsec:Summary-and-Interpretation}Summary and Interpretation}

In conclusion, we have obtained\footnote{One might wonder about the appearance of $h_{v}\left(v_{0}\right)$
in \eqref{eq:summary_G_f_v} and \eqref{eq:summary_F_v}, since the
true phase space variable is $H_{v}$, defined implicitly in \eqref{eq:DeltaH_v}
as a function of $h_{v}\left(v\right)$ and $h_{v}\left(v_{0}\right)$.
It is possible that there is an expression for these two constraints
in terms of $H_{v}$ instead of $h_{v}\left(v_{0}\right)$, but since
we only have an \textbf{implicit }definition for $H_{v}$ in terms
of its variation $\De H_{v}$, it is unclear how to obtain it. For
now, we simply assume that both $H_{v}$ and $h_{v}\left(v_{0}\right)$
are phase space variables. See also footnote \ref{fn:H_v-foot}.} Gauss constraints $\G_{c},\G_{v},\G_{f_{v}}$ and curvature constraints
$F_{c},F_{v},F_{f_{v}}$ for each cell $c$, disk $v^{*}$ and face
$f_{v}$:
\begin{equation}
\G_{c}\equiv\sum_{i=1}^{N_{c}}\left(\XXt_{c}^{c_{i}}+\XXt_{c}^{v_{i}}\right)=0,\label{eq:summary_G_c}
\end{equation}
\begin{equation}
\G_{v}\equiv\sum_{i=1}^{N_{v}}\XXt_{v}^{c_{i}}-\SS_{v}=0,\label{eq:summary_G_v}
\end{equation}
\begin{equation}
\G_{f_{v}}\equiv\sum_{i=1}^{N_{v}}\X_{c_{i}}^{c_{i+1}}-h_{v}^{-1}\left(v_{0}\right)\SS_{v}h_{v}\left(v_{0}\right)=0,\label{eq:summary_G_f_v}
\end{equation}
\begin{equation}
F_{c}\equiv\prod_{i=1}^{N_{c}}\Ht_{cc_{i}}\Ht_{cv_{i}}=1,\label{eq:summary_F_c}
\end{equation}
\begin{equation}
F_{v}\equiv\left(\prod_{i=1}^{N_{v}}\Ht_{vc_{i}}\right)h_{v}^{-1}\left(v_{0}\right)\e^{-\M_{v}}h_{v}\left(v_{0}\right)=1,\label{eq:summary_F_v}
\end{equation}
\begin{equation}
F_{f_{v}}\equiv\left(\prod_{i=1}^{N_{v}}H_{c_{i}c_{i+1}}\right)H_{c_{1}v}\e^{-\M_{v}}H_{vc_{1}}=1.\label{eq:summary_F_f_v}
\end{equation}
The Gauss constraint on the cell $c$ can also be written as
\begin{equation}
\sum_{c'\ni c}\XXt_{c}^{c'}=-\sum_{v\ni c}\XXt_{c}^{v}.\label{eq:G_c-split}
\end{equation}
It tells us that the sum of fluxes along the edges and arcs surrounding
$c$ is zero, as expected given that the interior of $c$ is flat.
Alternatively, we may say that the sum of fluxes along the edges is
prevented from summing to zero by the presence of the fluxes on the
arcs.

The Gauss constraint on the punctured disk $v^{*}$ can also be written
as
\begin{equation}
\sum_{c\in v}\XXt_{v}^{c}=\SS_{v}.\label{eq:G_v-split}
\end{equation}
It tells us that the sum of fluxes on the arcs of the disk is prevented
from summing to zero due to the torsion at the vertex $v$, as encoded
in the parameter $\SS_{v}$. Note that if $\SS_{v}=0$, that is, there
is no torsion at $v$, then the constraint becomes simply $\sum_{c\in v}\XXt_{v}^{c}=0.$

Importantly, notice that the sum $\sum_{v\ni c}\XXt_{c}^{v}$ on the
right-hand side of \eqref{eq:G_c-split} is over all the fluxes on
the arcs surrounding a particular cell $c$, while the sum $\sum_{c\in v}\XXt_{v}^{c}$
on the left-hand side of \eqref{eq:G_v-split} is over all the fluxes
on the arcs surrounding a particular disk $v^{*}$. While the sums
look alike at first sight, they are completely different and one cannot
be exchanged for the other.

The Gauss constraint on the face $f_{v}$ can also be written as
\[
\sum_{i=1}^{N_{v}}\X_{c_{i}}^{c_{i+1}}=h_{v}^{-1}\left(v_{0}\right)\SS_{v}h_{v}\left(v_{0}\right).
\]
It tells us that the sum of fluxes on the link forming the boundary
of the face is prevented from summing to zero due to the torsion at
the vertex $v$, as encoded in the parameter $\SS_{v}$.

The curvature constraint on the cell $c$ is
\[
\prod_{i=1}^{N_{c}}\Ht_{cc_{i}}\Ht_{cv_{i}}=1.
\]
It is analogous to the cell Gauss constraint, and imposes that the
product of holonomies along the boundary of the cell is the identity.

The curvature constraint on the punctured disk $v^{*}$ can also be
written as
\[
\prod_{c\in v}\Ht_{vc}=h_{v}^{-1}\left(v_{0}\right)\e^{\M_{v}}h_{v}\left(v_{0}\right).
\]
On the left-hand side, we have a loop of holonomies around the vertex
$v$. If $\M_{v}=0$, that is, there is no curvature at $v$, then
the right-hand side becomes the identity, as we would expect. Otherwise,
it is a quantity which depends on the curvature. The curvature constraint
on the disks is thus analogous to the Gauss constraint on the disks,
with torsion replaced by curvature.

Finally, the curvature constraint on the face $f_{v}$ can also be
written as
\[
\prod_{i=1}^{N_{v}}H_{c_{i}c_{i+1}}=H_{c_{1}v}\e^{\M_{v}}H_{vc_{1}}.
\]
It has the same meaning as the one on the disks, except that the loop
of holonomies around the vertex $v$ is now composed of links instead
of arcs.

\section{The Constraints as Generators of Symmetries}

\subsection{The Discrete Symplectic Form}

The discrete symplectic potential we have found is
\begin{align*}
\Theta & =\sum_{\left(cc'\right)}\left(\left(1-\lambda\right)\XXt_{c}^{c'}\cdot\De H_{c}^{c'}-\lambda\X_{c}^{c'}\cdot\De\Ht_{c}^{c'}\right)+\\
 & \qquad+\sum_{\left(vc\right)}\left(\left(1-\lambda\right)\XXt_{v}^{c}\cdot\De H_{v}^{c}-\lambda\X_{v}^{c}\cdot\De\Ht_{v}^{c}\right)+\\
 & \qquad+\sum_{v}\left(\X_{v}\cdot\delta\M_{v}-\left(\SS_{v}+\left[\M_{v},\X_{v}\right]\right)\cdot\De H_{v}\right).
\end{align*}
In the second line, we can use \eqref{eq:flux-arc-cv}, that is, $\XXt_{c}^{v}=-H_{cv}\XXt_{v}^{c}H_{vc}$,
to write
\[
\XXt_{v}^{c}\cdot\De H_{v}^{c}=\left(-H_{vc}\XXt_{c}^{v}H_{cv}\right)\cdot\left(\delta H_{vc}H_{cv}\right)=\XXt_{c}^{v}\cdot\De H_{c}^{v}.
\]
Thus, the labels $c$ and $v$ may be freely exchanged. Using the
identity $\delta\De H=\hf\left[\De H,\De H\right]$, we find that
the corresponding symplectic form $\Omega\equiv\delta\Theta$ is
\begin{align*}
\Omega & =\sum_{\left(cc'\right)}\left(\left(1-\lambda\right)\left(\delta\XXt_{c}^{c'}\cdot\De H_{c}^{c'}+\hf\XXt_{c}^{c'}\cdot\left[\De H_{c}^{c'},\De H_{c}^{c'}\right]\right)-\lambda\left(\delta\X_{c}^{c'}\cdot\De\Ht_{c}^{c'}+\hf\X_{c}^{c'}\cdot\left[\De\Ht_{c}^{c'},\De\Ht_{c}^{c'}\right]\right)\right)+\\
 & \qquad+\sum_{\left(vc\right)}\left(\left(1-\lambda\right)\left(\delta\XXt_{v}^{c}\cdot\De H_{v}^{c}+\hf\XXt_{v}^{c}\cdot\left[\De H_{v}^{c},\De H_{v}^{c}\right]\right)-\lambda\left(\delta\X_{v}^{c}\cdot\De\Ht_{v}^{c}+\hf\X_{v}^{c}\cdot\left[\De\Ht_{v}^{c},\De\Ht_{v}^{c}\right]\right)\right)+\\
 & \qquad+\sum_{v}\left(\delta\X_{v}\cdot\delta\M_{v}-\left(\delta\SS_{v}+\left[\delta\M_{v},\X_{v}\right]+\left[\M_{v},\delta\X_{v}\right]\right)\cdot\De H_{v}-\hf\left(\SS_{v}+\left[\M_{v},\X_{v}\right]\right)\cdot\left[\De H_{v},\De H_{v}\right]\right).
\end{align*}
We now look for transformations\footnote{The transformations will be given by the action of the Lie derivative
$\LLL_{\a}\equiv I_{\a}\delta+\delta I_{\a}$ where $I_{\a}$ is the
variational interior product with respect to $\a$. In the literature
the notation $\delta_{\a}$ is often used instead, but we avoid it
in order to prevent confusion with the variational exterior derivative
$\delta$.} with parameters $g_{c}\equiv\e^{\be_{c}},g_{v}\equiv\e^{\be_{v}},$
$\z_{c}$ and $\z_{v}$ such that:
\[
I_{\be_{c}}\Omega\propto-\be_{c}\cdot\delta\G_{c}\sp I_{\be_{v}}\Omega\propto-\be_{v}\cdot\delta\G_{v},
\]
\[
I_{\z_{c}}\Omega\propto-\z_{c}\cdot\De F_{c}\sp I_{\z_{v}}\Omega\propto-\z_{v}\cdot\De F_{v}.
\]
We will see that the proportionality coefficients will be $\lambda$-dependent.

\subsection{The Gauss Constraints as Generators of Rotations}

\subsubsection{The Gauss Constraint on the Cells}

Let us consider the rotation transformation with parameter $\be_{c}$
defined by
\[
\LLL_{\be_{c}}H_{cc'}=\be_{c}H_{cc'}\sp\LLL_{\be_{c}}H_{cv}=\be_{c}H_{cv}\sp\LLL_{\be_{c}}\XXt_{c}^{c'}=[\be_{c},\XXt_{c}^{c'}]\sp\LLL_{\be_{c}}\XXt_{c}^{v}=[\be_{c},\XXt_{c}^{v}],
\]
such that any other variables (in particular, those unrelated to the
particular $c$ of choice) are unaffected. Applying it to $\Omega$
and using the identity $I_{\be_{c}}\De H_{c}^{c'}=I_{\be_{c}}\De H_{c}^{v}=\be_{c}$,
we get:
\begin{align*}
I_{\be_{c}}\Omega & =\sum_{c'\ni c}\left(1-\lambda\right)\left([\be_{c},\XXt_{c}^{c'}]\cdot\De H_{c}^{c'}-\delta\XXt_{c}^{c'}\cdot\be_{c}+\XXt_{c}^{c'}\cdot[\be_{c},\De H_{c}^{c'}]\right)+\\
 & \qquad+\sum_{v\ni c}\left(1-\lambda\right)\left([\be_{c},\XXt_{c}^{v}]\cdot\De H_{c}^{v}-\delta\XXt_{c}^{v}\cdot\be_{c}+\XXt_{c}^{v}\cdot[\be_{c},\De H_{c}^{v}]\right).
\end{align*}
However, the first and last triple products in each line cancel each
other, and we are left with:
\begin{align*}
I_{\be_{c}}\Omega & =-\left(1-\lambda\right)\be_{c}\cdot\left(\sum_{c'\ni c}\delta\XXt_{c}^{c'}+\sum_{v\ni c}\delta\XXt_{c}^{v}\right)=-\left(1-\lambda\right)\be_{c}\cdot\delta\G_{c}.
\end{align*}
Hence this transformation is generated by the cell Gauss constraint
$\G_{c}$, given by \eqref{eq:summary_G_c}, as long as $\lambda\ne1$.

\subsubsection{The Gauss Constraint on the Disks}

Next we consider the rotation transformation with parameter $\be_{v}$
defined by
\[
\LLL_{\be_{v}}H_{vc}=\be_{v}H_{vc}\sp\LLL_{\be_{v}}\XXt_{v}^{c}=[\be_{v},\XXt_{v}^{c}]\sp\LLL_{\be_{v}}H_{v}=\left(1-\lambda\right)\be_{v}H_{v}\sp\LLL_{\be_{v}}\X_{v}=\left(1-\lambda\right)[\be_{v},\X_{v}],
\]
such that any other variables (in particular, those unrelated to the
particular $v$ of choice) are unaffected. Importantly, we choose
the 0-form $\be_{v}$ to be valued in the Cartan subalgebra, so it
commutes with $\M_{v}$ and $\SS_{v}$. Applying the transformation
to $\Omega$ and using the identities $I_{\be_{v}}\De H_{v}^{c}=\be_{v}$
and $I_{\be_{v}}\De H_{v}=\left(1-\lambda\right)\be_{v}$, we get:
\begin{align*}
I_{\be_{v}}\Omega & =\left(1-\lambda\right)\sum_{c\in v}\left([\be_{v},\XXt_{v}^{c}]\cdot\De H_{v}^{c}-\delta\XXt_{v}^{c}\cdot\be_{v}+\XXt_{v}^{c}\cdot\left[\be_{v},\De H_{v}^{c}\right]\right)+\\
 & \qquad+\left(1-\lambda\right)\left([\be_{v},\X_{v}]\cdot\delta\M_{v}-\left[\M_{v},[\be_{v},\X_{v}]\right]\cdot\De H_{v}\right)+\\
 & \qquad+\left(1-\lambda\right)\left(\left(\delta\SS_{v}+\left[\delta\M_{v},\X_{v}\right]+\left[\M_{v},\delta\X_{v}\right]\right)\cdot\be_{v}-\left(\SS_{v}+\left[\M_{v},\X_{v}\right]\right)\cdot\left[\be_{v},\De H_{v}\right]\right).
\end{align*}
Isolating $\be_{v}$ and using the fact that it commutes with $\M_{v}$
and $\SS_{v}$, we see that most terms cancel\footnote{In this calculation, we make use of the Jacobi identity:
\[
\left[\be_{v},\left[\M_{v},\X_{v}\right]\right]+\left[\M_{v},\left[\X_{v},\be_{v}\right]\right]=-\left[\X_{v},\left[\be_{v},\M_{v}\right]\right]=0.
\]
}, and we get:
\[
I_{\be_{v}}\Omega=-\left(1-\lambda\right)\be_{v}\cdot\left(\sum_{c\in v}\delta\XXt_{v}^{c}-\delta\SS_{v}\right)=-\left(1-\lambda\right)\be_{v}\cdot\G_{v}.
\]
Hence this transformation is generated by the disk Gauss constraint
$\G_{v}$, given by \eqref{eq:summary_G_v}, as long as $\lambda\ne1$.

\subsubsection{The Gauss Constraint on the Faces}

Lastly, we consider the rotation transformation with parameter $\be_{f_{v}}$
defined by
\[
\LLL_{\be_{f_{v}}}\Ht_{cc'}=-\be_{f_{v}}\Ht_{cc'}\sp\LLL_{\be_{f_{v}}}\X_{c}^{c'}=-[\be_{f_{v}},\X_{c}^{c'}]\sp\LLL_{\be_{f_{v}}}H_{v}=\lambda\bar{\be}_{f_{v}}H_{v}\sp\LLL_{\be_{f_{v}}}\X_{v}=\lambda[\bar{\be}_{f_{v}},\X_{v}],
\]
such that any other variables (in particular, those unrelated to the
particular $v$ of choice) are unaffected, and such that
\[
\be_{f_{v}}\equiv h_{v}^{-1}\left(v_{0}\right)\bar{\be}_{f_{v}}h_{v}\left(v_{0}\right),
\]
where $\bar{\be}_{f_{v}}$ is valued in the Cartan subalgebra. Applying
the transformation to $\Omega$, we get after a calculation analogous
to the one we did for the disks,
\begin{align*}
I_{\be_{f_{v}}}\Omega & =-\lambda\left(\be_{f_{v}}\cdot\sum_{c'\in c}\delta\X_{c}^{c'}-\bar{\be}_{f_{v}}\cdot\delta\SS_{v}\right)\\
 & =-\lambda\be_{f_{v}}\cdot\left(\sum_{c'\in c}\delta\X_{c}^{c'}-h_{v}^{-1}\left(v_{0}\right)\delta\SS_{v}h_{v}\left(v_{0}\right)\right).
\end{align*}
The variation of the Gauss constraint \eqref{eq:summary_G_f_v} is
\[
\delta\G_{f_{v}}=\sum_{i=1}^{N_{v}}\delta\X_{c_{i}}^{c_{i+1}}-h_{v}^{-1}\left(v_{0}\right)\left(\delta\SS_{v}+\left[\SS_{v},\De h_{v}\left(v_{0}\right)\right]\right)h_{v}\left(v_{0}\right),
\]
but since $\bar{\be}_{f_{v}}$ is in the Cartan we have $\bar{\be}_{f_{v}}\cdot\left[\SS_{v},\De h_{v}\left(v_{0}\right)\right]=0$,
so this simplifies to
\[
\be_{f_{v}}\cdot\delta\G_{f_{v}}=\be_{f_{v}}\cdot\left(\sum_{i=1}^{N_{v}}\delta\X_{c_{i}}^{c_{i+1}}-h_{v}^{-1}\left(v_{0}\right)\delta\SS_{v}h_{v}\left(v_{0}\right)\right).
\]
Thus, in conclusion,
\[
I_{\be_{f_{v}}}\Omega=-\lambda\be_{f_{v}}\cdot\delta\G_{f_{v}},
\]
and this transformation is generated by the face Gauss constraint
$\G_{v}$, given by \eqref{eq:summary_G_f_v}, as long as $\lambda\ne0$.

\subsection{The Curvature Constraints as Generators of Translations}

\subsubsection{The Curvature Constraint on the Cells}

For the curvature constraint on the cells, we would like to find a
translation transformation with parameter $\z_{c}$ such that
\[
I_{\z_{c}}\Omega=-\z_{c}\cdot\De F_{c}.
\]
First, we should calculate $\De F_{c}$. Recall that
\[
F_{c}\equiv\prod_{i=1}^{N_{c}}\Ht_{cc_{i}}\Ht_{cv_{i}}=1.
\]
To simplify the calculation, let us define $K_{i}\equiv\Ht_{cc_{i}}\Ht_{cv_{i}}$
such that we may write
\[
F_{c}=\prod_{i=1}^{N}K_{i}=K_{1}\cdots K_{N},
\]
where we omit the subscript $c$ on $N_{c}$ for brevity. Then
\begin{align*}
\delta F_{c} & =\delta K_{1}K_{2}\cdots K_{N}+K_{1}\delta K_{2}K_{3}\cdots K_{N}+\cdots+K_{1}\cdots K_{N-2}\delta K_{N-1}K_{N}+K_{1}\cdots K_{N-1}\delta K_{N}\\
 & =\De K_{1}K_{1}K_{2}\cdots K_{N}+K_{1}\De K_{2}K_{2}K_{3}\cdots K_{N}+\cdots+K_{1}\cdots K_{N-2}\De K_{N-1}K_{N-1}K_{N}+K_{1}\cdots K_{N-1}\De K_{N}K_{N},
\end{align*}
where $\De K_{i}\equiv\delta K_{i}K_{i}^{-1}$. Hence 
\begin{align*}
\De F_{c} & \equiv\delta F_{c}F_{c}^{-1}\\
 & =\De K_{1}+K_{1}\De K_{2}K_{1}^{-1}+\cdots+\left(K_{1}\cdots K_{N-2}\right)\De K_{N-1}\left(K_{1}\cdots K_{N-2}\right)^{-1}+\left(K_{1}\cdots K_{N-1}\right)\De K_{N}\left(K_{1}\cdots K_{N-1}\right)^{-1}\\
 & \equiv\sum_{i=1}^{N}\left(K_{1}\cdots K_{i-1}\right)\De K_{i}\left(K_{1}\cdots K_{i-1}\right)^{-1},
\end{align*}
where $K_{1}\cdots K_{i-1}\equiv1$ for $i=1$. For conciseness, we
may define $\chi_{i}$ such that $\chi_{1}\equiv1$ and, for $i>1$,
\[
\chi_{i}\equiv K_{1}\cdots K_{i-1}=\Ht_{cc_{1}}\Ht_{cv_{1}}\cdots\Ht_{cc_{i-1}}\Ht_{cv_{i-1}},
\]
and write
\[
\De F_{c}=\sum_{i=1}^{N}\chi_{i}\De K_{i}\chi_{i}^{-1}.
\]

Plugging in $K_{i}\equiv\Ht_{cc_{i}}\Ht_{cv_{i}}$ back, and using
the identity
\[
\De K_{i}=\De\Ht_{c}^{c_{i}}+\Ht_{cc_{i}}\De\Ht_{c}^{v_{i}}\Ht_{c_{i}c}
\]
we get
\begin{equation}
\De F_{c}=\sum_{i=1}^{N}\chi_{i}\left(\De\Ht_{c}^{c_{i}}+\Ht_{cc_{i}}\De\Ht_{c}^{v_{i}}\Ht_{c_{i}c}\right)\chi_{i}^{-1}.\label{eq:DeltaF_c}
\end{equation}
Now, if we transform only the dual fluxes $\X_{c}^{c'}$ and $\X_{c}^{v}$
(for a particular $c$), then we get
\[
I_{\z_{c}}\Omega=-\lambda\sum_{i=1}^{N_{c}}\left(\LLL_{\z_{c}}\X_{c}^{c_{i}}\cdot\De\Ht_{c}^{c_{i}}+\LLL_{\z_{c}}\X_{c}^{v_{i}}\cdot\De\Ht_{c}^{v_{i}}\right).
\]
Comparing with \eqref{eq:DeltaF_c}, we see that if we take
\[
\LLL_{\z_{c}}\X_{c}^{c_{i}}=\chi_{i}^{-1}\z_{c}\chi_{i}\sp\LLL_{\z_{c}}\X_{c}^{v_{i}}=\Ht_{c_{i}c}\chi_{i}^{-1}\z_{c}\chi_{i}\Ht_{cc_{i}},
\]
we will obtain
\[
I_{\z_{c}}\Omega=-\lambda\z_{c}\cdot\De F_{c},
\]
as required. Hence this transformation is generated by the cell curvature
constraint $F_{c}$, given by \eqref{eq:summary_F_c}, as long as
$\lambda\ne0$.

\subsubsection{The Curvature Constraint on the Disks}

As in the cell case, we would like to find a translation transformation
with parameter $\z_{v}$ such that
\[
I_{\z_{v}}\Omega=-\z_{v}\cdot\De F_{v},
\]
where 
\[
F_{v}\equiv\left(\prod_{i=1}^{N_{v}}\Ht_{vc_{i}}\right)h_{v}^{-1}\left(v_{0}\right)\e^{-\M_{v}}h_{v}\left(v_{0}\right)=1.
\]
First, we should calculate $\De F_{v}$. Let us define, omitting the
subscript $v$ on $N_{v}$ for brevity,
\[
K_{i}\equiv\Ht_{vc_{i}}\sp i\in\left\{ 1,\ldots,N\right\} ,
\]
\[
K_{N+1}\equiv h_{v}^{-1}\left(v_{0}\right)\e^{-\M_{v}}h_{v}\left(v_{0}\right),
\]
and
\[
\chi_{1}\equiv1\sp\chi_{i}\equiv K_{1}\cdots K_{i-1}.
\]
Then we may calculate similarly to the previous subsection
\[
F_{v}=\prod_{i=1}^{N+1}K_{i}\soosp\De F_{v}=\sum_{i=1}^{N+1}\chi_{i}\De K_{i}\chi_{i}^{-1}.
\]
Note that for $i=N+1$ we have
\[
\chi_{N+1}\equiv K_{1}\cdots K_{N}=F_{v}K_{N+1}^{-1}=F_{v}h_{v}^{-1}\left(v_{0}\right)\e^{\M_{v}}h_{v}\left(v_{0}\right),
\]
and since we are imposing $F_{v}=1$, we get simply
\[
\chi_{N+1}=h_{v}^{-1}\left(v_{0}\right)\e^{\M_{v}}h_{v}\left(v_{0}\right).
\]
Furthermore, using the fact that
\[
\De K_{N+1}=h_{v}^{-1}\left(v_{0}\right)\left(\e^{-\M_{v}}\De h_{v}\left(v_{0}\right)\e^{\M_{v}}-\De h_{v}\left(v_{0}\right)-\delta\M_{v}\right)h_{v}\left(v_{0}\right),
\]
we see that
\[
\chi_{N+1}\De K_{N+1}\chi_{N+1}^{-1}=h_{v}^{-1}\left(v_{0}\right)\left(\De h_{v}\left(v_{0}\right)-\e^{\M_{v}}\De h_{v}\left(v_{0}\right)\e^{-\M_{v}}-\delta\M_{v}\right)h_{v}\left(v_{0}\right).
\]
Therefore, we finally obtain the result
\[
\De F_{v}=\sum_{i=1}^{N_{v}}\chi_{i}\De\Ht_{v}^{c_{i}}\chi_{i}^{-1}+h_{v}^{-1}\left(v_{0}\right)\left(\De h_{v}\left(v_{0}\right)-\e^{\M_{v}}\De h_{v}\left(v_{0}\right)\e^{-\M_{v}}-\delta\M_{v}\right)h_{v}\left(v_{0}\right).
\]
Now, let us take
\[
\z_{v}\equiv h_{v}^{-1}\left(v_{0}\right)\bar{\z}_{v}h_{v}\left(v_{0}\right),
\]
where $\bar{\z}_{v}$ is a 0-form valued in the Cartan subalgebra,
and calculate $\z_{v}\cdot\De F_{v}$. We find that, since $\left[\bar{\z}_{v},\M_{v}\right]=0$,
the terms $\De h_{v}\left(v_{0}\right)-\e^{\M_{v}}\De h_{v}\left(v_{0}\right)\e^{-\M_{v}}$
cancel out and we are left with
\begin{equation}
\z_{v}\cdot\De F_{v}=\z_{v}\cdot\left(\sum_{i=1}^{N_{v}}\chi_{i}\De\Ht_{v}^{c_{i}}\chi_{i}^{-1}-h_{v}^{-1}\left(v_{0}\right)\delta\M_{v}h_{v}\left(v_{0}\right)\right).\label{eq:DeltaF_v}
\end{equation}
We may now derive the appropriate transformation. If we transform
only the segment flux $\X_{v}^{c}$ and the vertex flux $\X_{v}$
(for a particular $v$), then we get
\[
I_{\z_{v}}\Omega=-\lambda\sum_{i=1}^{N_{v}}\LLL_{\z_{v}}\X_{v}^{c_{i}}\cdot\De\Ht_{v}^{c_{i}}+\LLL_{\z_{v}}\X_{v}\cdot\left(\delta\M_{v}+\left[\M_{v},\De H_{v}\right]\right).
\]
Comparing with \eqref{eq:DeltaF_v}, we see that if we take
\[
\LLL_{\z_{v}}\X_{v}^{c_{i}}=\chi_{i}^{-1}\z_{v}\chi_{i}\sp\LLL_{\z_{v}}\X_{v}=\lambda\bar{\z}_{v},
\]
we will obtain, since $\bar{\z}_{v}\cdot\left[\M_{v},\De H_{v}\right]=0$,
\[
I_{\z_{v}}\Omega=-\lambda\z_{v}\cdot\left(\sum_{i=1}^{N_{v}}\chi_{i}\De\Ht_{v}^{c_{i}}\chi_{i}^{-1}-h_{v}^{-1}\left(v_{0}\right)\delta\M_{v}h_{v}\left(v_{0}\right)\right)=-\lambda\z_{v}\cdot\De F_{v},
\]
as required. Hence this transformation is generated by the disk curvature
constraint $F_{v}$, given by \eqref{eq:summary_F_v}, as long as
$\lambda\ne0$.

\subsubsection{The Curvature Constraint on the Faces}

We would now like to find a translation transformation with parameter
$\z_{f_{v}}$ such that
\[
I_{\z_{f_{v}}}\Omega=-\z_{f_{v}}\cdot\De F_{f_{v}},
\]
where 
\[
F_{f_{v}}\equiv\left(\prod_{i=1}^{N_{v}}H_{c_{i}c_{i+1}}\right)H_{c_{1}v}\e^{-\M_{v}}H_{vc_{1}}=1.
\]
As before, to calculate $\De F_{f_{v}}$ we define, omitting the subscript
$v$ on $N_{v}$ for brevity,
\[
K_{i}\equiv H_{c_{i}c_{i+1}}\sp i\in\left\{ 1,\ldots,N\right\} ,
\]
\[
K_{N+1}\equiv H_{c_{1}v}\e^{-\M_{v}}H_{vc_{1}},
\]
\[
\chi_{1}\equiv1\sp\chi_{i}\equiv K_{1}\cdots K_{i-1}.
\]
Then a similar calculation to the previous section gives
\[
\De F_{f_{v}}=\sum_{i=1}^{N_{v}}\chi_{i}\De H_{c_{i}}^{c_{i+1}}\chi_{i}^{-1}+H_{c_{1}v}\left(\De H_{v}^{c_{1}}-\e^{\M_{v}}\De H_{v}^{c_{1}}\e^{\M_{v}}-\delta\M_{v}\right)H_{vc_{1}},
\]
and if we take
\[
\z_{f_{v}}\equiv H_{c_{1}v}\bar{\z}_{f_{v}}H_{vc_{1}},
\]
where $\bar{\z}_{f_{v}}$ is a 0-form valued in the Cartan subalgebra,
we get
\begin{equation}
\z_{f_{v}}\cdot\De F_{f_{v}}=\z_{f_{v}}\cdot\left(\sum_{i=1}^{N_{v}}\chi_{i}\De H_{c_{i}}^{c_{i+1}}\chi_{i}^{-1}-H_{c_{1}v}\delta\M_{v}H_{vc_{1}}\right).\label{eq:DeltaF_f_v}
\end{equation}
We may now derive the appropriate transformation. If we transform
only the edge flux $\XXt_{c}^{c'}$ and the vertex flux $\X_{v}$
(for a particular $v$), then we get
\[
I_{\z_{f_{v}}}\Omega=\left(1-\lambda\right)\sum_{i=1}^{N_{v}}\LLL_{\z_{f_{v}}}\XXt_{c_{i}}^{c_{i+1}}\cdot\De H_{c}^{c'}+\LLL_{\z_{f_{v}}}\X_{v}\cdot\left(\delta\M_{v}+\left[\M_{v},\De H_{v}\right]\right).
\]
Comparing with \eqref{eq:DeltaF_f_v}, we see that if we take
\[
\LLL_{\z_{f_{v}}}\XXt_{c_{i}}^{c_{i+1}}=-\chi_{i}^{-1}\z_{f_{v}}\chi_{i}\sp\LLL_{\z_{v}}\X_{v}=\left(1-\lambda\right)H_{vc_{1}}\z_{f_{v}}H_{c_{1}v},
\]
we will obtain
\[
I_{\z_{f_{v}}}\Omega=-\left(1-\lambda\right)\z_{f_{v}}\cdot\left(\sum_{i=1}^{N_{v}}\chi_{i}\De H_{c_{i}}^{c_{i+1}}\chi_{i}^{-1}-H_{c_{1}v}\delta\M_{v}H_{vc_{1}}\right)=-\left(1-\lambda\right)\z_{f_{v}}\cdot\De F_{f_{v}},
\]
as required. Hence this transformation is generated by the face curvature
constraint $F_{v}$, given by \eqref{eq:summary_F_v}, as long as
$\lambda\ne0$.

\subsection{Conclusions}

We have found that the Gauss constraints $\G_{c},\G_{v},\G_{f_{v}}$
and curvature constraints $F_{c},F_{v},F_{f_{v}}$ for each cell $c$,
disk $v^{*}$ and face $f_{v}$, given by \eqref{eq:summary_G_c},
\eqref{eq:summary_G_v}, \eqref{eq:summary_G_f_v}, \eqref{eq:summary_F_c},
\eqref{eq:summary_F_v} and \eqref{eq:summary_F_f_v}, generate transformations
with rotation parameters $\be_{c},\be_{v},\be_{f_{v}}$ and translations
parameters $\z_{c},\z_{v},\z_{f_{v}}$ as follows:
\[
I_{\be_{c}}\Omega=-\left(1-\lambda\right)\be_{c}\cdot\delta\G_{c}\sp I_{\be_{v}}\Omega=-\left(1-\lambda\right)\be_{v}\cdot\delta\G_{v}\sp I_{\be_{f_{v}}}\Omega=-\lambda\be_{f_{v}}\cdot\delta\G_{f_{v}},
\]
\[
I_{\z_{c}}\Omega=-\lambda\z_{c}\cdot\De F_{c}\sp I_{\z_{v}}\Omega=-\lambda\z_{v}\cdot\De F_{v}\sp I_{\z_{f_{v}}}\Omega=-\left(1-\lambda\right)\z_{f_{v}}\cdot\De F_{f_{v}}.
\]
The Gauss constraint on the cell $c$ generates rotations of the holonomies
on the links $\left(cc'\right)^{*}$ and segments $\left(cv\right)^{*}$
connected to the node $c^{*}$ and the fluxes on the edges $\left(cc'\right)$
and arcs $\left(cv\right)$ surrounding $c$:
\[
\LLL_{\be_{c}}H_{cc'}=\be_{c}H_{cc'}\sp\LLL_{\be_{c}}H_{cv}=\be_{c}H_{cv}\sp\LLL_{\be_{c}}\XXt_{c}^{c'}=[\be_{c},\XXt_{c}^{c'}]\sp\LLL_{\be_{c}}\XXt_{c}^{v}=[\be_{c},\XXt_{c}^{v}],
\]
where $\be_{c}$ is a $\mfg^{*}$-valued 0-form.

The Gauss constraint on the disk $v^{*}$ generates rotations of the
holonomies on the segments $\left(vc\right)^{*}$ connected to the
vertex $v$ and the fluxes on the arcs $\left(vc\right)$ surrounding
$v^{*}$, as well as the holonomy and flux on the vertex $v$ itself:
\[
\LLL_{\be_{v}}H_{vc}=\be_{v}H_{vc}\sp\LLL_{\be_{v}}\XXt_{v}^{c}=[\be_{v},\XXt_{v}^{c}]\sp\LLL_{\be_{v}}H_{v}=\left(1-\lambda\right)\be_{v}H_{v}\sp\LLL_{\be_{v}}\X_{v}=\left(1-\lambda\right)[\be_{v},\X_{v}],
\]
where $\be_{v}$ is a 0-form valued in the Cartan subalgebra $\mfh^{*}$
of $\mfg^{*}$.

The Gauss constraint on the face $f_{v}$ generates rotations of the
fluxes on the links $\left(cc'\right)^{*}$ surrounding $f_{v}$ and
the holonomies on their dual edges $\left(cc'\right)$, as well as
the holonomy and flux on the vertex $v$ itself:
\[
\LLL_{\be_{f_{v}}}\Ht_{cc'}=-\be_{f_{v}}\Ht_{cc'}\sp\LLL_{\be_{f_{v}}}\X_{c}^{c'}=-[\be_{f_{v}},\X_{c}^{c'}]\sp\LLL_{\be_{f_{v}}}H_{v}=\lambda\bar{\be}_{f_{v}}H_{v}\sp\LLL_{\be_{f_{v}}}\X_{v}=\lambda[\bar{\be}_{f_{v}},\X_{v}],
\]
where $\bar{\be}_{f_{v}}$ is a 0-form valued in the Cartan subalgebra
$\mfh^{*}$ of $\mfg^{*}$ and $\be_{f_{v}}\equiv h_{v}^{-1}\left(v_{0}\right)\bar{\be}_{f_{v}}h_{v}\left(v_{0}\right)$.

The curvature constraint on the cell $c$ generates translations\footnote{Note that the curvature constraints do not transform any holonomies,
since the holonomies are unaffected by translations.} of the fluxes on the links $\left(cc'\right)^{*}$ and segments $\left(cv\right)^{*}$
connected to the node $c^{*}$:
\[
\LLL_{\z_{c}}\X_{c}^{c_{i}}=\chi_{i}^{-1}\z_{c}\chi_{i}\sp\LLL_{\z_{c}}\X_{c}^{v_{i}}=\Ht_{c_{i}c}\chi_{i}^{-1}\z_{c}\chi_{i}\Ht_{cc_{i}},
\]
where
\[
\chi_{1}\equiv1\sp\chi_{i}=\Ht_{cc_{1}}\Ht_{cv_{1}}\cdots\Ht_{cc_{i-1}}\Ht_{cv_{i-1}},
\]
and $\z_{c}$ is a $\mfg$-valued 0-form.

The curvature constraint on the disk $v^{*}$ generates translations
of the fluxes on the segments $\left(vc\right)^{*}$ connected to
the vertex $v$, as well as the flux on the vertex $v$ itself:
\[
\LLL_{\z_{v}}\X_{v}^{c_{i}}=\chi_{i}^{-1}\z_{v}\chi_{i}\sp\LLL_{\z_{v}}\X_{v}=\lambda\bar{\z}_{v},
\]
where
\[
\chi_{1}\equiv1\sp\chi_{i}\equiv\Ht_{vc_{i}}\cdots\Ht_{vc_{i-1}},
\]
$\bar{\z}_{v}$ is a 0-form valued in the Cartan subalgebra $\mfh$
of $\mfg$, and $\z_{v}\equiv h_{v}^{-1}\left(v_{0}\right)\bar{\z}_{v}h_{v}\left(v_{0}\right)$.

The curvature constraint on the face $f_{v}$ generates translations
of the fluxes on the edges $\left(cc'\right)$ dual to the links surrounding
the face $f_{v}$, as well as the flux on the vertex $v$ itself:

\[
\LLL_{\z_{f_{v}}}\XXt_{c_{i}}^{c_{i+1}}=-\chi_{i}^{-1}\z_{f_{v}}\chi_{i}\sp\LLL_{\z_{v}}\X_{v}=\left(1-\lambda\right)H_{vc_{1}}\z_{f_{v}}H_{c_{1}v},
\]
where
\[
\chi_{1}\equiv1\sp\chi_{i}\equiv H_{c_{1}c_{2}}\cdots H_{c_{i-1}c_{i}},
\]
and $\z_{f_{v}}$ is a 0-form valued in the Cartan subalgebra $\mfh$
of $\mfg$.

Importantly, in the case $\lambda=0$, the usual loop gravity polarization,
the curvature constraints on the cells and disks do not generate any
transformations since $I_{\z_{c}}\Omega=I_{\z_{v}}\Omega=0$. Similarly,
for the case $\lambda=1$, the dual polarization, the Gauss constraints
on the cells and disks do not generate any transformations since $I_{\be_{c}}\Omega=I_{\be_{v}}\Omega=0$.
Of course, the reason for this is that, as we noted earlier, these
constraints are formulated in the first place in terms of holonomies
and fluxes which only exist in a particular polarization. Thus for
$\lambda=0$ we must instead use the curvature constraint on the faces\footnote{Which is indeed what we did in \cite{FirstPaper}.},
and for $\lambda=1$ we must instead use the Gauss constraint on the
faces.

In the hybrid polarization with $\lambda=1/2$, all of the discrete
variables exist: there are holonomies and fluxes on both the links/edges
and the arcs/segments. Therefore, in this polarization all 6 types
of constraints may be consistently formulated using the available
variables, and all of them generate transformations.

\section{Summary and Outlook}

In this paper, we generalized the work of \cite{FirstPaper} to include
the most general possible discretization. We discovered a family of
polarizations of the discrete phase space, given by different values
of the parameter $\lambda$. Of these, the three cases of interest
are $\lambda=0$, $\lambda=1$ and $\lambda=1/2$.

In the $\lambda=0$ case, which is the one we discussed in \cite{FirstPaper},
the holonomies are on the links (and segments) and the fluxes are
on their corresponding edges (and arcs), as in the familiar case of
loop gravity. The Gauss constraints on the cells and disks generate
rotations for all of the discrete variables, while the curvature constraints
on the faces generate translations only for the fluxes on the edges
and vertices.

In the $\lambda=1$ case, the positions of the holonomies and fluxes
are reversed. The holonomies are on the edges (and arcs) and the fluxes
are on their corresponding links (and segments). The curvature constraints
on the cells and disks generate translations for all of the fluxes,
while the Gauss constraints on the faces generate rotations only for
the fluxes on the links, holonomies on the edges, and fluxes and holonomies
on the vertices.

Finally, in the $\lambda=1/2$ case, we have the variables for both
polarizations simultaneously. All 6 types of constraints exist, and
each of them generates its associated transformations.

Intuitively, we may now conclude that the $\lambda=0$ polarization
corresponds to usual 2+1D general relativity, while $\lambda=1$ (the
dual polarization) corresponds to teleparallel gravity. This intuition
is motivated by the fact that, as we have seen, in the $\lambda=1$
polarization the holonomies and fluxes switch places, and thus the
curvature and torsion (and their respective constraints) also switch
places.

Since 2+1D general relativity has curvature but zero torsion, and
teleparallel gravity has torsion but zero curvature, it makes sense
to claim that these polarizations are related. Indeed, this is why
we used the same parameter $\lambda$ in both \eqref{eq:Theta-general}
and \eqref{eq:Theta-general-2}. Since the choice $\lambda=1/2$ in
\eqref{eq:Theta-general} corresponds to Chern-Simons theory , we
may further claim that the $\lambda=1/2$ polarization in the discrete
case is a discretization of Chern-Simons theory. Thus:
\begin{itemize}
\item The polarization $\lambda=0$ corresponds to 2+1D general relativity,
\item The polarization $\lambda=1/2$ corresponds to Chern-Simons theory,
\item The polarization $\lambda=1$ corresponds to teleparallel gravity.
\end{itemize}
A discussion of quantization in different polarizations is provided
in \cite{clement}. There, it is shown that in the $\lambda=0$ case,
the Gauss constraint is imposed at the kinematical level while the
curvature constraint encodes the dynamics. In the $\lambda=1$, the
roles of the constraints are reversed. This again motivates a relation
between the $\lambda=1$ case and teleparallel gravity. The relation
of the $\lambda=1/2$ case to Chern-Simons theory is motivated in
\cite{Dupuis:2017otn}. We leave a more in-depth discussion and analysis
of the relations between the $\lambda=1$ case and teleparallel gravity,
and between the $\lambda=1/2$ case and Chern-Simons theory, to future
work.

Following our exhaustive study of discretization of 2+1D gravity,
it is our goal to adapt this discretization scheme to the physically
relevant case of 3+1D gravity. While in the 2+1D case there is only
one place where an integration may be performed in two different ways,
in the 3+1D case there are two such integrations, since we have one
more dimension. We expect to find both 3+1D general relativity and
3+1D teleparallel gravity as different polarizations of the discrete
phase space. The discretization in 3+1 dimensions will be presented
in an upcoming paper \cite{barak4d}.

\subsection{Acknowledgments}

The author would like to thank Laurent Freidel and Florian Girelli
for their mentorship, and the anonymous referee for helpful comments.
This research was supported in part by Perimeter Institute for Theoretical
Physics. Research at Perimeter Institute is supported by the Government
of Canada through the Department of Innovation, Science and Economic
Development Canada and by the Province of Ontario through the Ministry
of Research, Innovation and Science.

\bibliographystyle{Utphys}
\phantomsection\addcontentsline{toc}{section}{\refname}\bibliography{Dual-2P1-LQG-On-The-Edge}

\providecommand{\href}[2]{#2}\begingroup\raggedright\begin{thebibliography}{10}

\bibitem{2005physics...3046U}
A.~{Unzicker} and T.~{Case}, ``{Translation of Einstein's Attempt of a Unified
  Field Theory with Teleparallelism},'' {\em arXiv e-prints} (Mar, 2005)
  physics/0503046, \href{http://arxiv.org/abs/physics/0503046}{{\ttfamily
  arXiv:physics/0503046 [physics.hist-ph]}}.

\bibitem{Maluf:2013gaa}
J.~W. Maluf, ``{The teleparallel equivalent of general relativity},''
  \href{http://dx.doi.org/10.1002/andp.201200272}{{\em Annalen Phys.}
  {\bfseries 525} (2013) 339--357},
\href{http://arxiv.org/abs/1303.3897}{{\ttfamily arXiv:1303.3897 [gr-qc]}}.

\bibitem{Ferraro:2016wht}
R.~Ferraro and M.~J. Guzman, ``{Hamiltonian formulation of teleparallel
  gravity},'' \href{http://dx.doi.org/10.1103/PhysRevD.94.104045}{{\em Phys.
  Rev.} {\bfseries D94} no.~10, (2016) 104045},
\href{http://arxiv.org/abs/1609.06766}{{\ttfamily arXiv:1609.06766 [gr-qc]}}.

\bibitem{Rovelli:2004tv}
C.~Rovelli, {\em {Quantum gravity}}.
\newblock Cambridge Monographs on Mathematical Physics. Cambridge Univ. Pr.,
  Cambridge, UK,
2004.
\newblock

\bibitem{thiemann_2007}
T.~Thiemann, \href{http://dx.doi.org/10.1017/CBO9780511755682}{{\em Modern
  Canonical Quantum General Relativity}}.
\newblock Cambridge Monographs on Mathematical Physics. Cambridge University
  Press, 2007.

\bibitem{Henneaux:1992ig}
M.~Henneaux and C.~Teitelboim, {\em {Quantization of gauge systems}}.
\newblock
1992.
\newblock

\bibitem{Carlip:1998uc}
S.~Carlip, {\em {Quantum gravity in 2+1 dimensions}}.
\newblock Cambridge University Press,
2003.
\newblock

\bibitem{Ashtekar:1991kc}
A.~Ashtekar and C.~J. Isham, ``{Representations of the holonomy algebras of
  gravity and nonAbelian gauge theories},''
  \href{http://dx.doi.org/10.1088/0264-9381/9/6/004}{{\em Class. Quant. Grav.}
  {\bfseries 9} (1992) 1433--1468},
\href{http://arxiv.org/abs/hep-th/9202053}{{\ttfamily arXiv:hep-th/9202053
  [hep-th]}}.

\bibitem{clement}
C.~Delcamp, L.~Freidel, and F.~Girelli, ``{Dual loop quantizations of 3d
  gravity},''
\href{http://arxiv.org/abs/1803.03246}{{\ttfamily arXiv:1803.03246 [gr-qc]}}.

\bibitem{Dittrich:2014wpa}
B.~Dittrich and M.~Geiller, ``{A new vacuum for Loop Quantum Gravity},''
  \href{http://dx.doi.org/10.1088/0264-9381/32/11/112001}{{\em Class. Quant.
  Grav.} {\bfseries 32} no.~11, (2015) 112001},
\href{http://arxiv.org/abs/1401.6441}{{\ttfamily arXiv:1401.6441 [gr-qc]}}.

\bibitem{Dittrich:2014wda}
B.~Dittrich and M.~Geiller, ``{Flux formulation of loop quantum gravity:
  Classical framework},''
  \href{http://dx.doi.org/10.1088/0264-9381/32/13/135016}{{\em Class. Quant.
  Grav.} {\bfseries 32} no.~13, (2015) 135016},
\href{http://arxiv.org/abs/1412.3752}{{\ttfamily arXiv:1412.3752 [gr-qc]}}.

\bibitem{Freidel:2010aq}
L.~Freidel and S.~Speziale, ``{Twisted geometries: A geometric parametrisation
  of SU(2) phase space},''
  \href{http://dx.doi.org/10.1103/PhysRevD.82.084040}{{\em Phys. Rev.}
  {\bfseries D82} (2010) 084040},
\href{http://arxiv.org/abs/1001.2748}{{\ttfamily arXiv:1001.2748 [gr-qc]}}.

\bibitem{FirstPaper}
L.~Freidel, F.~Girelli, and B.~Shoshany, ``{2+1D Loop Quantum Gravity on the
  Edge},'' \href{http://dx.doi.org/10.1103/PhysRevD.99.046003}{{\em Phys. Rev.
  D} {\bfseries 99} (Feb, 2019) 046003},
  \href{http://arxiv.org/abs/1811.04360}{{\ttfamily arXiv:1811.04360 [gr-qc]}}.

\bibitem{Dupuis:2017otn}
M.~Dupuis, L.~Freidel, and F.~Girelli, ``{Discretization of 3d gravity in
  different polarizations},''
  \href{http://dx.doi.org/10.1103/PhysRevD.96.086017}{{\em Phys. Rev.}
  {\bfseries D96} no.~8, (2017) 086017},
\href{http://arxiv.org/abs/1701.02439}{{\ttfamily arXiv:1701.02439 [gr-qc]}}.

\bibitem{Horowitz:1989ng}
G.~T. Horowitz, ``{Exactly Soluble Diffeomorphism Invariant Theories},''
\href{http://dx.doi.org/10.1007/BF01218410}{{\em Commun. Math. Phys.}
  {\bfseries 125} (1989) 417}.

\bibitem{Alekseev:1993rj}
A.~{\relax Yu}. Alekseev and A.~Z. Malkin, ``{Symplectic structure of the
  moduli space of flat connection on a Riemann surface},''
  \href{http://dx.doi.org/10.1007/BF02101598}{{\em Commun. Math. Phys.}
  {\bfseries 169} (1995) 99--120},
\href{http://arxiv.org/abs/hep-th/9312004}{{\ttfamily arXiv:hep-th/9312004
  [hep-th]}}.

\bibitem{Alekseev:1994pa}
A.~{\relax Yu}. Alekseev, H.~Grosse, and V.~Schomerus, ``{Combinatorial
  quantization of the Hamiltonian Chern-Simons theory},''
  \href{http://dx.doi.org/10.1007/BF02099431}{{\em Commun. Math. Phys.}
  {\bfseries 172} (1995) 317--358},
\href{http://arxiv.org/abs/hep-th/9403066}{{\ttfamily arXiv:hep-th/9403066
  [hep-th]}}.

\bibitem{Alekseev:1994au}
A.~{\relax Yu}. Alekseev, H.~Grosse, and V.~Schomerus, ``{Combinatorial
  quantization of the Hamiltonian Chern-Simons theory II},''
  \href{http://dx.doi.org/10.1007/BF02101528}{{\em Commun. Math. Phys.}
  {\bfseries 174} (1996) 561--604},
\href{http://arxiv.org/abs/hep-th/9408097}{{\ttfamily arXiv:hep-th/9408097
  [hep-th]}}.

\bibitem{Meusburger:2003hc}
C.~Meusburger and B.~J. Schroers, ``{The quantisation of Poisson structures
  arising in Chern-Simons theory with gauge group $G \ltimes
  \mathfrak{g}^*$},'' \href{http://dx.doi.org/10.4310/ATMP.2003.v7.n6.a3}{{\em
  Adv. Theor. Math. Phys.} {\bfseries 7} no.~6, (2003) 1003--1043},
\href{http://arxiv.org/abs/hep-th/0310218}{{\ttfamily arXiv:hep-th/0310218
  [hep-th]}}.

\bibitem{Meusburger:2005mg}
C.~Meusburger and B.~J. Schroers, ``{Phase space structure of Chern-Simons
  theory with a non-standard puncture},''
  \href{http://dx.doi.org/10.1016/j.nuclphysb.2006.01.014}{{\em Nucl. Phys.}
  {\bfseries B738} (2006) 425--456},
\href{http://arxiv.org/abs/hep-th/0505143}{{\ttfamily arXiv:hep-th/0505143
  [hep-th]}}.

\bibitem{Meusburger:2003ta}
C.~Meusburger and B.~J. Schroers, ``{Poisson structure and symmetry in the
  Chern-Simons formulation of (2+1)-dimensional gravity},''
  \href{http://dx.doi.org/10.1088/0264-9381/20/11/318}{{\em Class. Quant.
  Grav.} {\bfseries 20} (2003) 2193--2234},
\href{http://arxiv.org/abs/gr-qc/0301108}{{\ttfamily arXiv:gr-qc/0301108
  [gr-qc]}}.

\bibitem{Meusburger:2008dc}
C.~Meusburger and B.~J. Schroers, ``{Generalised Chern-Simons actions for 3d
  gravity and kappa-Poincare symmetry},''
  \href{http://dx.doi.org/10.1016/j.nuclphysb.2008.06.023}{{\em Nucl. Phys.}
  {\bfseries B806} (2009) 462--488},
\href{http://arxiv.org/abs/0805.3318}{{\ttfamily arXiv:0805.3318 [gr-qc]}}.

\bibitem{Freidel:2011ue}
L.~Freidel, M.~Geiller, and J.~Ziprick, ``{Continuous formulation of the Loop
  Quantum Gravity phase space},''
  \href{http://dx.doi.org/10.1088/0264-9381/30/8/085013}{{\em Class. Quant.
  Grav.} {\bfseries 30} (2013) 085013},
\href{http://arxiv.org/abs/1110.4833}{{\ttfamily arXiv:1110.4833 [gr-qc]}}.

\bibitem{Freidel:2013bfa}
L.~Freidel and J.~Ziprick, ``{Spinning geometry = Twisted geometry},''
  \href{http://dx.doi.org/10.1088/0264-9381/31/4/045007}{{\em Class. Quant.
  Grav.} {\bfseries 31} no.~4, (2014) 045007},
\href{http://arxiv.org/abs/1308.0040}{{\ttfamily arXiv:1308.0040 [gr-qc]}}.

\bibitem{Donnelly:2016auv}
W.~Donnelly and L.~Freidel, ``{Local subsystems in gauge theory and gravity},''
  \href{http://dx.doi.org/10.1007/JHEP09(2016)102}{{\em JHEP} {\bfseries 09}
  (2016) 102},
\href{http://arxiv.org/abs/1601.04744}{{\ttfamily arXiv:1601.04744 [hep-th]}}.

\bibitem{Geiller:2017xad}
M.~Geiller, ``{Edge modes and corner ambiguities in 3d Chern-Simons theory and
  gravity},'' \href{http://dx.doi.org/10.1016/j.nuclphysb.2017.09.010}{{\em
  Nucl. Phys.} {\bfseries B924} (2017) 312--365},
\href{http://arxiv.org/abs/1703.04748}{{\ttfamily arXiv:1703.04748 [gr-qc]}}.

\bibitem{Geiller:2017whh}
M.~Geiller, ``{Lorentz-diffeomorphism edge modes in 3d gravity},''
  \href{http://dx.doi.org/10.1007/JHEP02(2018)029}{{\em JHEP} {\bfseries 02}
  (2018) 029},
\href{http://arxiv.org/abs/1712.05269}{{\ttfamily arXiv:1712.05269 [gr-qc]}}.

\bibitem{Rovelli:2013fga}
C.~Rovelli, ``{Why Gauge?},''
  \href{http://dx.doi.org/10.1007/s10701-013-9768-7}{{\em Found. Phys.}
  {\bfseries 44} no.~1, (2014) 91--104},
\href{http://arxiv.org/abs/1308.5599}{{\ttfamily arXiv:1308.5599 [hep-th]}}.

\bibitem{Freidel:2015gpa}
L.~Freidel and A.~Perez, ``{Quantum gravity at the corner},''
\href{http://arxiv.org/abs/1507.02573}{{\ttfamily arXiv:1507.02573 [gr-qc]}}.

\bibitem{Freidel:2016bxd}
L.~Freidel, A.~Perez, and D.~Pranzetti, ``{Loop gravity string},''
  \href{http://dx.doi.org/10.1103/PhysRevD.95.106002}{{\em Phys. Rev.}
  {\bfseries D95} no.~10, (2017) 106002},
\href{http://arxiv.org/abs/1611.03668}{{\ttfamily arXiv:1611.03668 [gr-qc]}}.

\bibitem{Freidel:2018pvm}
L.~Freidel and E.~R. Livine, ``{Bubble Networks: Framed Discrete Geometry for
  Quantum Gravity},''
\href{http://arxiv.org/abs/1810.09364}{{\ttfamily arXiv:1810.09364 [gr-qc]}}.

\bibitem{Kirillov}
A.~Kirillov, {\em {Lectures on the Orbit Method}}.
\newblock American Mathematical Society, 2004.

\bibitem{Rempel:2015foa}
T.~Rempel and L.~Freidel, ``{Interaction Vertex for Classical Spinning
  Particles},'' \href{http://dx.doi.org/10.1103/PhysRevD.94.044011}{{\em Phys.
  Rev.} {\bfseries D94} no.~4, (2016) 044011},
\href{http://arxiv.org/abs/1507.05826}{{\ttfamily arXiv:1507.05826 [hep-th]}}.

\bibitem{Teleparallel}
M.~Dupuis, F.~Girelli, A.~Osumanu, and W.~Wieland, ``{First-order formulation
  of teleparallel gravity and dual loop gravity},''
\href{http://arxiv.org/abs/1906.02801}{{\ttfamily arXiv:1906.02801 [gr-qc]}}.

\bibitem{barak4d}
B.~Shoshany, ``Spin networks and cosmic strings in 3+1 dimensions,''
  \href{http://dx.doi.org/10.1088/1361-6382/ab778e}{{\em Classical and Quantum
  Gravity} {\bfseries 37} no.~8, (Mar, 2020) 085019},
  \href{http://arxiv.org/abs/1911.07837}{{\ttfamily arXiv:1911.07837 [gr-qc]}}.

\end{thebibliography}\endgroup

\end{document}